\documentclass[aop]{imsart}
\usepackage{multicol}
\usepackage{enumerate}
\usepackage[dvipsnames]{xcolor}
\usepackage{makeidx}
\makeindex

\usepackage{savesym,amsmath,ulem,cancel}\def\em{\it}
\savesymbol{subequations}
\usepackage{cases}
\usepackage{ifthen,bm}
\usepackage{dsfont}
\usepackage{pgf}
\usepackage{latexsym, amsfonts, amssymb}
\usepackage{alltt,latexsym, verbatim}
\usepackage{float}
\usepackage{amsfonts}
\usepackage{xspace}
\usepackage{chicago}

\newtheorem{theorem}{Theorem}[section]
\newtheorem{lem}{Lemma}[section]
\newtheorem{pro}{Proposition}[section]
\newtheorem{cor}{Corollary}[section]
\newtheorem{conj}{Conjecture}[section]

\newtheorem{rem}{Remark}[section]
\newtheorem{com}{Comments}[section]
\newtheorem{ex}{Example}[section]
\newtheorem{defi}{Definition}[section]
\newtheorem{hyp}{Assumption}[section]

\numberwithin{equation}{section}

\newcommand{\bt}{\begin{theorem}}\newcommand{\et}{\end{theorem}}
\newcommand{\bl}{\begin{lem}}\newcommand{\el}{\end{lem}}
\newcommand{\bp}{\begin{pro}}\newcommand{\ep}{\end{pro}}
\newcommand{\bcor}{\begin{cor}}\newcommand{\ecor}{\end{cor}}
\newcommand{\bconj}{\begin{conj}}\newcommand{\econj}{\end{conj}}

\newcommand{\bd}{\begin{defi} \rm }\newcommand{\ed}{\end{defi} }
\newcommand{\brem }{\begin{rem} \rm }\newcommand{\erem }{\end{rem}}
\newcommand{\bcom}{\begin{com} \rm }\newcommand{\ecom }{\end{com}}
\newcommand{\brems }{\begin{rem} \rm }\newcommand{\erems }{\end{rem}}
\newcommand{\bex}{\begin{ex} \rm }\newcommand{\eex}{\end{ex}}
\newcommand{\bhyp}{\begin{hyp} \rm }\newcommand{\ehyp}{\end{hyp}}

\def\proof{\noindent \textbf{\emph{\textbf{Proof}.$\qqq$}}}
\def\finproof {\hfill $\Box$ \vskip 5 pt }

\def \Int{\displaystyle\int}

\def \be{\begin{eqnarray}}
\def \ee{\end{eqnarray}}
\def \b*{\begin{eqnarray*}}
\def \e*{\end{eqnarray*}}

\def \[{[\,\!\![}
\def \]{]\,\!\!]}

\def \1{{\bf 1}}

\def \proof{{\noindent \bf Proof. }}

\def\F{{\cal T}}\def\F{{\cal G}}\def\F{{\cal F}}

\def\ttt{t \in [0,\Ts]}
\newcommand{\bea}{\begin{eqnarray*}}
\newcommand{\eea}{\end{eqnarray*}}
\newcommand{\beqa}{\begin{eqnarray}}
\newcommand{\eeqa}{\end{eqnarray}}

\def\N{N}\def\N{{\mathbb N}}
\def\R{{\mathbb R}}
\def\proof{\noindent {\it Proof. $\, $}}
\def\finproof {\hfill $\Box$ \vskip 5 pt }
\def\I{\mathds{1}}
\def\sp{\,,\ \,}
\def\l{\label}

\def\bal{\begin{aligned}}
\def\eal{\end{aligned}}

\def\ttt{{t \in [0,\Ts]}}

\def\mym{m}

\def\xitau2{\xi_{(\thetau)}}\def\xitau2{\xi}
\def\xiitau2{\xi^i_{(\thetau)}}\def\xiitau2{\xi^i}
\def\xintau2{\tilde{\xi}_{(\thetau)}}\def\xintau2{\tilde{\xi}}

\def\Ltau2{\tilde{\xi}(\tau_{\mym},\tau_1)}
\def\chiitau2{P^i_{\tau}}

\def\cI{\mathcal{I}}

\newcommand{\beq}{\begin{eqnarray*}}
\newcommand{\eeq}{\end{eqnarray*}}

\def\mym{-}

\def\theg{g}

\def\tilde{\widetilde}
\def\cadlag{c\`adl\`ag }

\def\emph{}

\def\gg{{\mathbb G}}\def\gg{{\mathbb F}}

\def\ff{\mathbb{F}}

\def\ff{\mathbb{F}^*}
\def\F{{\cal F}^*}

\def\gg{{\cal G}}\def\gg{\mathbb{G}}

\def\ff{{\cal F}}\def\ff{\mathbb{F}}
\def\F{{\cal F}}

\def\E{\mathbb{E}}

\def\Rf{\mathfrak{R}}\def\Rf{\mathfrak{r}}
\def\Rf{\mathfrak{R}}\def\Rf{\mathfrak{r}}\def\Rf{\mathfrak{r}^b}

\def\thisR{R}\def\thisR{\rho}\def\thisR{R}\def\thisR{\rho^b}

\def\thetau{\vartheta}
\def\bthet a{\theta}

\def\Ts{\bar{T}}\def\Ts{\mathcal{T}}\def\Ts{T}

\def\Ts{\mathcal{T}}\def\Ts{\mathsf{T}}\def\Ts{\bar{T}}\def\Ts{T}

\def\ind{\mathds{1}}
\def\e{z}\def\e{e}

\def\theU{U}

\newcommand{\beql}[1]{\beqa\label{#1}\begin{aligned}}
\newcommand{\eeql}{\eal\eeqa}
\newcommand{\bel}{\bea\bal}
\newcommand{\eel}{\end{aligned}\eea}

\def\eee{\end{document}}
\def\ind {1\!\!1}\def\ind{\mathds{1}}
\def\I{\ind}

\def\F{{\cal F}}

\def\ff{{\mathbb F}}

\def\gg{{\mathbb G}}

\def\P{\mathbb P}
 \def\Q{\mathbb Q}

\def\EP{{\mathbb E}^{\mathbb P}}

\def\EQ{{\mathbb E}^{{\mathbb Q}}}

\def\E {{\mathbb E} }

\def\R{{\mathbb R}}
\def\N{{\mathbb N}}

\def\finproof {\hfill $\Box$ \vskip 5 pt }

\def\bal{\begin{aligned}}
\def\eal{\end{aligned}}

\def\finproof {\hfill $\Box$ \vskip 5 pt }

\def\Proba{\Proba}\def\Proba{\mathbb{Q}}

\def\sp{,\ \, }

\def\ttt{{t \in [0,T]}}

\def\xitau2{\xi_{(\thetau)}}
\def\xiitau2{\xi^i_{(\thetau)}}
\def\xintau2{\tilde{\xi}_{(\thetau)}}
\def\Ltau2{\tilde{\xi}(\tau_0,\tau_1)}
\def\chiitau2{P^i_{\tau}}

\def\thetau{\tau}

\def\cD{\mathcal{D}}
\def\cE{\mathcal{E}}
\def\cF{\mathcal{F}}

\def\cI{\mathcal{I}}

\def\cM{\mathcal{M}}

\def\cQ{\mathcal{Q}}

\def\cS{\mathcal{S}}

\def\bQ{\mathbb{Q}}

\def\l{\label}
\def\emph{}

\def\thisR{\thisR }

\def\eee{\end{document}}
\def\textsl{}

\def\ttt{{t \in [0,\Ts]}}

\def\Ts{\mathcal{T}}\def\Ts{\mathsf{T}}\def\Ts{\bar{T}}\def\Ts{T}

\def\proof{\noindent {\it {\textbf{Proof}}}.$\;\,$}
\def\finproof {\hfill $\Box$ \vskip 5 pt }\def\finproof{\rule{4pt}{6pt}}
\def\finproofs{\finproof\medskip}

\def \Int{\displaystyle\int}

\def \be{\begin{eqnarray}}
\def \ee{\end{eqnarray}}
\def \b*{\begin{eqnarray*}}
\def \e*{\end{eqnarray*}}

\def \[{[\,\!\![}
\def \]{]\,\!\!]}

\def \1{{\bf 1}}

\def\F{{\cal T}}\def\F{{\cal G}}\def\F{{\cal F}}

\def\ttt{t \in [0,\Ts]}

\def\N{N}\def\N{{\mathbb N}}
\def\R{{\mathbb R}}

\def\I{\mathds{1}}

\def\bal{\begin{aligned}}
\def\eal{\end{aligned}}

\def\mym{m}

\def\xitau2{\xi_{(\thetau)}}\def\xitau2{\xi}
\def\xiitau2{\xi^i_{(\thetau)}}\def\xiitau2{\xi^i}
\def\xintau2{\tilde{\xi}_{(\thetau)}}\def\xintau2{\tilde{\xi}}

\def\Ltau2{\tilde{\xi}(\tau_{\mym},\tau_1)}
\def\chiitau2{P^i_{\tau}}

\def\cI{\mathcal{I}}

\def\b{\textcolor{blue}}%\def\s{\textcolor{blue}}

\def\mym{-}

\def\theg{g}

\def\tilde{\widetilde}
\def\cadlag{c\`adl\`ag }

\def\emph{}

\def\usual{satisfying the usual conditions\xspace}
\def\representative{reduction\xspace}

\def\ff{\mathbb{F}}

\def\ff{\mathbb{F}^{\star}}
\def\F{{\cal F}^{\star}}

\def\gg{{\cal G}}\def\gg{\mathbb{G}}

\def\ff{{\cal F}}\def\ff{\mathbb{F}}
\def\F{{\cal F}}

\def\E{\mathbb{E}}

\def\Rf{\mathfrak{R}}\def\Rf{\Rf}\def\Rf{R_{f}}\def\Rf{\bar{R}_{b}}

\def\thisR{R}\def\thisR{\thisR }\def\thisR{R_b}

\def\thetau{\vartheta}

\def\Ts{\bar{T}}\def\Ts{\mathcal{T}}\def\Ts{T}

\def\Ts{\mathcal{T}}\def\Ts{\mathsf{T}}\def\Ts{\bar{T}}\def\Ts{T}
\def\equiv{=}
\def\paragraph{\noindent\textbf}

\def\thec{c}

\def\qqq{\quad\quad\quad}

\def\thec{c}

\def\eee{\bibliographystyle{chicago}\bibliography{B}\end{document}} \def\eee{\end{document}}
\def\bthet{\boldsymbol\theta}\def\bthet{\mathbf{k}}

\def\theeta{\hat\tau}\def\theeta{\hat\eta}\def\theeta{\zeta}\def\theeta{\eta}

\def\theQ{P}
\def\thisQ{Q}
\def\thisR{R}

\def\thenu{\tau}\def\thenu{\nu}\def\thenu{\rho}\def\thenu{x}\def\thenu{\mathfrak{p}}\def\thenu{\mathsf{p}}

\def\qr#1{\eqref{#1}}

\def\sr#1{Sect.~\ref{#1}}

\newcommand{\indi}[1]{\I_{\{{#1}\}}}
\newcommand{\bethe}{\bt}
\newcommand{\ethe}{\et}

\newcommand{\iend}{\end{itemize}}
\newcommand{\ok}{\rule{4pt}{6pt}}\renewcommand{\ok}{\finproof}
\newcommand{\desb}{\begin{description}}
\newcommand{\dese}{\end{description}}

\newcommand{\dcb}{\begin{array}{lll}}
\newcommand{\dce}{\end{array}}
\newcommand{\ebe}{\begin{enumerate}[1)]}
\newcommand{\dbe}{\end{enumerate}}

\newcommand{\ibegin}{\begin{itemize}\setlength{\baselineskip}{19pt}\setlength{\parskip}{7pt}}

\def\cro#1{\langle #1\rangle}

\makeatletter
\newenvironment{systeme*}{\arraycolsep=1.4pt\left\{\begin{array}{l}}{\end{array}\right.}

\makeatother

\def\cro#1{{{\boldsymbol\langle} #1{\boldsymbol\rangle}}}
\def\croc#1{{\langle#1\rangle}}

\def\ftime{\theta}

\def\thisg{g}

\def\wG{\widehat{\mathtt{G}}}\def\wG{\widehat{G}}
\def\thisU{U}

\def\ttG{\mathtt{G}}
\def\theU{U}\def\theU{\thesigma}\def\theU{\ftime }\def\theU{\tau}

\def\tSigma{\Sigma}\def\tSigma{\ttS}
\def\theM{M}\def\theM{R}

\def\theV{V}\def\theV{\ftime}\def\theV{\nu}\def\theV{\sigma}

\def\then{n}
\def\thisY{Y}
\def\fo{\mathbb{F}.o}\def\fo{o}
\def\fp{\mathbb{F}.p}\def\fp{p}
\def\thisM{R}\def\thisM{M}

\def\thisN{N}\def\thisN{P}

\def\monp{\mathsf{p}}\def\monp{p}
\def\monq{\mathsf{q}}\def\monq{q}

\def\bp{\bar{p}}\def\bp{{\bf p}}

\def\qr#1{\eqref{#1}}

\def\sr#1{Sect.~\ref{#1}}\def\sr#1{Section~\ref{#1}}

\def\Jt{J^\star}\def\Jt{\mathsf{J}^\star}
\def\gr{[\![}\def\gr{[}
\def\rg{]\!]}\def\rg{]}

\def\thatF{F}\def\thatF{\mathtt{G}}

\def\thisGp{\theF}

\def\subset{\subseteq}

\def\cM{Q}\def\cM{\mathtt{M}}\def\cM{\mathtt{Q}}
\def\Alpha{\mathtt{D}}
\def\pAlph{\mathtt{A}}
\def\ttS{\mathtt{S}}
\def\pSigma{\hat{\tSigma}}\def\pSigma{{^{p}}\Sigma}\def\pSigma{{^{p}}\!\ttS}

\def\mnqs{\mathcal{E}(\frac{1}{\pSigma }\centerdot \cM)}
\def\monqs{\mathcal{E}(\ind_{\{\pSigma >0\}}\frac{1}{\pSigma }\centerdot \cM)}
\def\In{\cup_n[0,\varsigma_n]}\def\In{\{\ttS_{-}>0\}}
\def\Int{\cup_n[0,T\wedge\varsigma_n]}\def\Int{\In}

\def\therho{\rho}\def\therho{\rho}\def\therho{\mathfrak{q}}\def\therho{\mathsf{q}}
\def\ttJ{{\ind_{[0,\theta)}}}\def\ttJ{\mathtt{J}}

\newtheorem{question}{Question}[section]
\newcommand {\bquestion}{\begin{question}}
\newcommand {\equestion}{\end{question}}

\def\mathtt{\mathsf}
\def\ttH{\mathtt{H}}\def\ttH{{\ind_{[\theta,\infty)}}}
\def\Jt{\ind_{(0,\theta]}}

\def\girs#1{\mathfrak{Q}(#1)}

\def\cro#1{{\langle#1\rangle}}
\def\thenu{\bar{p}}
\def\therho{\bar{q}}
\def\thatF{G}
\def\cI{I}

\def\theV{\tau'}

\usepackage[margin=1in]{geometry}
\begin{document}
\begin{frontmatter}

\title{Invariance Times\protect\thanksref{T1}}
\runtitle{Invariance Times}
% and Probability Measures
\thankstext{T1}{This research benefited from the support of the ``Chair Markets in Transition'',
%under the aegis of Louis Bachelier laboratory, a joint initiative of \'Ecole polytechnique, Universit\'e d'\'Evry Val d'Essonne and 
F\'ed\'eration Bancaire Fran\c caise, and of the ANR project 11-LABX-0019.
The authors are grateful to the associate editor and both referees for insightful
%stimulating 
comments. They also
%would also like to 
thank Monique Jeanblanc for her remarks on a preliminary version of the manuscript.}

\begin{aug}
 \author{\fnms{St\'{e}phane} \snm{ Cr\'{e}pey}\corref{}\ead[label=e1]{stephane.crepey@univ-evry.fr}} \and
 \author{\fnms{Shiqi} \snm{Song}\ead[label=e2]{shiqi.song@univ-evry.fr}}

 \runauthor{S. Cr\'{e}pey and S. Song}

 \affiliation{Universit\'e d'\'Evry Val d'Essonne \\
Laboratoire de Math\'ematiques et Mod\'elisation d'\'Evry and UMR CNRS 8071\\91037 \'Evry Cedex, France}

 \address{S. Cr\'{e}pey and S. Song\\Universit\'e d'\'Evry Val d'Essonne \\
Laboratoire de Math\'ematiques et Mod\'elisation d'\'Evry and UMR CNRS 8071\\91037 \'Evry Cedex, France\\ 
     \printead{e1,e2}}

\end{aug}

\begin{abstract}
On a probability space $(\Omega,\mathcal{A},\mathbb{Q})$ we consider two filtrations 
%$\mathbb{F}$ and $\mathbb{G}$ with *
$\mathbb{F}\subset \mathbb{G}$ and a $\mathbb{G}$ stopping time $\theta$ such that the $\mathbb{G}$ predictable processes coincide with $\mathbb{F}$ predictable processes on $(0,\theta]$. In this setup it is well-known that, for any $\mathbb{F}$ semimartingale $X$, 
the process $X^{\theta-}$ ($X$ stopped ``right before $\theta$'') is a $\mathbb{G}$ semimartingale.
Given a positive constant $T$, 
we call $\theta$ an invariance time if there exists a probability measure $\mathbb{P}$ equivalent to $\mathbb{Q}$ on $\mathcal{F}_T$ such that, for any $(\mathbb{F},\mathbb{P})$ local martingale $X$, $X^{\theta-}$ is a $(\mathbb{G},\mathbb{Q})$ local martingale. 
We characterize invariance times in terms of the $(\mathbb{F},\mathbb{Q})$ Az\'ema supermartingale of $\theta$ and we derive a mild and tractable invariance time sufficiency condition. We discuss invariance times in mathematical finance and BSDE applications.
%In this case, we call $\theta$ an invariance time and $\mathbb{P}$ an invariance measure.
%In other words, we investigate the condition under which the Jeulin-Yor enlargement formula can be compensated by the Girsanov measure change formula. 
%There exists a long tradition to connect the Jeulin-Yor with the Girsanov formula.
%The condition $(A)$ is a new problem in this literature.  
\end{abstract}



\end{frontmatter}

\section{Introduction}\label{s:intro}

On a probability space $(\Omega,\mathcal{A},\mathbb{Q})$ we consider two filtrations 
$\mathbb{F}\subset \mathbb{G}$ of sub-$\sigma$-algebras of $\mathcal{A}$ and a $\mathbb{G}$ stopping time $\theta$. We suppose the condition (B) that the $\mathbb{G}$ predictable processes coincide with the $\mathbb{F}$ predictable processes on $(0,\theta]$. In this setup it is well-known that, for any $(\mathbb{F},\mathbb{Q})$ local martingale $X$, 
the process $X^{\theta-}$ ($X$ stopped ``right before $\theta$'') is a $(\mathbb{G},\mathbb{Q})$ special semimartingale, whose drift part can be deduced from the \citeN{JeulinYor78} formula. 
In this paper, given a positive constant $T$, we study the condition (A) that there exists a probability measure $\mathbb{P}$ equivalent to $\mathbb{Q}$ on $\mathcal{F}_T$ such that, for any $(\mathbb{F},\mathbb{P})$ local martingale $X$, the process $X^{\theta-}$ is a $(\mathbb{G},\mathbb{Q})$ local martingale. If such a probability measure $\mathbb{P}$ exists, we call $\theta$ an invariance time and $\mathbb{P}$ an invariance measure.  

One way of dealing with the above martingale invariance 
%property preservation 
problem (we will propose two approaches to it in the paper)
is to investigate the
%Thus, our consists in finding 
conditions under which the enlargement of filtration Jeulin-Yor formula can be compensated by the Girsanov formula of an equivalent change of probability measure. 
%We argue in the paper that, from this point of view, the ``right'' stopping operator to consider is indeed 
%%${\cdot}^{\theta-}$ rather than ${\cdot}^{\theta}.$}
%%object to consider is indeed 
%$X^{\theta-},$ rather than $X^{\theta}.$
%%alt..nonstandard
The Jeulin-Yor formula, initially proved by projection computations, has long been supposed to be affiliated with the Girsanov formula. \citeN{Yoeurp1985} gives a formal proof that the Jeulin-Yor formula can be obtained by the so-called generalized Girsanov formula (between non absolutely continuous probability measures). Actually, according to 
\citeANP{Song87} (\citeyearNP{Song87}, \citeyearNP{Song13}), 
not only the Jeulin-Yor formula, but
most of the formulas in enlargement of filtration, can be retrieved by 
the generalized Girsanov formula.
However, the methodologies of the above-mentioned papers
are not applicable to the study of the condition (A), notably because they operate in an enlarged probability space and with a non absolutely continuous probability measure. Hence, a new approach is required for our problem.

Besides their theoretical interest, invariance times are closely linked to questions raised from recent mathematical finance developments. Since the occurrence of the global financial crisis, the theoretical research on random times has seen a vigorous revival in relation with the field of default risk modeling, viewed under different angles: no arbitrage, pricing and hedging of credit derivatives, counterparty risk, insider trading, etc..
%New classes of random times have emerged.
See \citeN{NikeghbaliYor05}, 
\citeANP{JeanblancSong11b} (\citeyearNP{JeanblancSong11b}, \citeyearNP{JeanblancSong12}), 
\citeANP{AksamitChoulliDengJeanblanc13} (\citeyearNP{AksamitChoulliDengJeanblanc13}, \citeyearNP{AksamitChoulliDengJeanblanc14}),
 \citeN{Kardaras14}, \citeN{Song16a},
%\citeANP{Song13b} (\citeyearNP{Song13b}, \citeyearNP{Song16a}), 
\citeN{LiRutkowski14},
%\citeN{JeanblancSong11b}, 
\citeN{FontanaJeanblancSong14} and \citeN{AcciaioFontanaKardaras},
among others. 
Invariance times yield a class of models of default times
for which intensity-based defaultable asset pricing formulas can be obtained beyond the classical but restrictive ``immersion'' setup (see \citeANP{Crepey13a1} (\citeyearNP{Crepey13a1}, 
\citeyearNP{CrepeySong15})).
%, used for counterparty risk applications
%in \citeANP{Crepey13a1} (\citeyearNP{Crepey13a1}, 
%\citeyearNP{CrepeySong15}), , where conditional independence between the default time and the market reference (or underlying exposure) filtration is assumed.

%are 
%simplifying the pricing equations of counterparty risk,
%beyond the classical but restrictive immersion setup.
%The results in the present paper provide the exact conditions under which the corresponding
%reduced-form approach is possible.
%%\b{In particular, they hold in the wrong-way risk setups of \citeN{CrepeySong15}.}
%%%concrete setups considered in which is exploited for counterparty risk modeling in \citeN{CrepeySong15}.
%%%Note that Theorem \ref{casuffi} is not only essential to understand the random times of \shortciteN{CollinDufresneGoldsteinHugonnier2004} in relation with invariance times, it also shows that invariance times have a broad range of applicability, which is exploited for counterparty risk modeling in \citeN{CrepeySong15}.

Our main results are Theorem \ref{part1} and Theorems \ref{A2bis}-\ref{c:mart}, which give necessary and sufficient conditions, stated in terms of 
the Az\'ema supermartingale of $\theta$, for $\mathbb{P}$ to be an invariance measure and $\theta$ to be an invariance time, respectively. These theorems are established in
Sections \ref{firstproofofThmpart1} and \ref{firstproofofThm} by reduction, a methodology introduced in \citeN{Song16a} for transferring properties in $\mathbb{G}$ into properties in $\mathbb{F}$. 
However, the ensuing proofs 
%of Theorem \ref{part1} 
do not directly explain how a Girsanov drift can compensate a Jeulin-Yor drift. Given the importance of that matter for the purpose of this work, we provide in Appendix \ref{proofinGJY} an alternative proof of Theorem \ref{part1} based on that compensation.

The second main contribution of this work is Theorem \ref{sfcnd}.
%are Theorems \ref{sfcnd} and \ref{ncsfcnd}.
When applied to a defaultable asset, 
a basic no-arbitrage pricing formula 
explicitly involves the
default time $\theta$, 
whereas it is only the intensity of $\theta$ that can be retrieved, by calibration, from market data.
To tackle this issue, 
\citeN{DuffieSchroderSkiadas96}
have established a defaultable asset pricing formula stated in terms of the intensity process (assumed to exist) of $\theta$. From a financial interpretation point of view, their intensity-based formula also shows that credit risk can
be valued as a shift in interest rates. However, the tractability of this formula is subject to a technical no-jump condition at time $\theta$. 
In a progressive enlargement of filtration setup satisfying a restrictive immersion assumption,
this no-jump condition is satisfied. 
Alternatively, \shortciteN{CollinDufresneGoldsteinHugonnier2004}
 have proposed a reinterpretation of the 
\shortciteN{DuffieSchroderSkiadas96}'s formula exempt from no-jump condition, under the so-called
survival measure.
%, following the terminology introduced for a similar object in
%\citeANP{Schoenbucher99} (\citeyearNP{Schoenbucher99}, 
%\citeyearNP{Schoenbucher04}).
As explained in Section \ref{s:applic}, Theorem \ref{sfcnd}
provides a connection between the two approaches 
 by showing that, under mild conditions, the restriction to $\cF_T$ 
of the survival measure yields an invariance measure.
Theorem \ref{sfcnd} also provides a tractable invariance time sufficiency condition.

Several additional results are useful for the establishment of Theorem \ref{sfcnd}. In particular, according to Theorems \ref{A2bis}-\ref{c:mart}, the positivity of a
tentative $\Q$ to $\P$ measure change density on $[0,T]$,
as well as the true martingale property on $[0,T]$ of the martingale part 
$\mathcal{Q}$ in the predictable multiplicative decomposition of the Az\'ema supermartingale $\ttS$ of $\theta$, are key conditions for $\theta$ to be an invariance time. In Theorem \ref{t:cerise}, we characterize
the above positivity
 by the predictability of the time of first zero of $\ttS$ on the time interval $[0,T]$. 
%Theorem \ref{sfcnd} provides a mild sufficiency condition 
% for the true martingale property of $\mathcal{Q}$.
Theorem 
%\ref{sfcnd} and 
\ref{ncsfcnd} 
provides
% a mild sufficiency condition 
a necessary and sufficient condition for the true martingale property of $\mathcal{Q}$.

Complementary results are provided for a better 
understanding 
%appreciation 
of invariance times. 
The starting point of this paper is the condition (B), which is discussed in \sr{cB}.
As recalled in \sr{s:bsdecr}, the condition (A) 
%and the conjecture of the characterizations of Theorems \ref{A2bis}-\ref{c:mart} 
appears naturally in the study of counterparty risk.
Theorem \ref{c:pq} 
characterizes in terms of $\ttS$
the local martingales under the invariance measure $\mathbb{P}.$ 
This characterization plays a key role in \citeN{Crepey13a1}.
%It is also used for establishing the bijection of Theorem \ref{invarTH} between $(\gg,\Q)$ local martingales stopped before $\theta$ and $(\ff,\P)$ local martingales (assuming $\theta$ totally inaccessible and $\ttS_T$ positive).
Section \ref{diffsituas} studies invariance times in several situations, comparing them with the so-called pseudo-stopping times, the stopping times in \shortciteN{CollinDufresneGoldsteinHugonnier2004} and showing how invariance times are involved in a variety of applications.

\subsection{Basic Notation and Terminology}\label{ss:san}

The real line, half-line and the nonnegative integers are respectively denoted by $\R,$ $\R_+$ {and $\mathbb{N};$ 
$\mathcal{B}({\R})$ and $\mathcal{B}({\R}_+)$ are the Borel $\sigma$ fields on $\R$ and $\R_+$;
\index{l@$\boldsymbol\lambda$}$\boldsymbol\lambda$ is the Lebesgue measure on $\R_+$.}
Unless otherwise stated, {a function (or process) is real-valued}; order relationships between random variables (respectively processes) are meant almost surely (respectively in the indistinguishable sense); a time interval is random. 
We do not explicitly mention the domain of definition of a function (or process)
when it is implied by the measurability, e.g.~we write
``a $\mathcal{B} ({\R})$ measurable function $h$ (or $h(x)$)'' rather than
``a $\mathcal{B} ({\R})$ measurable function $h$ defined on ${\R}$''.
For a function $h(\omega,x)$ defined on a product space $\Omega\times E$, we usually write $h(x)$ without $\omega$ (or $h_t$ in the case of a stochastic process). 

We employ the tools and terminology of the general theory of processes and of filtrations as given in the books by \citeN{DM75}
and \citeN{HeWangYan92}. Footnotes are used for referring to comparatively standard results.  
For any semimartingale $X$ and any predictable process $L$ integrable with respect to $X$, the corresponding stochastic integral is denoted by $\int_0^{\cdot} L_t dX_t=\int_{(0,\cdot]}L_t dX_t=L\centerdot X$, with the usual precedence convention \index{é@$\centerdot$}$KL\centerdot X=(KL)\centerdot X$ if $K$ is another predictable process such that $KL$ is integrable with respect to $X$. The stochastic exponential of a semimartingale $X$ is denoted by $\mathcal{E}(X)$ (in particular, $\mathcal{E}(X)_0=1$). We denote by $\bar{X}=\frac{1}{X_-}\centerdot X$ (whenever it exists) the so-called stochastic logarithm of a positive semimartingale $X$, such that $X=X_0\mathcal{E}(\bar{X})$. 
%Define X:
%X_t = t - [t], if t is not an integer
%X_t = 1, if t is an integer
%X est une semimartingale
%Mais, X n'a pas de log.
Given semimartingales $X$ and $X'$, the bracket process $[X,X']$ and its predictable counterpart $\cro{X,X'}$ are defined as in \shortciteN[Definition 8.2]{HeWangYan92}. In particular, we use the convention $[X,X']_0=0$.

For any c\`adl\`ag process $X$, for any random time $\tau $ (nonnegative random variable), $\Delta_\tau X$ represents the jump of $X$ at $\tau .$
Following \citeN{DM75} and \shortciteN{HeWangYan92},
we use the convention that \index{é@${\cdot}_{0-}$}$X_{0-}=X_0$ (hence $\Delta_0 X=0$) 
and we write \index{é@${\cdot}^{\tau -}$}$X^{\tau}$ and \index{é@${\cdot}^{\tau -}$}$X^{\tau -}$ for the process $X$ stopped at $\tau$ and before $\tau$, respectively, i.e.
\beql{right}
X^{\tau }=X\ind_{[0,\tau )} +X_{\tau} \ind_{[\tau ,+\infty)} \sp X^{\tau -}=X\ind_{[0,\tau )} +X_{\tau -} \ind_{[\tau ,+\infty)}.
\eeql 
We call compensator of a stopping time $\tau$ the compensator of the process $\ind_{[\tau,\infty)}$.\footnote{See \shortciteN[Definition 5.21]{HeWangYan92} for 
the notion of compensator of a locally integrable nondecreasing process.}    We say that a stopping time $\tau$ is totally inaccessible (respectively, has an intensity) if
it is positive and if its
compensator is continuous (respectively, absolutely continuous) on $[0,\tau]$. \index{é@$\tau _{\cdot}$}
 Given a filtration $\mathbb{G}=(\mathcal{G}_t)_{t\in\mathbb{R}_+}$
and a $\mathbb{G}$ stopping time $\tau,$ 
 for $A$ in $\mathcal{G}_\tau $, we denote by
$
\tau _A
$ 
the $\mathbb{G}$ stopping time\footnote{Cf.~Theorem 3.9 in \shortciteN{HeWangYan92}.} $\ind_A
\tau + \ind_{A^c} \infty$. 

We work with semimartingales on a predictable set of interval type $\cI$ as defined in \shortciteN[Sect.~VIII.3]{HeWangYan92}. In particular, {$X$ is a local martingale on $\cI$ (respectively $Y=L\centerdot X$ on $\cI$) 
means that
\beql{e:intint}
X^{\tau_n} \mbox{ local martingale (respectively }
%$\cI$ 
Y^{\tau_n}=L\centerdot (X^{\tau_n}) 
\mbox{)}
\eeql 
holds for at least one, or equivalently any, nondecreasing sequence of stopping times such that $\cup [0,\tau_n]=\cI$.
The existence of such a sequence is ensured by \shortciteN[Theorem 8.18 3)]{HeWangYan92}. From a computational point of view, stochastic calculus on predictable intervals reduces to standard stochastic calculus on $\R_+$ for each of the processes stopped at $\tau_n.$ But a process $L$ can be integrable with respect to $X$ on $\cI$ without being locally integrable on $\R_+$ if 
the stochastic integrals $Y^{\tau_n}$ explode as $n\to\infty.$ }

Given a filtration $\mathbb{G}$ and a probability measure $\mathbb{Q},$ we denote by 
${\cal S}_{\cI}(\mathbb{G} ,\mathbb{Q})$ and $\mathcal{M}_{\cI}(\mathbb{G},\mathbb{Q})$ the respective sets of $(\mathbb{G},\mathbb{Q})$ semimartingales and local martingales on a predictable interval $\cI$ (or $\R_+ ,$ when no interval $I$ is mentioned in the notation). 
The $\mathbb{G}$ predictable and optional $\sigma$ fields are denoted by
$\mathcal{P}(\mathbb{G})$\index{p@$\mathcal{P}(\mathbb{F})$} and 
$\mathcal{O}(\mathbb{G})$\index{o@$\mathcal{O}(\mathbb{F})$}.

Throughout the paper, $\Omega$ is a space equipped with a $\sigma$ field \index{a@$\mathcal{A}$}$\mathcal{A}$, \index{q@$\mathbb{Q}$}$\mathbb{Q}$ is a probability measure on $\mathcal{A},$ $\mathbb{G}=(\mathcal{G}_t)_{t\in\mathbb{R}_+}$ is a filtration 
\index{g@$\mathbb{F}$} of sub $\sigma$ fields of $\mathcal{A}$,
\index{t@$\ftime$}$\ftime $ is a $\mathbb{G}$
stopping time and $\mathbb{F}=(\mathcal{F}_t)_{t\in\mathbb{R}_+}$ is
a subfiltration of $\mathbb{G}$.
Both filtrations $\mathbb{F}$ and $\mathbb{G}$ are supposed to satisfy the usual conditions.
Regarding the filtration $\ff,$ we have to deal with two probability measures $\mathbb{Q}$ and $\mathbb{P}$. Accordingly, letters 
of the family ``q'' and ``p'' are respectively used for $(\ff,\Q)$ and $(\ff,\P)$ local martingales.
By default $\mathbb{E}$ refers to the $\mathbb{Q}$ 
expectation, whereas the 
$\mathbb{P}$ 
expectation is denoted by \index{e@$\mathbb{E}^{\mathbb{P}}$}$\mathbb{E}^{\mathbb{P}}.$

\section{Preliminaries}
%\section{Preparatory Results}
%\section{Preliminary Results}

\subsection{The Condition (B)}\label{cB}

We consider the following:\\

\noindent\textbf{Condition (B)}. 
For any ${\gg}$ predictable process $L$, there exists an ${\ff}$ predictable process $L'$, called the $\ff$ predictable \representative\footnote{Also known as pre-default process of $L$ in the credit risk literature (see e.g.~\citeN{Bielecki2009}).} of $L$, such that $\Jt L = \Jt L' $.
%\\ 

%\noindent 
\brem\label{r:indis} 
The equality $\Jt L = \Jt L' $ is in the sense of indistinguishability. But, as $\mathbb{F}$ satisfies the usual conditions, we can find a version of $L'$ such that the equality holds everywhere.\footnote{Cf.~\shortciteN[Theorem 4.26)]{HeWangYan92} and Lemma \ref{optionalsplitting} 2).}
\erem

The condition (B) corresponds to a relaxation of the classical progressive enlargement of filtration setup, where the bigger filtration $\gg$ is given as
%coincides with 
the progressive enlargement by $\theta$ of a reference filtration $\ff.$ 
As compared with this classical case, the possibility to
use a bigger filtration $\gg$ in
%additional flexibility offered by 
the condition (B) 
%As recapped in \sr{s:dmo}, the additional flexibility offered by the condition (B)
is material, for instance, in the dynamic Marshall-Olkin copula model of \citeN{CrepeySong15} 
%to deal with a dynamic Marshall-Olkin copula model of counterparty risk on credit derivatives 
(see \sr{ss:dcm} below).

As an immediate consequence of the condition (B), we have:\footnote{Cf.~\shortciteN[Corollary 3.23 2)]{HeWangYan92}.}
$$
\{0<\ftime <\infty\}\cap\mathcal{G}_{\ftime -}=\{0<\ftime <\infty\}\cap\mathcal{F}_{\ftime -},
$$
where we recall\footnote{Cf.~\shortciteN[Definition 3.3, equation (3.3)]{HeWangYan92}.} that
$$
\mathcal{G}_{\ftime -}=\mathcal{G}_{0}\vee \sigma\left\{B \cap\{t<\ftime \}, B\in\mathcal{G}_{t},t\in\R_+ \right\}\sp \mathcal{F}_{\ftime -}=\mathcal{F}_{0}\vee \sigma\left\{A \cap\{t<\ftime \}, A\in\mathcal{F}_{t},t\in\R_+ \right\}.
$$
%and likewise for $\mathcal{F}_{\ftime -}$.
But we can say more. We introduce the right-continuous\footnote{And complete under our assumption that $\ff$ satisfies the usual conditions.}
filtration $\overline{\mathbb{F}}=(\overline{\mathcal{F}}_t)_{t\in\mathbb{R}_+}$, where
\begin{equation}\label{gbar}
\overline{\mathcal{F}}_t
=
\big\{B \in {\mathcal{A}}: \exists A \in\mathcal{F}_t, B \cap\{t<\ftime \}=A \cap\{t<\ftime \}\big\}
\end{equation}
(see \shortciteN[Chapitre XX, n$^\circ$75]{DellacherieMaisonneuveMeyer92}). 
\bl\label{c:b} 
$\mathbb{F},$ $\mathbb{G}$ and $\theta$ satisfy the condition {(B)} if and only if $\mathbb{G}$ is a subfiltration of $\overline{\mathbb{F}}.$ 
\el

\proof Suppose the condition {(B)}. For any $t\in\mathbb{R}_+$ and $B\in\mathcal{G}_{t}$, $\ind_{B}\ind_{(t,\infty)}$ is a $\mathbb{G}$ predictable process, having a bounded $\mathbb{F}$ predictable \representative $K$ such that $
\Jt \ind_{B}\ind_{(t,\infty)}=\Jt K\ind_{(t,\infty)}.
$
Then $\ind_{B}\ind_{\{t<s\leq \ftime \}}=K_s\ind_{\{t<s\leq \ftime \}}$, 
hence $$
\liminf_{s\downarrow t}\ind_{B}\ind_{\{t<s\leq \ftime \}}=\liminf_{s\downarrow t}K_s\ind_{\{t<s\leq \ftime \}}.
$$
But $\liminf_{s\downarrow t}\ind_{B}\ind_{\{t<s\leq \ftime \}}=\ind_{B}\ind_{\{t< \ftime \}}$ and $\liminf_{s\downarrow t}K_s\ind_{\{t<s\leq \ftime \}}=(\liminf_{s\downarrow t}K_s)\ind_{\{t<\ftime \}}$, which proves $B\in\overline{\mathcal{F}}_{t}$.
Conversely (cf.~Lemma 1 in \citeN{JeulinYor78}),
suppose that $\mathbb{G}$ is a subfiltration of $\overline{\mathbb{F}}$. For any $t>0$, for any $B \in\mathcal{G}_{t}$, let $A\in\mathcal{F}_{t}$ satisfy $B\cap\{t<\ftime \}=A\cap\{t<\ftime \}$, so that
$$
\Jt \ind_{B}\ind_{(t,\infty)}
=
\Jt \ind_{A}\ind_{(t,\infty)}.
$$
Note that $\ind_{A}\ind_{(t,\infty)}$ is an $\mathbb{F}$ predictable process. For any $B\in\mathcal{G}_{0}$, $\Jt \ind_{B}\ind_{\{0\}}=0,$ which is an $\mathbb{F}$ predictable process. 
Since the processes
 $\ind_{B}\ind_{(t,\infty)}$ ($t>0$, $B\in\mathcal{G}_{t}$) and $\ind_{B}\ind_{\{0\}}$ ($B\in\mathcal{G}_{0}$) 
generate the $\mathbb{G}$ predictable $\sigma$-algebra,\footnote{Cf.~Theorem 3.21 in \shortciteN{HeWangYan92}.} this proves the condition {(B)}. \finproofs

%\subsection{Reduction of Filtration Toolbox}
\textbf{The condition (B) is assumed everywhere} in the sequel of the paper. Let \index{o@$^o$}${^{\fo }}\!\cdot$ and \index{p@$^p$}${^{p}}\!\cdot$ denote the $(\mathbb{F},\mathbb{Q})$ optional and predictable projections and let
$\croc{\cdot,\cdot}$ and $[\cdot,\cdot]$ 
denote the $(\mathbb{F},\mathbb{Q})$ optional and predictable brackets.
We introduce, denoted by straight capital letters, a number of processes related to $\theta$.
We write \index{j@$\ttJ$}$\ttJ=\ind_{[0,\theta)}$, hence $\ttJ_- =\indi{0<\theta}\ind_{[0,\theta]}$.
The fundamental tool to work with the condition (B) is the Az\'ema supermartingale $\ttS={^{\fo }}\! \mathtt{J} $ of $\theta$, 
%which is defined by $\ttS $\index{s@$\ttS$},
i.e. $\ttS_t =\mathbb{Q}(\theta>t\big|\cF_t),$ $t\in \R_+,$
with
canonical Doob-Meyer decomposition $\ttS= \ttS_0+ \cM-\Alpha,$
where $\cM$ is an $(\mathbb{F},\Q)$ martingale starting from 0
while $\Alpha$ is the
$(\mathbb{F},\Q)$ dual predictable projection of ${\ind_{\{0<\ftime\}}}\ttH$.
The most classical properties of $\ttS$ useful for this work are recalled in \sr{s:az}.

The proofs of the progressive of enlargement results in \citeN{JeulinYor78} or Chapitre XX in \citeN{DellacherieMaisonneuveMeyer92} 
only require that $\mathbb{G}$ is a subfiltration of $\overline{\mathbb{F}}.$
Hence, in view of Lemma \ref{c:b}, all these results hold under the condition (B). 
The next lemma gathers the main ones that we need in the sequel.

\bl\label{optionalsplitting} %Under the condition (B):
\ebe
\item
For any $\mathbb{G}$ stopping time $\theU$, there exists an $\mathbb{F}$ stopping time $\theV$, which we call the $\mathbb{F}$ \representative of $\theU$, such that ${\{\theU<\ftime \}}={\{\theV<\ftime \}}\subseteq {\{\theU=\theV\}}$.
\item
Let $(E,\mathcal{E})$ be a measurable space.
Any
 $\mathcal{P}(\mathbb{G})\otimes\mathcal{E}$ (respectively $\mathcal{O}(\mathbb{G})\otimes\mathcal{E}$) measurable
function $g_t(\omega,x)$ admits a $\mathcal{P}(\mathbb{F})\otimes\mathcal{E}$ (respectively $\mathcal{O}(\mathbb{F})\otimes\mathcal{E}$) \representative, i.e.~a $\mathcal{P}(\mathbb{F})\otimes\mathcal{E}$
(respectively
$\mathcal{O}(\mathbb{F})\otimes\mathcal{E}$) measurable function
 $g'_t(\omega,x)$ such that
$\ind_{(0,\theta]} g =\ind_{(0,\theta]} g' $
(respectively $\ind_{[0,\theta)} g =\ind_{[0,\theta)} g' $) holds almost everywhere. 
Moreover, these relations can be made
{to hold everywhere} by the choice of suitable versions.

%We denote by ${\cdot}^{\prime}$ a generic reduction operator (denoting an arbitrary everywhere reduction) applicable to stopping times, processes and random functions.
\item We have
%(Jeulin-Yor formula as of \shortciteN{DellacherieMaisonneuveMeyer92})
%%For any $\thisQ\in\b{\mathcal{M} (\mathbb{F},\mathbb{Q})}$, 
\beql{e:jyt}
\thisQ\in \mathcal{M} (\mathbb{F},\mathbb{Q}) \Longrightarrow 
\thisQ^{\ftime -}- \frac{\ttJ_-}{\ttS _-}
\centerdot \croc{\ttS,\thisQ}
%^{(\ff,\Q)}
\in \mathcal{M}(\mathbb{G},\mathbb{Q}).
%\sp\forall\thisQ\in \mathcal{M} (\mathbb{F},\mathbb{Q}).
\eeql 
\item {\def\K{K}\def\K{M'} For any $K\in {\cal S}_{\{\ttS_{-}>0\}}(\mathbb{F},\mathbb{Q})$ such that $\ttS_-\centerdot K+[\ttS,K]\in\mathcal{M}_{\{\ttS_{-}>0\}}(\mathbb{F},\mathbb{Q}),$ we have $K^{\theta-}\in\mathcal{M}(\mathbb{G},\mathbb{Q}).$ 

Conversely, for any $M\in\mathcal{M}(\mathbb{G},\mathbb{Q})$ with $\Delta_\theta \thisM =0$ on $\{\theta < \infty\},$ any $\ff$ optional reduction $\K$ of $M$
is in ${\cal S}_{\{\ttS_{-}>0\}}(\mathbb{F},\mathbb{Q}),$ {${\ind_{\{\ttS_{-}>0\}}} \K_- $ is an $\mathbb{F}$ predictable \representative of $\thisM _-$ and} $\ttS_-\centerdot \K+[\ttS,\K]\in\mathcal{M}_{\{\ttS_{-}>0\}}(\mathbb{F},\mathbb{Q}).$}

\item The Az\'ema supermartingale $\ttS$ admits the predictable multiplicative decomposition $\ttS=\ttS_0\cQ\cD$, 
for the 
finite variation predictable factor
$\cD=  \mathcal{E}(-\indi{\ttS_- >0}\frac{1}{\ttS_-}\centerdot \Alpha)$
and the
local martingale factor $\cQ$ 
 defined
by the pointwise limit
\beql{e:multdeclim}
\cQ
=
\lim_{n\rightarrow\infty}\mathcal{E}(\frac{1}{\pSigma}\centerdot\cM)^{\zeta_{n}},
\eeql
where $(\zeta_n)_{n\in\mathbb{N}}$ is the sequence that appears in \qr{e:Sp}. In particular,
on the random set $\{\pSigma >0\},$ we have $\cQ=\mathcal{E}(\frac{1}{\pSigma }\centerdot \cM ),$ 
$\cD= \mathcal{E}(-\frac{1}{\ttS_-}\centerdot \Alpha)$ 
and
\beql{e:multdec}%\monqs
%\mathcal{E}(\frac{1}{\pSigma }\centerdot \cM ) 
%\sp \cD= \mathcal{E}(-\frac{1}{\ttS_-}\centerdot \Alpha)
%\sp
\mathcal{E}(\frac{1}{^p\!\ttS}\centerdot \Alpha) \mathcal{E}(-\frac{1}{\ttS_{-}}\centerdot \Alpha) =
1 .
\eeql
%\b{(this holds in any predictable multiplicative decomposition $\ttS=\ttS_0\cQ\cD$ on $\{\pSigma >0\}$)
%%cf Jacod
%.}
If $\ttS_T$ is 
positive, then $\cQ=\mnqs>0$ holds 
on $[0,T]$.
%\item The Az\'ema supermartingale $\ttS$ admits a predictable multiplicative decomposition $\ttS=\ttS_0\cQ\cD$ 
%on $\R_+$, 
%for some local martingale factor $\cQ$ and finite variation predictable factor $\cD$
%both starting from 1.
%On the random set $\{\pSigma >0\},$ we have
%\beql{e:multdec}\cQ=%\monqs
%\mathcal{E}(\frac{1}{\pSigma }\centerdot \cM ) 
%\sp \cD= \mathcal{E}(-\frac{1}{\ttS_-}\centerdot \Alpha)
%\sp
%\mathcal{E}(\frac{1}{^p\!\ttS}\centerdot \Alpha) \mathcal{E}(-\frac{1}{\ttS_{-}}\centerdot \Alpha) =
%1. 
%\eeql
%In particular, if $\ttS_T$ is 
%%almost surely 
%positive, then $\cQ=\mnqs>0$ holds
%%almost surely 
%on $[0,T]$.
%
%In this paper, we use the decomposition 
%$\ttS=\ttS_0\cQ\cD$ as of \citeN{Song16a}, with $\cQ$ and $\cD$ respectively defined on $\R_+$
%\b{by the pointwise limits}
%\beql{e:multdeclim}
%\cQ
%=
%\lim_{n\rightarrow\infty}\mathcal{E}(\frac{1}{\pSigma}\centerdot\cM)^{\zeta_{n}}\sp
%\cD= \lim_{n\rightarrow\infty} \mathcal{E}(-\frac{1}{\ttS_-}\centerdot \Alpha)^{\zeta_{n}},
%\eeql
%where $(\zeta_n)_{n\in\mathbb{N}}$ is the sequence that appears in \qr{e:Sp}.
\dbe
\el
 
\proof 
1) is proven in Chapitre XX, n$^\circ$75 a) of \shortciteN{DellacherieMaisonneuveMeyer92}; 
3) is proven in Chapitre XX, n$^\circ$77 b) of \shortciteN{DellacherieMaisonneuveMeyer92}; 4) is proven in \citeN[Lemmas 6.5 and 6.8]{Song16a}. 

5) is proven in \citeN[Lemmas 3.9 and 3.10]{Song16a}. 
Specifically, \qr{e:multdec} is \citeN[Lemma 3.9]{Song16a}.
In \qr{e:multdeclim} the limits exist and the local martingale property of $\mathcal{E}(\frac{1}{\pSigma }\centerdot \cM )^{\zeta_{n}}$ passes to the limit $\mathcal{Q}$ by virtue of
\citeN[Lemma 3.10]{Song16a}. 
Note that $\frac{1}{\ttS_-}\centerdot \Alpha$ (resp. $\frac{1}{\pSigma }\centerdot %\c
\cM $) is well defined on $\In ,$ by \qr{e:S} (resp. on $\{\pSigma >0\}$,
by 
\qr{e:Sp}). 

Regarding
2), we first consider 
the question for processes, i.e. without the measurable space $(E,\mathcal{E})$.
Let $\chi $ be a bounded $\mathcal{G}_\infty$ measurable random variable and $Y$ be the (c\`adl\`ag) $\mathbb{G}$ martingale with {terminal variable} $\chi $. By a classical result\footnote{Known as the key lemma in the credit risk literature (see e.g.~\shortciteN[Lemma 3.1.2]{Bielecki2009}).} (see e.g. 
\shortciteN[Chapitre XX n$^\circ$75 (75.2)]{DellacherieMaisonneuveMeyer92}), 
{for every $t\in\mathbb{R}_+$}, we have the almost sure identity
$$
\ttJ_t Y_t=\ttJ_t \mathbb{E}[\chi |\mathcal{G}_t]=\frac{\ttJ_t}{\ttS _t}\mathbb{E}[\ttJ_t \chi |\mathcal{F}_t]
=\ttJ_t X_t,
$$
where
$$
X_t= \frac{{^{\fo }}\!(\ttJ \chi )_t}{\ttS _t}\ind_{\{\ttS_t>0\}}.
$$
By \shortciteN[Chapitre VI n$^\circ$47]{DM75}, the process ${^{\fo }}\!(\ttJ \chi )$ is c\`adl\`ag, so that, actually,
$
\ttJ X = \ttJ Y
$
holds in the indistinguishable sense.
This proves the existence of an $\mathbb{F}$ optional reduction for the $\mathbb{G}$ martingale $Y$.

Let $\mathfrak{C}$ denote the class of all the bounded $\mathcal{B}(\R_+)\otimes\mathcal{G}_\infty$ measurable functions $L$ such that ${^{{\mathbb{G}\cdot o}}}L$ admits an $\mathbb{F}$ optional reduction, where ${^{{\mathbb{G}\cdot o}}}\!\cdot$ denotes the $(\mathbb{G},\mathbb{Q})$ optional projection.
We can verify that $\mathfrak{C}$ is a functional monotone class in the sense of the monotone class theorem 1.4 in \shortciteN{HeWangYan92}. It results from above that $\mathfrak{C}$ contains all the random variables $\ind_{(a,b]}\chi $, where $a,b\in\mathbb{R}_+$ and $\chi $ is a bounded $\mathcal{G}_\infty$ measurable random variable. Therefore, by the monotone class theorem, $\mathfrak{C}$ contains all the bounded $\mathcal{B}(\R_+)\otimes\mathcal{G}_\infty$-measurable random variables.
In particular, every bounded $\mathbb{G}$ optional process
admits an
% (bounded) 
$\mathbb{F}$ optional reduction.
By taking limits, the result is extended to general $\mathbb{G}$ optional processes.

Since $\mathbb{F}$ satisfies the usual conditions, any evanescent measurable process is $\mathbb{F}$ predictable (Theorem 4.26 in \shortciteN{HeWangYan92}). For any $\mathbb{G}$ optional process $L$ with $\mathbb{F}$ optional reduction $K$, there exists a negligible set $O$ such that $\ttJ_- K= \ttJ_- L$ holds everywhere outside $O$. Therefore, defining $\tilde{K}=K-(K-L)\ind_{O}$, the process $\tilde{K}$ is $\mathbb{F}$ optional and satisfies $\ttJ_- \tilde{K}= \ttJ_- L$ everywhere. 
%This shows that the $\mathbb{F}$ optional reduction of $L$ can be defined so that the reduction relation holds everywhere.

We have thus proved the optional version of the part 2) of the lemma in the case of processes. A standard reasoning by monotone class theorem proves the result in the presence of the measurable space $(E,\mathcal{E})$. 

The predictable version of the part 2) of the lemma can be proved similarly. In fact, the predictable version for processes (i.e. without the measurable space $(E,\mathcal{E})$) is precisely the condition (B)
(cf.~the remark \ref{r:indis}).
\finproofs

Recall the respective $\ff$ predictable and optional Girsanov formulas\footnote{Cf.~\shortciteN[Theorem 12.18]{HeWangYan92}.} 
involving the 
density process $q$ of the measure change from the probability measure $\mathbb{Q}$ to some $\mathbb{Q}$ absolutely continuous probability measure $\mathbb{P}$:
\beql{e:gp}
&
%\mbox{For any bounded $\b{Q}\in\mathcal{M}(\mathbb{F},\mathbb{Q}),$ }
Q \mbox{ bounded in }\mathcal{M}(\mathbb{F},\mathbb{Q}) \Longrightarrow
Q - \frac{1}{q _-}\centerdot \croc{q,Q}
%^{(\mathbb{F},\mathbb{Q})}
\in\mathcal{M}(\mathbb{F},\mathbb{P}),
%\sp \forall \mbox{ bounded } X\in\mathcal{M}(\mathbb{F},\mathbb{Q}),  
\eeql
respectively
\beql{e:go}
& \qqq P\in\mathcal{M}(\mathbb{F},\mathbb{P})\Longleftrightarrow P\in\mathcal{S}_{\{q_{-}>0\}}(\mathbb{F},\mathbb{Q}) \mbox{ and }q_-\centerdot P+[q,P]\in\mathcal{M}_{\{q_{-}>0\}}(\mathbb{F},\mathbb{Q}).
\eeql 
Observe that the Jeulin-Yor formula \qr{e:jyt} and Lemma \ref{optionalsplitting} 4) are formal
analogs, in the field of progressive enlargement of filtration, of these respective Girsanov measure change formulas, 
the Az\'ema supermartingale $\ttS$ playing the role of the measure change density $q$. Starting from the so-called generalized Girsanov formulas (between possibly non absolutely continuous probability measures)
and representing the Az\'ema supermartingale $\ttS$ as a ``generalized density'',
these formal analogies can be turned into proofs of the corresponding enlargement of filtration formulas (see \citeN{Yoeurp1985} and \citeANP{Song87} (\citeyearNP{Song87}, \citeyearNP{Song13})).

%\brem\label{rem:jy} 
Note that the classical formulation of the Jeulin-Yor formula is stated in terms of $Q^{\theta},$
instead of $Q^{\theta-}$ in
\qr{e:jyt},
as
\beql{e:jytbis} 
\thisQ\in \mathcal{M} (\mathbb{F},\mathbb{Q}) \Longrightarrow
 \thisQ^{\ftime }- \frac{\ttJ_-}{\ttS _-}
 \centerdot (\croc{\ttS,\thisQ}
%^{(\ff,\Q)}
+B)\in \mathcal{M}(\mathbb{G},\mathbb{Q}),
\eeql
where $B$ is the $(\ff,\Q)$ dual predictable projection of the process $\Delta_{\theta} Q \indi{\theta\leq \cdot}$ 
(cf. \citeN[Theorem 1 and Lemma 4 b)]{JeulinYor78}). However,
as visible in the proof of Theorem 1 in \citeN{JeulinYor78},
%(which first establishes \qr{e:jytbis}), 
the bracket 
$\cro{\ttS,Q}$
%$\cro{\ttS,Q}^{(\ff,\mathbb{Q})}$ 
in the Jeulin-Yor formula \eqref{e:jyt}
is intrinsically linked with $Q^{\theta-},$ rather than with $Q^{\theta}$.
%\erem

The next	 result shows that
$\ff$ predictable and optional reductions of $\gg$ predictable and optional processes are uniquely defined on the random intervals
$\{\ttS_{-}>0\}$ (which contains $\theta$ on $\{0<\theta<\infty\}$, cf. \qr{yortheta}) and $\{\ttS>0\}$, respectively. 

\bl\label{l:Spos} Two $\ff$ predictable (respectively optional) processes 
$K$ and $\tilde{K}$ 
undistinguishable on $[0,\theta]$ (respectively $[0,\theta)$) are undistinguishable on $\{\ttS_{-}>0\}$ (respectively $\{\ttS>0\}$).
\el 
\proof
Otherwise, the predictable section theorem\footnote{Cf.~\shortciteN[Theorem 4.8]{HeWangYan92}.} (considering the predictable case in the lemma)
would imply the existence of an $\ff$ predictable stopping time $\sigma$ such that $\mathbb{E} [\ind_{K_{\sigma}\neq \tilde{K}_{\sigma}}\ttS_{\sigma-}\ind_{\{\sigma<\infty\}} ]> 0$, in contradiction with
$$
\dcb
\mathbb{E}\left[\ind_{K_{\sigma}\neq \tilde{K}_{\sigma}}\ttS_{\sigma-}\ind_{\{\sigma<\infty\}}\right] 
=\mathbb{E}\left[\ind_{K \neq \tilde{K} }\ttS_{-} \centerdot (\ind_{\{\sigma>0\}}\ind_{[\sigma,+\infty)})\right]\\
=\mathbb{E}\left[\ind_{K \neq \tilde{K} }\mathsf{J}_{-} \centerdot (\ind_{\{\sigma>0\}}\ind_{[\sigma,+\infty))} \right]=\mathbb{E}\left[\ind_{K_{\sigma}\neq \tilde{K}_{\sigma}}\mathsf{J}_{\sigma-}\ind_{\{\sigma<\infty\}}\right] 
=0.
\dce
$$
The optional version of the lemma can be proven similarly.
\finproof \\

%\subsection{About the Random Intervals  
%$\{\pSigma >0\}$ and $\{\ttS_{-} >0\}$}\label{ss:ri}
For the (first) proof of Theorem \ref{part1} below, we need a 
refined comparison
%description 
of 
the random intervals 
%$\{\ttS >0\},$ 
$\{\pSigma >0\}$ and $\{\ttS_{-} >0\}$. 
Let \beql{e:theeta}\index{e@$\eta$}
\varsigma=\inf\{s>0;\ttS_s=0\}\sp\theeta
 =\inf\{s>0; {\pSigma }_{s}=0, \,\ttS_{s-}>0\}.
\eeql

\bl\label{Seta} 
We have
\beql{e:theetasuite} 
&\theeta 
=\inf\{s>0; \ttS_{s-}=\Delta_s\Alpha>0\}
=\inf\{ {s\in\Int}; \mathcal{E}(-\frac{1}{\ttS_-}\centerdot \Alpha)_s=0\} .
\eeql
Moreover, we have $\theeta \geq\varsigma$, $\theeta =\varsigma$ 
on $\{\theeta <\infty\}$ and
\begin{equation}\label{pS>0}
\{\ttS_- >0\}\setminus \{\pSigma >0\}=[\theeta].
\end{equation}
In particular, $\eta$ is $\mathbb{F}$ predictable.
\el

\proof
The first equality in \qr{e:theetasuite} results from \qr{e:SS}.
The stochastic exponential $\mathcal{E}(-\frac{1}{\ttS_-}\centerdot \Alpha )$ 
vanishes at $t$ in $\Int$ if and only if
\bel
& \Delta_t\Big(-\frac{1}{\ttS_-}\centerdot \Alpha \Big)
=
 \frac{-1}{\ttS_{t-}}\Delta_t \Alpha = -1 
\sp \mbox{ i.e.~}\ttS_{t-}= \Delta_t \Alpha.
\eel 
Hence,
\bel
&\inf\{ s\in\Int ; \mathcal{E}(-\frac{1}{\ttS_-}\centerdot \Alpha)_s=0\}=\inf\{s\in\Int;\ttS_{s-}= \Delta_s \Alpha \}\\&\qqq
= \inf\{s\in\Int;\ttS_{s-}= \Delta_s \Alpha >0\}= \inf\{s>0;\ttS_{s-}= \Delta_s \Alpha >0\}=\theeta,
\eel 
which proves the second equality in \qr{e:theetasuite}. The remainder of the lemma is the consequence of Lemma 3.2 (cf.~also Lemmas 3.3 and 3.6) in \citeN{Song16a}. \finproof

\bl\label{QS=0} We have
$\ind_{\{\pSigma=0\}}\centerdot \cM=0$.
\el

\proof
By (\ref{e:SS}), we have $\ttS_-\geq \pSigma$. Hence$$
\dcb
\ind_{\{\pSigma=0\}}\centerdot \cM
&=&
\ind_{(\varsigma,\infty)}\centerdot \cM
+
\ind_{(0,\varsigma]}\ind_{\{\ttS_-=0\}}\centerdot \cM+
\ind_{\{\pSigma=0, \ttS_->0\}}\centerdot \cM \\

&=&
\ind_{(\varsigma,\infty)}\centerdot \cM
+
\ind_{\{\ttS_{\varsigma-}=0\}}\Delta_{\varsigma}\cM\ind_{[\varsigma,\infty)}+
\Delta_{\eta}\cM \ind_{[\varsigma,\infty)},
\dce
$$
by Lemma \ref{Seta}. The first term is null because $\cM$ is constant on $(\varsigma,\infty)$ (cf. \qr{e:sss}). The second term is null because of \citeN[Lemmas 3.4 and 3.7]{Song16a}. The third term is null because (cf.~(\ref{e:SS}) and Lemma \ref{Seta}) 
$ \ind_{\{\eta<\infty\}} \Delta_{\eta}\cM
=
 \ind_{\{\eta<\infty\}} \ind_{\{\ttS_{\varsigma-}>0,\pSigma_\varsigma=0\}}(\ttS_{\varsigma}-\pSigma_\varsigma)=0.
$ 
\finproof

\subsection{Toward the Condition (A): Counterparty Risk BSDEs Motivation} \label{s:bsdecr} 

\def\theg{f}\def\theg{g'}
\def\varphi{\gamma'}
\def\thisGp{\widehat{\ttG}'}\def\thisGp{G'}\def\thisGp{F}\def\thisGp{\thatF'}

\def\wG{\widehat{G}}\def\wG{G}\def\wG{\thatF}
\def\theeta{\eta}
 
This section, for motivation mainly, can be skipped at no harm from the
point of the theoretical developments of Section \ref{cA}. More on applications
will be delivered in \sr{diffsituas}.
We consider:
\begin{itemize}
%\item[-] a $\mathbb{G}$ semimartingale $P,$ assumed without jump at $\theta$ for simplicity here, representing the risk-free value process of an underlying portfolio of financial derivatives between a bank and a counterparty,
% \item[-] {a counterparty default time $\ftime $ totally inaccessible with $\mathbb{G}$ compensator $\gamma\centerdot \boldsymbol\lambda$,}
\item[-] An exposure at default, or ``recovery'' of a bank upon the default of its counterparty, of the form $\ind_{\{\ftime <T\}}\thatF_{\ftime}$, where $T>0$ is some maturity, $\theta$ represents the default time of the counterparty
% of the portfolio) and \beql{e:P}\thatF=\max(P_{-},0),\eeql 
and $\thatF$ is a $\mathbb{G}$ predictable process (for simplicity of presentation here, see however the comment following \qr{e:Gg}),
\item[-] A $\mathcal{P}(\mathbb{G})\otimes\mathcal{B}(\mathbb{R})$ funding cost coefficient \emph{$\thisg_t(\omega,x)$ of the bank.} 
 \end{itemize}
Assuming $\theta$ endowed with an intensity $\gamma,$
the \emph{counterparty risk backward
stochastic differential equation (BSDE)},
 which prices the exposure at default $\thatF_{\ftime}$ at $\theta$ (if $<T$) and the funding costs $g$ until $\theta\wedge T$, can be formulated as 
the following BSDE for some process $Z$ in $\cS(\mathbb{G},\mathbb{Q})$:
%\b{poss? turn $T-$ to $T$}
\begin{equation}\label{equat5}
\left\{
\dcb
Z_{T-}\ind_{\{T\leq\theta\}}=0, \\
\\
Z^{\theta\wedge T-} +\int_0^{\cdot\wedge \theta\wedge T}\left( g_s(Z_{s-}) +( {\thatF}_s - Z_{s-}) \gamma_s\right) ds \in \mathcal{M}(\mathbb{G},\mathbb{Q})
.
\dce\right.
\end{equation} 
For the sake of conciseness, we present the counterparty risk
%pricing 
BSDE under this slightly unusual appearance, which is the equation (3.8) in \citeN[Theorem 3.1]{Crepey13a1},
%under an appearance which is rather unsual,
%(with an implicit ``no-jump condition'' at $\theta\wedge T$), 
because this formulation is the most convenient for the discussion that follows. 
Moreover, we only state the problem in its most
basic
% primitive 
form here. To be in line with applications, the assumptions in \citeN{Crepey13a1} cover more general BSDEs with 
\beql{e:Gg}
\thatF =\thatF_t(x,u)\sp g=g_t(x,u),\eeql 
where the additional argument $u$ 
corresponds to integrands in a stochastic integral representation of the martingale part of $Z$. In particular, dependencies of $\thatF$ as of \qr{e:Gg}, where the dependency in $t$ is not necessarily of
predictable type,
%to which the following discussion is extended in \citeN[Section 4]{Crepey13a1},
make the corresponding form of \qr{equat5} a 
%truly 
nonstandard BSDE.

In essence, the \shortciteN{DuffieSchroderSkiadas96}'s approach to \qr{equat5} would consist in forgetting about $\theta$ there (or ``sending $\theta$ to infinity''), which results in a simpler equation 
``without $\theta$'', where $\theta$ is only indirectly represented through its intensity $\gamma$. 
One then tentatively sets $Z=\tilde{Z}^{\theta-}$, where $\tilde{Z}$ is
a solution $\tilde{Z}$ to the simpler equation 
without $\theta$.
However, this only yields a solution $Z$ to \qr{equat5} if $\tilde{Z}$ does not jump at $\theta.$ In the 
basic reduced-form setup 
discussed in \sr{rem:bis} below,
%the concluding paragraph of this section,
%%\sr{s:bsdecr},
%%%(cf. the remark \ref{rem:bis} below)},
this no-jump condition is satisfied. 
But, apart from this restrictive situation, the no-jump condition is unverifiable and it does not hold in general.

One way out of this 
%(cf. also \sr{s:applic} below) 
is to
%In order to obtain a more tractable form of \qr{equat5}, we 
introduce a ``reduced BSDE'', given a smaller filtration $\ff$ such that the condition (B) is satisfied.
For any \cadlag process $X$ on $\R_+$ (or any predictable set of interval type), we write 
\beql{e:nu}
&\overline{X}
=X + \left(\theg_\cdot(X _{-})+({ \thisGp }
-X _{ -}){\varphi }\right) \centerdot \boldsymbol\lambda.
\eeql
%\b{where $\theg$, $\thisGp$ and $\varphi$ stand for $\ff$ predictable reductions of $g$, $\thatF$ and $\gamma$, respectively}.
%${\cdot}^{\prime}$ stands for an $\ff$ predictable reduction of each concerned random process or function.
Observe that, assuming the BSDE (\ref{equat5}) has a solution $Z $ and letting 
$\thisU=Z'$
%$\thisU$ 
denote an $\ff$ optional reduction of $Z$, the martingale term in the BSDE (\ref{equat5}) satisfies \index{z@$Z$}
\beql{e:ZU}
&Z ^{\ftime \wedge T-}_t+\int_0^{t\wedge \ftime \wedge T}\big(\thisg_s(Z _{s-})+(\wG _s -Z _{s-})\big){\gamma_s} ds\\
&\qqq =
\thisU ^{\ftime \wedge T-}_t+\int_0^{t\wedge \ftime \wedge T}\big(\theg_s(\thisU _{s-})+(\thisGp_s -\thisU _{s-}){\varphi_s}\big) ds\\&\qqq 
=\overline{\thisU}^{\ftime \wedge T-}_t=(\overline{\thisU}^{ T-})^{\ftime -}_t .
\eeql 

This suggests to solve the BSDE (\ref{equat5}) with Lemma \ref{optionalsplitting} 4). 
Namely,
we consider the
following BSDE for some process $U$ in $\cS_{\Int}(\mathbb{F},\mathbb{Q}) $:
\begin{equation}\label{equat6}
\dcb
U_{T-} 
\ttS_{T-} =0\sp
\emph{\ttS_-\centerdot \overline{\thisU}^{ T-}+[\ttS,\overline{\thisU}^{ T-}]}\in \mathcal{M}_{\Int}(\mathbb{F},\mathbb{Q}) 	
.
\dce 
\end{equation} 
Based on Lemma \ref{optionalsplitting} 4), 
%and an integrability study, 
the following result is proved in \citeN{Crepey13a1}. 
\begin{pro}\label{SW}
The BSDEs (\ref{equat5}) and (\ref{equat6}) are equivalent, in the following sense:
\hfill\break{\rm -}
If $Z$ is a solution to the BSDE (\ref{equat5}), then 
$U=Z'$ 
is a solution to the BSDE (\ref{equat6}); 
\hfill\break{\rm -}
Conversely, if $U$ is a solution to the BSDE (\ref{equat6}), then $Z=U^{\ftime-}$ is a solution to the BSDE (\ref{equat5}).
\end{pro}
In \qr{equat6}, $\theta$ is only indirectly represented, through 
$\varphi$ in $\overline{\thisU}$ (cf. \qr{e:nu}).
In this sense, passing from
\qr{equat5} to \qr{equat6} removes $\theta$ from the equation.
However, beyond the simple case where $\ttS$ is continuous and nonincreasing so that $[\ttS,\cdot]=0$,
this comes at the expense of an additional bracket in the martingale condition of \qr{equat6}. 

To untie the Gordian knot that we are facing here,
%As a shortcut out of this,
%another way to go, 
let us suppose 
for a moment 
the condition (A) stated in the beginning of \sr{cA} below,
for some invariance measure $\P$. 
Under this condition, in view of \qr{e:ZU}, any solution $U$ in
$\cS_{\Int}(\mathbb{F},\mathbb{P})$ to
\begin{equation}\label{equat7} 
\thisU _{T-} \ttS_{T-} =0 \sp
\emph{\overline{\thisU}^{T-} \in
\mathcal{M}_{\Int}(\mathbb{F},\mathbb{P})}
\end{equation} 
yields a solution $Z=U^{\ftime-}$ to (\ref{equat5})
(since $\ttS_{\theta	-}>0$, cf. 
\qr{yortheta}).

As compared with \qr{equat5}, this approach allows 
getting rid of $\theta$ in the equation
%the $\ff$ reduction $\varphi$ of its intensity $\gamma$, cf. \qr{e:nu})
and it no longer comes at the expense of
a more complicated martingale condition.
% with respect to \qr{equat5}.
In fact, the martingale condition in \qr{equat7} is essentially the same as
the one in \qr{equat5}, modulo reduction. 
From this point of view, 
the condition (A) and the related invariance measure $\mathbb{P}$ appear as ``deus ex machina'' for dealing with \qr{equat5}. See \sr{ex:cap} for a concrete application.

However, such an approach
%``extended reduced-form approach'' 
postulating the condition (A) raises two important issues:
\begin{enumerate}
\item How strong is the condition (A)?
\item Do we have equivalence, under the condition (A), between (\ref{equat7}) and (\ref{equat5}), and not only (\ref{equat7}) implies (\ref{equat5})? (for instance, for the application mentioned in \sr{ex:cap} below,
one
really needs the equivalence).
\end{enumerate}
These two questions were our initial motivation for the introduction and study of the condition (A). 
Regarding the first one, a complete characterization of the condition (A) and a mild sufficiency condition 
for it 
are established as 
Theorems \ref{A2bis}-\ref{c:mart} 
and Theorem \ref{sfcnd}, respectively.
The second question is given a positive answer in
\citeN[Theorems 3.1 and 4.3]{Crepey13a1}, based on Theorem \ref{c:pq} below
% the study of \sr{ss:carmap} below 
regarding local martingales under an invariance measure $\mathbb{P}$.
%Regarding the first question, in view of Proposition \ref{t:exptransf}, a natural conjecture is that the condition (A) is related to the $(\ff,\Q)$ true martingale property of the process $\cQ$ on $[0,T].$ 
%This conjecture is addressed in the sequel of the paper, where the corresponding characterization of the condition (A) and a mild sufficiency condition 
%for it 
%are established as 
%Theorems \ref{c:mart}
%%Theorems \ref{A2bis} 
%and Theorem \ref{sfcnd}, respectively.
%The second question is given a positive answer in
%\citeN[Theorems 3.1 and 4.3]{Crepey13a1}, based on Theorem \ref{c:pq} below
%% the study of \sr{ss:carmap} below 
%regarding local martingales under an invariance measure $\mathbb{P}$.\\

\subsubsection{Basic reduced-form setup} 
\label{rem:bis}
The 
immersion property, first introduced under the name of (${\cal H}$) hypothesis in \citeN[page 284]{BremaudYor78}, means that all $\mathbb{F}$ local martingales are ${\mathbb{G}}$ local martingales. 
%The
%pseudo-stopping time property of \citeN{NikeghbaliYor05} means that all $%
%{\mathbb{F}}$ local martingales {{stopped at ${\theta}$}}
%are ${\mathbb{G}}$ local martingales.
In the historically much considered case of a Brownian filtration $\ff$
(see e.g. \citeN{MansuyYor06}),
the immersion property implies that $\ttS$ is a finite variation and predictable process. In this case,
``$\emph{\overline{\thisU}^{ T-}} \in \mathcal{M}_{\Int}(\mathbb{F},\mathbb{Q})$'' suffices to
ensure the more involved martingale condition in \qr{equat6}, by Yoeurp's lemma.\footnote{See \shortciteN[Exercise 9.4 1)]{HeWangYan92}.} 
% so that $[\ttS,\cdot]$} is a local martingale, by 
%reduces to 
%$\emph{\overline{\thisU}^{ T-}} \in \mathcal{M}_{\Int}(\mathbb{F},\mathbb{Q})$.
Accordingly,
%For this reason, 
people tend
%one tends 
to identify immersion as the ``easy'' case where 
an enlargement of filtration approach
%a simple reduced-form approach 
without measure change 
is successful for dealing with pricing equations such as \qr{equat5} 
(see 
the comments before Section 3 in \shortciteN{DuffieSchroderSkiadas96} 
or in \shortciteN[page 1379]{CollinDufresneGoldsteinHugonnier2004}
and see the comments following (3.22) and (H.3) or the remarks following Proposition 6.1 in \citeN{BieleckiRutkowski00}).
However, on the one hand, immersion is unnecessarily strong for that purpose, because whatever happens after $\theta$ is irrelevant here. 
On the other hand, even in the Brownian case, immersion
is not enough to grant the converse implication from \qr{equat5} to \qr{equat6}. In fact, the key property granting the equivalence between \qr{equat5} and \qr{equat6}, including in models with jumps, is not immersion, but rather the property that $\ttS$ is continuous and nonincreasing (see the comments following Proposition \ref{SW}), 
which corresponds to the case of a 
pseudo-stopping time avoiding $\ff$ stopping times (cf. \sr{s:inv}). 
 In the sequel we call ``basic reduced-form setup'', by contrast with
the ``extended reduced-form setup'' provided by the invariance times of this paper, the 
case
where 
$\ttS$ has no martingale component (i.e. $\cM=0$),
$\theta$ has an intensity (hence $\ttS=\ttS_0 + \Alpha$ is continuous, by Lemma \ref{l:inten})
and $\gg$ is $\ff$ progressively enlarged by $\theta.$ The simplest situation of this kind is the Cox process framework
(see \shortciteN[Chapter 3]{Bielecki2009}),
%\b{(see \shortciteN[Page 1379]{CollinDufresneGoldsteinHugonnier2004})}, 
in which case immersion holds, but this does not necessarily need to be the case even in a basic reduced-form setup.

\section{Invariance Measures and Invariance Times} \label{cA} 
%\section{The Condition (A)} \label{cA} 

For a given triplet $(\mathbb{F}, \mathbb{G},\theta)$ satisfying the condition (B),
% of Section $\ref{cB}$, 
for a given
positive constant $T$, we introduce the following: \\

\noindent\textbf{Condition \index{a@(A)}(A)}. 
There exists 
a probability measure 
\index{p@$\mathbb{P}$}$\mathbb{P}$ equivalent to $\mathbb{Q}$ on {${\cal F}_{T}$} 
such that, for any $({\ff},\mathbb{P})$ local martingale $\thisN$,
{$\thisN^{\theta-}$}
is a $(\gg,\mathbb{Q})$ local martingale {on $[0,T]$,} 
i.e. $$\thisN\in\mathcal{M}({\ff},\mathbb{P})\Longrightarrow \thisN^{\theta-}\in\mathcal{M}_{[0,T]}({\gg},\mathbb{Q}).$$
%$$ \thisN^{\theta-}\in\mathcal{M}_{[0,T]}({\gg},\mathbb{Q})\sp \forall\thisN\in\mathcal{M}({\ff},\mathbb{P}).$$
%$$\forall\thisN\in\mathcal{M}({\ff},\mathbb{P})\sp \thisN^{\theta-}\in\mathcal{M}_{[0,T]}({\gg},\mathbb{Q}) .$$ 
If this condition is satisfied, we call the random time $\theta$ an invariance time and the related probability measure $\mathbb{P}$ an invariance measure.
\\

If $\theta$ is $\gg$ predictable and $\ff=\gg,$ then
$\mathbb{P}=\mathbb{Q}$ is an invariance measure.
But we are mostly interested in the case where
$\theta$ has a nontrivial totally inaccessible part and $\theta$ is not an $\ff$ stopping time. 
The possibility to change the measure in the condition (A) is 
 material, for instance, in the dynamic Gaussian copula model of \citeN{CrepeySong15} 
(see \sr{ss:dcm} below).

%\brem 
%As seen above, the condition (A) appears naturally in the motivating example of \sr{s:bsdecr}. 
To the best of our knowledge, the condition (A) has not been considered before in the probabilistic literature. 
In relation with it,
%As \b{far reminiscent} to it, 
one may think of
 the density hypothesis {of \citeN{Jacod87}}, initially formulated in
a setup of initial enlargement and reconsidered in a progressive enlargement setup in 
 \citeN{JeanblancLeCam09b}, 
under which 
there exists a
 measure change to a probability that makes the reference filtration $\ff$ and
the random time $\theta$ independent. However, under an invariance measure $\mathbb{P}$,
$\ff$ and $\theta$ do not need to be independent. The spirit of invariance times is not to 
extend the case of independence by a measure change.
%boild down, reduce
% the density hypothesis does not deal with martingale invariance properties

Stopping before $\theta$ in the condition (A), rather than at $\theta$
in the case of pseudo-stopping times (cf. \sr{s:inv}),
 appears naturally in the motivating application of \sr{s:bsdecr}. 
On top of that,
there are (at least) two reasons for stopping before $\theta$ rather than at $\theta$ in the condition (A).
First, in view of the optional version of Lemmas \ref{optionalsplitting} 2) 
%and \ref{l:Spos}}
(under the condition (B) which is assumed throughout the paper),
{$\thisN^{\theta-}$}
 is 
%uniquely 
determined by the information of $\mathbb{F}$, which is not the case of {$\thisN^{\theta}$}.
Second, as explained after the equation \qr{e:jytbis},
the bracket 
$\cro{\ttS,Q}$ 
%$\cro{\ttS,Q}^{(\ff,\mathbb{Q})}$ 
in the Jeulin-Yor formula \eqref{e:jyt}
is intrinsically linked with $Q^{\theta-},$ rather than with {$Q^{\theta}$}.
%\erem

This section is a theoretical study of the condition (A). 
Sections \ref{firstproofofThmpart1} and \ref{firstproofofThm} 
establish the invariance measure and invariance time characterizations of
Theorems \ref{part1} and \ref{A2bis}-\ref{c:mart}. 
\sr{s:pos} studies the positivity of the stochastic exponential $\monqs$ that appears as the tentative density process of the measure change in Theorem \ref{A2bis}.
\sr{ss:mar} is about the true martingale property of 
the multiplicative martingale part $\mathcal{Q}$ of $\ttS$, which, under the condition (A), will be seen in Theorems %\ref{part1} and 
\ref{A2bis}-\ref{c:mart} to coincide with $\monqs$ on $[0,T]$.
%\sr{s:applic} provides an invariance sufficiency condition and makes the connection between the invariance measure and the so-called survival measure. 
\sr{ss:carmap} yields a characterization of local martingales under an invariance measure.
%%\sr{ss:exptransf} derives a transfer formula between $\mathbb{P}$ and $\mathbb{Q}$ expectations. 

%$\mathbb{Q}$
%and $\mathbb{P}$ 
%expectations are denoted by \index{e@$\mathbb{E}^{\mathbb{Q}}$}\b{$\mathbb{E}=\mathbb{E}^{\mathbb{Q}}$} and by \index{e@$\mathbb{E}^{\mathbb{P}}$}$\mathbb{E}^{\mathbb{P}}$. 

\subsection{Az\'ema Supermartingale Characterization of Invariance Measures}\label{firstproofofThmpart1}

The condition (A) is an existence condition for invariance measures. Before characterizing this existence, we consider in this section the conditions for a given probability measure $\mathbb{P}$, equivalent to $\mathbb{Q}$ on $\mathcal{F}_{T},$
to be an invariance measure. 
 
Given a probability measure \index{p@$\mathbb{P}$}$\mathbb{P}$
equivalent to $\mathbb{Q}$ on {${\cal F}_{T}$},
we denote by \index{q@$\monq$}\index{p@$\monp$}$\monq$ the $(\mathbb{F},\mathbb{Q})$ martingale of the density functions $\left.\frac{d\mathbb{P}}{d\mathbb{Q}}\right|_{\mathcal{F}_{t\wedge T}}, t\in\mathbb{R}_+$.
We also introduce $\monp =\frac{1}{\monq }$ and the
stochastic logarithms \index{p@$\thenu$}$\thenu$ and $\therho$\index{q@$\therho$} such that
\beql{e:pq}
\monp = {\monp_0}\mathcal{E}(\thenu ),\ \monq = {\monq_0}\mathcal{E}(\therho)\sp \thenu_0=\therho_0=0.
\eeql
In particular, $\monp$ and $\thenu $ (respectively $\monq$ and $\therho$) are
$(\mathbb{F},\mathbb{P})$ (respectively $(\mathbb{F},\mathbb{Q})$) local martingales on $[0,T].$
%in $\mathcal{M}_{[0,T]}(\mathbb{F},\mathbb{P})$ (respectively $\mathcal{M}_{[0,T]}(\mathbb{F},\mathbb{Q})$). 

\bl\label{cond3738} 
We consider a probability measure \index{p@$\mathbb{P}$}$\mathbb{P}$
equivalent to $\mathbb{Q}$ on {${\cal F}_{T}$} with the notation introduced in (\ref{e:pq}). 
The two conditions that follow are equivalent:
\begin{eqnarray}
\label{e:qf}
& q = q_0 \mathcal{E}(\frac{1}{\pSigma}\centerdot \cM) \mbox{ on } \{^p\!\ttS>0\} \cap [0,T] ,
\\ 
&
\label{e:cor} 
\pSigma \centerdot \therho =\cM \mbox{ on } [0,T] .
\end{eqnarray}
If they hold, then $\ind_{\{\pSigma>0\}}\frac{1}{\pSigma }$
 is $(\mathbb{F},\mathbb{Q})$ integrable with respect to $\cM$ on $[0,T].$ 
%Consider a probability measure \index{p@$\mathbb{P}$}$\mathbb{P}$
%equivalent to $\mathbb{Q}$ on {${\cal F}_{T}$} with notations in (\ref{e:pq}). The two conditions below are equivalent.
%\begin{eqnarray}
%\label{e:qf}
%& q = q_0 \mathcal{E}(\frac{1}{\pSigma}\centerdot \cM) \mbox{ on } \{^p\!\ttS>0\} \cap [0,T].
%\\ 
%&
%\label{e:cor} 
%\pSigma \centerdot \therho =\cM -\cM _0\mbox{ on } [0,T].
%\end{eqnarray}
%And they imply that $\ind_{\{\pSigma>0\}}\frac{1}{\pSigma }$
% is $\cM$ integrable on $[0,T]$ with respect to $(\mathbb{F},\mathbb{Q})$. 
\el

\proof 
Recalling \qr{e:intint} regarding the notion of a stochastic integral on the predictable interval
$\cI=\{^p\!\ttS>0\} \cap [0,T],$ 
we can interpret \qr{e:qf} by its versions stopped at each of the $\zeta_n$,
where $(\zeta_n)_{n\in\mathbb{N}}$ is the sequence that appears in \qr{e:Sp}.
%Recall that 
%$\{\pSigma >0\}$ is a predictable interval and, according to \citeN[(6.24) and (6.28)]{Jacod1979}, there exists
%a nondecreasing sequence
%\index{z@$\zeta_n$}
%$(\zeta_n)_{n\in\mathbb{N}}$ of $\ff$ stopping times such that
%\beql{e:Sp}
%\{\pSigma>0\}\cup\gr0\rg=\cup_{n}[0,\zeta_n] \mbox{ and $\frac{1}{\pSigma}$ is bounded on $(0,\zeta_n]$ for every $n$}.
%\eeql
%Hence, recalling \qr{e:intint} regarding the notion of a stochastic integral on the predictable interval
%$\cI=\{^p\!\ttS>0\} \cap [0,T],$ 
%we can interpret \qr{e:qf} by its versions stopped at each of the $\zeta_n,\,n\in\mathbb{N}$. 
Therefore, given the relationship between the stochastic exponential and the stochastic logarithm, 
\qr{e:qf} is equivalent to
\beql{qqn}
\ind_{(0,\zeta_n \wedge T ]}\centerdot \therho=\ind_{(0,\zeta_n \wedge T ]}\frac{1}{\pSigma }\centerdot \cM\sp\forall n\in\N ,
\eeql 
which in turn is equivalent to 
$$
\pSigma \ind_{(0,\zeta_n \wedge T ]}\centerdot \therho=\ind_{(0,\zeta_n \wedge T ]}\centerdot \cM\sp\forall n\in\N .
$$
In virtue of the dominated convergence theorem for stochastic integrals,\footnote{Cf.~\shortciteN[Theorem 9.30]{HeWangYan92}.}
this is equivalent to$$
\pSigma \centerdot \therho
=
\pSigma \ind_{\{\pSigma>0\}}\centerdot \therho
=\ind_{\{\pSigma>0\}}\centerdot \cM \mbox{ on $[0,T]$},
$$
which is equivalent to \qr{e:cor} because of Lemma \ref{QS=0}. The first part of the lemma is proved.
The second part holds because $\ind_{\{\pSigma>0\}}\frac{1}{\pSigma }$
 is $(\mathbb{F},\mathbb{Q})$ integrable with respect to $\pSigma \centerdot \therho$ on $[0,T]$.~\finproof\\

\noindent
A similar reasoning can be used for proving the following result. 

\bl\label{lem:intmult} The process $\ind_{\{\pSigma>0\}}\frac{1}{\pSigma}$ is 
$(\mathbb{F},\mathbb{Q})$ integrable with respect to $\cM$
on $[0,T]$ 
if and only if the multiplicative martingale part $\cQ$ of $\ttS$ as of \qr{e:multdeclim} is a Dol\'eans-Dade exponential on $[0,T]$, in which case
the identity $\cQ=\mathcal{E}(\ind_{\{\pSigma>0\}}\frac{1}{\pSigma}\centerdot\cM)$ holds on $[0,T]$.
\el

\proof 
By \qr{e:multdeclim}, we have$$
\ind_{[0,\zeta_n]}\frac{1}{\pSigma }\centerdot \cM
=
\ind_{[0,\zeta_n]}\ind_{\{\cQ_->0\}}\frac{1}{\cQ_-}\centerdot \cQ\sp \forall n\in \mathbb{N} .
$$
Hence, by \shortciteN[Theorem 9.2]{HeWangYan92},
if $\ind_{\{\cQ_->0\}}\frac{1}{\cQ_-}\centerdot \cQ$ exists on $[0,T]$, then the process $\ind_{\{\pSigma>0\}}\frac{1}{\pSigma}$ is 
$(\mathbb{F},\mathbb{Q})$ integrable with respect to $\cM$ on $[0,T]$. In addition, by the dominated convergence theorem for stochastic integrals, we have on $[0,T]:$
$$
\ind_{\{\pSigma>0\}}\frac{1}{\pSigma}\centerdot\cM
=
\ind_{\{\pSigma>0\}}\ind_{\{\cQ_->0\}}\frac{1}{\cQ_-}\centerdot \cQ
=
\ind_{\{\cQ_->0\}}\frac{1}{\cQ_-}\centerdot \cQ ,
$$
as $\mbox{supp}(d[\cQ,\cQ])\subset \{\pSigma>0\}$, by
\qr{e:multdeclim} and Lemma \ref{QS=0}. 

Conversely, suppose that $\ind_{\{\pSigma>0\}}\frac{1}{\pSigma}$ is 
$(\mathbb{F},\mathbb{Q})$ integrable with respect to $\cM$ on $[0,T]$. 
Then, on $[0,T],$ we have by \qr{e:multdeclim} the pointwise limits
$$
\cQ
=
\lim_{n\rightarrow\infty}\mathcal{E}(\frac{1}{\pSigma}\centerdot\cM)^{\zeta_{n}}
=
\lim_{n\rightarrow\infty}\monqs^{\zeta_{n}}
=
\monqs,
$$
by stochastic dominated convergence.~\finproof\\

%\bl\label{lem:intmult} The process $\ind_{\{\pSigma>0\}}\frac{1}{\pSigma}$ is 
%$(\mathbb{F},\mathbb{Q})$ integrable with respect to $\cM$
%on $[0,T]$ 
%if and only if the martingale part $\cQ$ of $\ttS$ in its predictable multiplicative decomposition (cf. Lemma \ref{optionalsplitting} 5)) is a Dol\'eans-Dade exponential on $[0,T]$, in which case
%$\cQ=\mathcal{E}(\ind_{\{\pSigma>0\}}\frac{1}{\pSigma}\centerdot\cM)$ holds on $[0,T]$.
%\el
%\b{\proof Cf. Lemma \ref{cond3738} 
%\finproof\\}

\noindent 
Theorem \ref{part1} below relates the conditions introduced in Lemma \ref{cond3738} to the invariance measure property. 
The proof that follows is based on a reduction of all the computations from $\mathbb{G}$ to $\mathbb{F}$. The basic idea is
to make use of Lemma \ref{optionalsplitting} 4)
for establishing an equivalence between the invariance measure property and an SDE (\ref{e:con}) for the process $p$. 
Being based on this reduction 
methodology, this proof 
does not directly explain how a Girsanov drift can compensate a Jeulin-Yor drift. Given the importance of that matter for the purpose of this paper, we provide in \sr{proofinGJY}
an alternative proof of Theorem \ref{part1} based on this compensation.
 
\bt\label{part1}
A probability measure 
\index{p@$\mathbb{P}$}$\mathbb{P}$ equivalent to $\mathbb{Q}$ on ${\cal F}_{T}$ 
is an invariance measure
%satisfies the condition (A) 
if and only if
\qr{e:qf} (i.e. \qr{e:cor}) holds.
\et

\proof 
%Let $\mathcal{M}_{[0,T]}(\mathbb{F},\mathbb{P})$ (respectively $\mathcal{M}_{[0,T]}(\mathbb{F},\mathbb{Q})$) denote the set of all $(\mathbb{F},\mathbb{P})$ (respectively $(\mathbb{F},\mathbb{Q})$) local martingales on $[0,T]$. 
The invariance measure property for $\mathbb{P}$ is equivalent to $$(P^{\theta-})^T=(P^T)^{\theta-}\in\mathcal{M}(\mathbb{G},\mathbb{Q})
\sp
\forall P\in\mathcal{M}_{[0,T]}(\mathbb{F},\mathbb{P})
.$$ By Lemma \ref{optionalsplitting} 4), this holds if and only if $$
%P\in\mathcal{M}_{[0,T]}(\mathbb{F},\mathbb{P})\Longrightarrow
%%\forall P\in\mathcal{M}_{[0,T]}(\mathbb{F},\mathbb{P}),\
\ttS_-\centerdot P^T+[\ttS,P^T]
\in \mathcal{M}_{\In}(\mathbb{F},\mathbb{Q})\sp \forall P\in\mathcal{M}_{[0,T]}(\mathbb{F},\mathbb{P}).
%\mbox{ is an $(\mathbb{F},\mathbb{Q})$ local martingale on $\In $. }
$$
We proceed to transform this property into an SDE (\ref{e:con}) for $p$. 
The integration by parts formula and the Doob-Meyer decomposition of $\ttS$ yield
$$
P^T\ttS=
P^T_-\centerdot \ttS + \ttS_-\centerdot P^T + [P^T,\ttS] =P^T_-\centerdot \cM -P^T_-\centerdot \Alpha + \ttS_-\centerdot P^T + [P^T,\ttS].$$
Hence, the preceding property is equivalent to
\beql{e:previ}
%P\in\mathcal{M}_{[0,T]}(\mathbb{F},\mathbb{P})\Longrightarrow
P^T\ttS+P^T_-\centerdot \Alpha\in \mathcal{M}_{\In}(\mathbb{F},\mathbb{Q})\sp \forall P\in\mathcal{M}_{[0,T]}(\mathbb{F},\mathbb{P}).
%\forall P\in\mathcal{M}_{[0,T]}(\mathbb{F},\mathbb{P}),\
%P^T\ttS+P^T_-\centerdot \Alpha
%\mbox{ is an $(\mathbb{F},\mathbb{Q})$ local martingale on $\In $. }
\eeql
Note that
$
\mathcal{M}_{[0,T]}(\mathbb{F},\mathbb{P})
=
\{Qp; \ Q\in\mathcal{M}_{[0,T]}(\mathbb{F},\mathbb{Q}) \}.
$\footnote{Cf.~\shortciteN[Theorem 12.12]{HeWangYan92}.}
For any $Q\in\mathcal{M}_{[0,T]}(\mathbb{F},\mathbb{Q})$, an application of the integration by parts formula to $Q ^T( \monp^T_-\centerdot \Alpha)$ yields
$$
Q ^T(\monp^T\ttS+\monp^T_-\centerdot \Alpha)
=
(Q \monp)^T\ttS+(Q \monp)^T_-\centerdot \Alpha+(\monp^T_-\centerdot \Alpha)\centerdot Q ^T+ [Q ^T, \monp^T_-\centerdot \Alpha].
$$
In the right hand side, by Yoeurp's lemma,\footnote{See \shortciteN[Exercise 9.4 1)]{HeWangYan92}.} 
the bracket $[Q ^T, \monp^T_-\centerdot \Alpha]$ is 
in $\mathcal{M}(\mathbb{F},\mathbb{Q}),$
%an $(\mathbb{F},\mathbb{Q})$ local martingale,
 as is also $(\monp^T_-\centerdot \Alpha)\centerdot Q ^T$. Hence,
(\ref{e:previ}) is equivalent to
$$
%Q\in\mathcal{M}_{[0,T]}(\mathbb{F},\mathbb{Q})\Longrightarrow
Q ^T(\monp^T\ttS+\monp^T_-\centerdot \Alpha)\in\mathcal{M}_{\In}(\mathbb{F},\mathbb{Q})\sp \forall Q\in\mathcal{M}_{[0,T]}(\mathbb{F},\mathbb{Q}).
$$
%$$
%\forall Q\in\mathcal{M}_{[0,T]}(\mathbb{F},\mathbb{Q}),\
%Q ^T(\monp^T\ttS+\monp^T_-\centerdot \Alpha)
%\mbox{ is an $(\mathbb{F},\mathbb{Q})$ local martingale on $\In $. }
%$$
By Lemma \ref{orthonull}, this in turn is equivalent to
\beql{e:con}
\monp^T\ttS+\monp^T_-
\centerdot \Alpha =\monp_0\ttS_0
\mbox{ on $ \In\cap [0,T]$.}\eeql
Noting that
$\monp\ttS+\monp_-
\centerdot \Alpha
=
\monp\ttS+(\monp\ttS)_-\frac{1}{\ttS_-}\centerdot \Alpha$,
we recognize in \qr{e:con} the linear SDE for the stochastic exponential 
of $(-\frac{1}{\ttS_-}\centerdot \Alpha)$ with the initial condition $\monp_0\ttS_0$ on $\Int \cap[0,T]$, i.e.
\qr{e:con} is equivalent to
\beql{e:ief}\monp\ttS=\monp_0\ttS_0\mathcal{E}(-\frac{1}{\ttS_-}\centerdot \Alpha)\mbox{
on }\Int \cap[0,T] .\eeql
Recall $\{\ttS_- >0\}\setminus \{\pSigma >0\}=[\theeta]$ (cf. \qr{pS>0} and \qr{e:theeta}).
Actually, the identity \qr{e:ief} is equivalent to the analogous identity 
on the smaller set $\{^p\!\ttS>0\} \cap[0,T] $. To understand why, note that,
if $\theeta $ is finite, then $\ttS_{\theeta }=0$, 
whereas
\qr{e:theetasuite} yields
$\mathcal{E}(-\frac{1}{\ttS_-}\centerdot \Alpha)_\theeta=0,$
so that one has the trivial equality
$$
(\monp\ttS)_\theeta =\monp_0\ttS_0\mathcal{E}(-\frac{1}{\ttS_-}\centerdot \Alpha)_\theeta=0.$$ 
%independently of the invariance measure property or not regarding $\mathbb{P}$. 
Hence, the identity \qr{e:ief}
is equivalent to
\begin{equation}\label{pSA}\mbox{
$\monp\ttS=\monp_0\ttS_0\mathcal{E}(-\frac{1}{\ttS_-}\centerdot \Alpha)$
on $\{\pSigma >0\} \cap[0,T],$}\end{equation} 
i.e.,
in view of Lemma \ref{optionalsplitting} 5), to
$$\monp\ttS_0 \mathcal{E}(-\frac{1}{\ttS_-}\centerdot \Alpha) \mathcal{E}(\frac{1}{\pSigma }\centerdot 
 \cM )=\monp_0\ttS_0\mathcal{E}(-\frac{1}{\ttS_-}\centerdot \Alpha) \mbox{ on } \{\pSigma >0\} \cap[0,T],$$
which is equivalent to \qr{e:qf}, because $\ttS_0\mathcal{E}(-\frac{1}{\ttS_-}\centerdot \Alpha)$ is positive on
$\{\pSigma >0\},$ by \qr{pS>0}.
\finproof

\subsection{Az\'ema Supermartingale Characterization of Invariance Times}\label{firstproofofThm}

In this section we provide two (closely related) Az\'ema supermartingale characterizations of the 
condition (A). 
Given Theorem \ref{part1}, which designates $\monqs$
as the tentative $\ff$ density process of $\frac{d\mathbb{P}}{d\mathbb{Q}}$ for some invariance measure $\mathbb{P}$, all needs be done for 
verifying the condition (A)
 is to check the positivity and true martingale property of $\monqs$. 
%candidate density process.

\bt\label{A2bis} 
The condition (A) holds if and only if 
\beql{thehyp}
\left\{
\dcb
 \ind_{\{\pSigma>0\}}\frac{1}{\pSigma }\mbox{
 is $(\mathbb{F},\mathbb{Q})$ integrable with respect to $\cM$ on $[0,T]$ and
}\vspace{6pt}&\\
 \monqs
\mbox{
is
a positive $(\mathbb{F},\mathbb{Q})$ true martingale on $ [0,T].$}& 
%with respect to $(\mathbb{F},\mathbb{Q}).$}&
\dce
\right.
\eeql
If this is satisfied, then an invariance
measure $\mathbb{P}$ is defined by the $\mathbb{Q}$ density $\monqs_T$ on $\cF_T$ and
any invariance measure $\mathbb{P}$ is such that
\beql{e:yf} 
q_T=q_0\monqs_T \mathcal{E}(\ind_{\{\pSigma =0\}}\centerdot \therho )_T.
\eeql
\et

\proof 
Suppose the existence of an invariance measure $\mathbb{P}$. By Theorem \ref{part1}, 
the $\ff$ density process $q$ of $\frac{d\mathbb{P}}{d\mathbb{Q}}$ satisfies \qr{e:qf}.
By Lemma \ref{cond3738}, the $\mathbb{F}$ predictable process $\ind_{\{\pSigma>0\}}\frac{1}{\pSigma }$
is $(\mathbb{F},\mathbb{Q})$ integrable with respect to $\cM$ on $[0,T].$ 

In order to show that the process $\mathcal{Q}=\mathcal{E}(\ind_{\{\pSigma>0\}}\frac{1}{\pSigma}\centerdot \cM)$ (cf. Lemma \ref{lem:intmult}) is a positive $(\mathbb{F},\mathbb{Q})$ 
true 
martingale on $[0,T],$ it is enough to represent it as the $(\mathbb{F},\mathbb{Q})$ conditional expectation process of a positive
and $\mathbb{Q}$ integrable $\mathcal{F}_{T}$ measurable random variable.
By \qr{e:qf} and \qr{e:Sp}, for any $n\in\mathbb{N}$ and $t\in[0,T]$, we have
\beql{e:compu}
&q_0 \mathcal{E}(\ind_{\{\pSigma>0\}}\frac{1}{\pSigma}\centerdot \cM)^{\zeta_n\wedge T}_t
=
q_0 \mathcal{E}(\frac{1}{\pSigma}\centerdot \cM)^{\zeta_n\wedge T}_t
\\&\qqq=
q^{\zeta_n\wedge T}_t
=
\mathbb{E}[q_{\zeta_n\wedge T}|\mathcal{F}_{t}]
=
\mathbb{E}\Big[\ \mathbb{E}[q_T|\mathcal{F}_{\zeta_n\wedge T}] \ \Big|\mathcal{F}_{t}\Big]. 
\eeql
%\b{where the $\zeta_n$ have been introduced in \qr{e:Sp}}. 
%We now compute the limits of these processes when $n\uparrow \infty$.
On the one hand, noting that
%$\mathcal{Q}=\mathcal{E}(\ind_{\{\pSigma>0\}}\frac{1}{\pSigma}\centerdot \cM)$, hence
$\mathcal{Q}=1+\mathcal{Q}_-  \ind_{\{\pSigma>0\}}\frac{1}{\pSigma}\centerdot \cM$,
the dominated convergence theorem for stochastic integrals yields
$$
\dcb
&&\lim_{n\rightarrow\infty}\mathcal{Q}^{\zeta_n\wedge T}
=
1+\lim_{n\rightarrow\infty}\ind_{(0,\zeta_n\wedge T]}\centerdot \mathcal{Q}\\

&&\qqq=
1+\lim_{n\rightarrow\infty}\ind_{(0,\zeta_n\wedge T]}\mathcal{Q}_-\ind_{\{\pSigma>0\}}\frac{1}{\pSigma}\centerdot \cM

=
1+\mathcal{Q}_-\ind_{\{\pSigma>0\}}\frac{1}{\pSigma}\centerdot \cM
=
\mathcal{Q}.
\dce
$$
On the other hand, $$
%\mathbb{E}[q_T|\vee_{n\in\mathbb{N}}(\mathcal{F}_{\zeta_n\wedge T})]=
%L^1
\lim_{n\rightarrow\infty}\mathbb{E}[q_T|\mathcal{F}_{\zeta_n\wedge T}]=
\mathbb{E}\big[q_T\big|\vee_{n\in\mathbb{N}}(\mathcal{F}_{\zeta_n\wedge T})\big]
\mbox{ holds in $L^1$.}
$$
Hence, for $\ttt,$ we obtain by passing to the limit in \qr{e:compu}
\begin{equation}\label{Eunifmart}
\mathcal{E}(\ind_{\{\pSigma>0\}}\frac{1}{\pSigma}\centerdot \cM)_t
=
\mathbb{E}\Big[\mathbb{E}\big[ \frac{q_T}{q_0} |\vee_{n\in\mathbb{N}} \mathcal{F}_{\zeta_n\wedge T} \big] \ \Big|\mathcal{F}_{t}\Big],
\end{equation} 
where $ \mathbb{E}\big[ \frac{q_T}{q_0} \big|\vee_{n\in\mathbb{N}} \mathcal{F}_{\zeta_n\wedge T} \big]$ is a positive
and $\mathbb{Q}$ integrable $\mathcal{F}_{T}$ measurable random variable.
As a consequence, the process $\mathcal{E}(\ind_{\{\pSigma>0\}}\frac{1}{\pSigma}\centerdot \cM)$ is a positive $(\mathbb{F},\mathbb{Q})$ uniformly integrable martingale on $[0,T]$, which proves (\ref{thehyp}).

Conversely, supposing (\ref{thehyp}), we can define a probability measure $\mathbb{P}$ by an $\ff$ density process of $\frac{d\mathbb{P}}{d\mathbb{Q}}$ given as $
q = \mathcal{E}(\ind_{\{\pSigma>0\}}\frac{1}{\pSigma}\centerdot \cM)
$
on $[0,T]$, hence $\therho=\ind_{\{\pSigma>0\}}\frac{1}{\pSigma}\centerdot \cM$ on $[0,T]$. Given Lemma \ref{QS=0}, this establishes (\ref{e:cor}). By Theorem \ref{part1}, $\mathbb{P}$ is therefore an invariance measure, which proves the condition $(A)$. 

Assuming the condition (A), \qr{e:yf} is the consequence of \qr{e:cor} (which implies $\ind_{\{\pSigma>0\}}\therho=\ind_{\{\pSigma>0\}}\frac{1}{\pSigma}\centerdot \cM$)
and of the formula $\mathcal{E}(\therho)=\mathcal{E}(\ind_{\{\pSigma>0\}}\therho)\mathcal{E}(\ind_{\{\pSigma=0\}}\therho)$ (cf.~
\citeN[Proposition 6.4]{Jacod1979}).~\finproof\\

\noindent
In view of Lemma \ref{lem:intmult}, Theorem \ref{A2bis} can be restated in the following form.

\bt\label{c:mart}
The condition (A) holds if and only if the multiplicative martingale part $\cQ$ of $\ttS$ as of \qr{e:multdeclim} 
%\b{\footnote{Cf. \qr{e:multdec}.}}
 is a positive $(\ff,\Q)$ true martingale on $[0,T]$. In this case
we have $\cQ=\monqs$ on $[0,T].$~\finproof
%an invariance
%measure $\mathbb{P}$ is defined by the $\mathbb{Q}$ density $\cQ_T$ on $\cF_T$.
% and
%any invariance measure $\mathbb{P}$ is such that
%\beql{e:yfbis} 
%q=q_0\cQ_T \mathcal{E}(\ind_{\{\pSigma =0\}}\centerdot \therho )_T.
%\eeql 
\et

%\proof In view of Lemma \ref{lem:intmult}, this readily follows from Theorem \ref{A2bis}.~\ok\\
%% The theorem is the consequence of this remark.

%\subsection{Extreme Cases}
%Extreme
In what follows we examine in the light of 
Theorems \ref{part1} and \ref{A2bis}-\ref{c:mart}
the extreme cases where 
$\mathbb{F}=\mathbb{G}$ or $\mathbb{P}=\mathbb{Q}$ in the condition (A). 
 
\begin{pro}\label{c:mc}
\ebe
\item $\mathbb{Q}$ itself is an invariance measure
for all $T>0$ 
if and only if $\cM=0$.
\item If $\mathbb{F}=\mathbb{G}$ and $\theta$ is totally inaccessible with $\Q(\theta\le T)$ positive, then the condition (A) cannot hold on $[0,T]$. 
\dbe
\ep

\proof 1) In the case where $\mathbb{P}=\mathbb{Q}$, we have $q={q_0}$ on $[0,T]$, i.e. $\therho=0$ on $[0,T].$
Hence, in view of Theorem \ref{part1} and of \qr{e:cor}, $\mathbb{P}$ is an invariance measure for all $T>0$ if and only if $\cM=0$ on $[0,T].$\vspace{6pt}

\noindent
2) In the case where $\mathbb{F}=\mathbb{G}$ and $\theta$ is totally inaccessible, we have by \qr{e:SS} and Lemma \ref{l:inten}:
$$
\dcb
\ttS=\ttJ\sp\Alpha \mbox{ is continuous}\sp {\pSigma}=\ttJ_{-} \mbox{ and } \cM=\ttJ+\Alpha-1 . 
\dce
$$
Hence, using the stochastic exponential formula,\footnote{See e.g. \shortciteN[Theorem 9.39]{HeWangYan92}.}
\bel
&\monqs _t =
\mathcal{E} (\cM )_t=e^{\cM_t 
%-\cM_0
}\prod_{s\leq t}(1+ \Delta_s\cM)e^{-\Delta_s\cM} 
=e^{\ttJ_t +\Alpha_t -1} \ttJ_t 
%=e^{\ttJ_t +\Alpha_t -1} \ttJ_t e^{(1-\ttJ_t)}
=e^{\Alpha_t} \ttJ_t,
\eel 
which 
%is a $(\gg,\Q)$ true martingale on $[0,T]$ (by the key lemma of credit risk), but 
vanishes at $\theta$ on $\{\theta\leq T\}$.
Therefore, in view of Theorem \ref{A2bis}, the condition (A) cannot hold on $[0,T]$ unless $\Q(\theta\le T)=0$. \finproof 

\bex \label{e:posexp}
Let $\gg$ be the
augmentation of the natural filtration of the jump process at an exponential time $\theta$
relative to some probability measure $\Q.$

For $\ff=\gg$ (so that the condition (B) holds trivially),	
Proposition \ref{c:mc} 2) shows that the condition (A) does not hold. This can also be recovered directly from the definitions.
In fact, for any probability measure $\mathbb{P}$
equivalent to $\mathbb{Q}$ on ${\cal F}_{T},$ each $(\ff,\mathbb{P})$ local martingale $P$ has to be an $(\ff,\mathbb{P})$ stochastic integral with respect to the compensated jump process of $\theta$ (cf.~\shortciteN[Theorem 13.20]{HeWangYan92}). 
Thus, the process
$P^{\theta-}$ is absolutely continuous, 
hence it is not a $(\gg,\Q)$ local martingale on $[0,T]$
unless it is constant on $[0,T]$. Therefore, $\P$ is not an invariance measure for $\ff=\gg.$

For $\ff$ trivial, any
$\gg$ predictable process coincides with a Borel function before $\theta,$ so that the condition (B) is satisfied.
The constants are the only $(\ff=\{\emptyset,\Omega\},\Q)$ local martingales, so that 
$\Q$ itself is an invariance measure and the condition (A) is satisfied. 
Consistent with this conclusion in regard of {Theorem \ref{A2bis}}, $\ttS$ is deterministic (equal to the survival function of $\theta$), $\cM=0$ and $\monq \equiv 1$, hence {\qr{e:qf} is satisfied}. 

In conclusion, an exponential time $\theta$ in its own filtration $\gg$
%\b{??INV VS (A) 
is an invariance time for $\ff$ trivial but not for $\ff=\gg.$ 
\eex

\def\theM{M} 
\def\then{n}
\def\thisX{X}\def\thisX{N}\def\thisX{P}
\def\thisY{Y}\def\thisY{M}\def\thisY{Q} 
\def\thec{c}
\def\thed{d}
\def\bQ{\mathbb{\theQ}}\def\bQ{}

\def\fo{\mathbb{F}.o}\def\fo{o}
\def\fp{\mathbb{F}.p}\def\fp{p} 

\def\EQ{\mathbb{E}^{\mathbb{Q}}}\def\EQ{\mathbb{E}}
\def\EP{\widetilde{\mathbb{E}}}\def\EP{\mathbb{E}^{\mathbb{P}}}
\def\cro#1{{{\boldsymbol\langle} #1{\boldsymbol\rangle}}}\def\cro#1{\langle #1\rangle^{\mathbb{P}}}
\def\croc#1{\langle #1\rangle^{\mathbb{Q}}}\def\croc#1{{\langle#1\rangle}}
\def\cpq#1{{\langle#1\rangle}}

%\section{\b{Invariance Sufficiency Condition}}

\subsection{Positivity of $\monqs$ on $[0,T]$}\label{s:pos}
\def\theV{\rho}\def\theV{\sigma}

%In Theorem \ref{casuffi},
%we will apply Theorem \ref{A2bis} to study the relationship between the survival measure in \shortciteN{CollinDufresneGoldsteinHugonnier2004} and the invariance measure of this paper. 
The positivity condition of $\monqs$ on $[0,T]$ is a key element 
in the characterization 
%\qr{thehyp} 
of
 Theorem \ref{A2bis}. There exist general results on the positivity of a stochastic exponential.\footnote{Cf.~\shortciteN[Lemma 9.40]{HeWangYan92}.} But, for our proof of Theorem \ref{sfcnd} below, we need a different characterization. 
%To conclude this section,
In this section 
we show that the positivity of $\monqs$ (assuming that $\ind_{\{\pSigma>0\}}\frac{1}{\pSigma }$
is $(\mathbb{F},\mathbb{Q})$ integrable with respect to $\cM$ on $[0,T]$) can be characterized in terms of the time $\varsigma$ of first zero
of $\ttS$ (cf.~\qr{e:Ibis}). 
Specifically, the positivity of $\mathcal{E}(\ind_{\{\pSigma >0\}}\frac{1}{{\pSigma }}\centerdot \cM)$
reduces to
the predictability of the $\ff$ stopping time $\varsigma_{\{\varsigma\leq T\}}$. 

\bl\label{sig0}
Let $\theV$ be an $\ff$ predictable stopping time. Then, on the set $\{\sigma<\infty\}$, we have $\pSigma _\theV=0$ if and only if $\theV\geq \varsigma$.
\el

\proof By definition of the predictable projection and by nonnegativity of $\ttS,$ on the set $\{\sigma<\infty\}$, 
it holds:
$$\pSigma _\theV=0\Longleftrightarrow
 \mathbb{E}[\ttS_\theV|\mathcal{F}_{\theV-}]=0\Longleftrightarrow\ttS_\theV=0\Longleftrightarrow\theV \geq \varsigma.~\finproof$$

\bl\label{l:yan} The $\mathbb{F}$ stopping times
$\varsigma_{\{\varsigma<\infty,\ttS_{\varsigma-}=0\}}$ and $\varsigma_{\{\varsigma<\infty,\pSigma _\varsigma=0\}}$ are
predictable.
\el

\proof 
The assertion regarding $\varsigma_{\{\varsigma<\infty,\ttS_{\varsigma-}=0\}}$
comes from the proof of Theorem 9.41 in \shortciteN{HeWangYan92}. For $\varsigma_{\{\varsigma<\infty,\pSigma _\varsigma=0\}}$, it suffices to note that$$
\varsigma_{\{\varsigma<\infty, \pSigma _\varsigma=0\}}
=
\varsigma_{\{\varsigma<\infty,\ttS_{\varsigma-}=0\}}
\wedge
\varsigma_{\{\varsigma<\infty,\ttS_{\varsigma-}>0,\pSigma _\varsigma=0\}}
=
\varsigma_{\{\varsigma<\infty,\ttS_{\varsigma-}=0\}}
\wedge
\eta,
$$
where $\eta$ is the $\mathbb{F}$ predictable time introduced in Lemma \ref{Seta}. \ok

\bethe\label{t:cerise} 
Assuming that
$\ind_{\{\pSigma>0\}}\frac{1}{\pSigma }$ is $(\mathbb{F},\mathbb{Q})$ integrable with respect to $\cM$ on $[0,T],$ 
the following conditions 
%\b{(almost surely)} 
are equivalent to each other: 
\begin{enumerate}[\rm (i)]
\item $\monqs>0$ on {$[0,T]$}, 
\item $\pSigma _\varsigma=0$ on $\{\varsigma\leq T\}$,
\item $\varsigma_{\{\varsigma\leq T\}}$ is a predictable stopping time.
\end{enumerate}
\ethe 

\proof Note that
\begin{equation*}
\ind_{\{\pSigma _t>0\}}\Delta_t\Big( \frac{1}{\pSigma }\centerdot \cM \Big)
=
\ind_{\{\pSigma _t>0\}}\frac{1}{\pSigma _t}\Delta_t\cM = \ind_{\{\pSigma _t>0\}} \big(\frac{\ttS_t}{\pSigma _t}-1\big) \sp \forall t\in[0,T],
\end{equation*}
where \qr{e:SS} was used in the last equality.
Hence, $\mathcal{E}(\ind_{\{\pSigma >0\}}\frac{1}{\pSigma }\centerdot \cM )$ is positive on $[0,T]$ if and only if
\begin{equation}\label{ppss}
%\ind_{\{\pSigma _t>0\}}\Delta_t\Big( \frac{1}{\pSigma }\centerdot \cM \Big)
%=\ind_{\{\pSigma _t>0\}}\frac{1}{\pSigma _t}\Delta_t\cM=
 \ind_{\{\pSigma _t>0\}} \big(\frac{\ttS_t}{\pSigma _t}-1\big) >-1 \sp\forall t\in[0,T].
\end{equation}
Recalling 
\qr{e:sss} and \qr{yorthetaprel}, the only way
\qr{ppss} can break down is if $\varsigma\leq T$ and 
$\pSigma _\varsigma>0$, which shows the equivalence between (i) and (ii).

To prove the equivalence between (ii) and (iii), let $\xi=\varsigma_{\{\varsigma<\infty,\pSigma _\varsigma=0\}},$ which is predictable by Lemma \ref{l:yan}.
 If \begin{equation}\label{stst}
\pSigma _\varsigma=0 \mbox{ on } \{\varsigma\leq T\},
\end{equation}
then $$\varsigma_{\{\varsigma\leq T\}}=(\varsigma_{\{\varsigma<\infty,\pSigma _\varsigma=0\}})_{\{\varsigma\leq T\}}=\xi_{\{\varsigma\leq T\}}
$$
and 
$$
\{\varsigma\leq T\}=\{\varsigma_{\{\varsigma<\infty,\pSigma _\varsigma=0\}}\leq T\}=\{\xi\leq T\} 
$$ 
{is $\cF_{\xi-}$ measurable. Hence, by \shortciteN[Theorem 3.29 7)]{HeWangYan92}, $\varsigma_{\{\varsigma\leq T\}}=\xi_{\{\varsigma\leq T\}}$ is predictable. Conversely, if $\varsigma_{\{\varsigma\leq T\}}$ is predictable, as $\varsigma_{\{\varsigma\leq T\}}\geq \varsigma$, the condition (\ref{stst}) holds by Lemma \ref{sig0}. \ok\\

\noindent
In particular, if $\ttS_T$ is positive almost surely, then 
$\varsigma_{\{\varsigma\leq T\}}=\infty,$ which is a predictable stopping time.
Therefore
$\mnqs$ is positive on $[0,T]$, consistently with the last statement in
Lemma \ref{optionalsplitting} 5).
\begin{ex}\rm
Consider $\ff=\gg$ and $\theta$ given as a totally inaccessible stopping time such that
$\mathbb{Q}(\theta\le T)>0$. Then $\varsigma=\theta$, so that the stopping time $\varsigma_{\{\varsigma\leq T\}}$ is not predictable. Accordingly,
as seen in the proof of Proposition \ref{c:mc}, 
$\monqs$
vanishes 
at $\theta$ on $\{\theta\leq T\}$.
\end{ex}

\subsection{True Martingale Property of $\cQ$}\label{ss:mar}

% of $\monqs,$
The second key ingredient in the characterization of Theorem \ref{A2bis} is the $(\mathbb{F},\mathbb{Q})$ true martingale property of $\monqs$ on $ [0,T]$ (on top of its positivity). Regarding the true martingale property of a Dol\'eans-Dade exponential, one immediately thinks of Novikov-Kazamaki type conditions (see \citeN{LarssonRuf14} for a survey). However, under the condition (B), the true martingale property of $\monqs$ or, more precisely, of the multiplicative martingale part $\cQ$ of $\ttS$ as of \qr{e:multdeclim} (cf. Theorems \ref{A2bis}-\ref{c:mart}),
can be better studied by means of Az\'ema supermartingale computations.
Note that, in the case of any nonnegative $(\ff,\Q)$ local martingale (hence supermartingale) $Q$, the $(\ff,\Q)$ true martingale property of $Q$ on $[0,T]$ is equivalent to $\E Q_T=\E Q_0$.

The following example shows that there exists times $\theta$ for which $\cQ$ is not a true martingale.

\bex\label{ex:invBessel}
Let $X$ be the inverse of a three dimensional $(\ff,\Q)$ Bessel process\footnote{Cf. , \citeN[Chapter V, Exercice (2.13)]{RevuzYor99}.} starting from $1.$ 
Define $X^*_{t}=\sup_{0\leq s\leq t}X_{s},$ for $t\geq 0$, and let $\theta=\sup\{t\geq 0: X_{t}=X^*_{t}\}$. According to 
\citeN{NikeghbaliYor2006},
$\ttS=\frac{X}{X^*}$ is the Az\'ema supermartingale of $\theta$. We have $\cQ=X$, which is not an $(\ff,\Q)$ true martingale on any non empty interval $[0,T]$. 
\eex
 
Our ensuing study of the true martingale property of $\cQ$ on $[0,T]$ is based on the following properties. Note that $\mathcal{E}(\frac{1}{^p\!\ttS}\centerdot \Alpha)$ is well-defined on the set $\{\pSigma>0\}$ and, since $\theta\in\{\ttS_->0\}$ on $\{0<\theta<\infty\}$ (cf. \qr{yortheta}), we have $[0,\theta)\subset \{^p\!\ttS>0\}$.

\bl\label{trsf} 
\ebe
\item 
The $(\mathbb{F},\mathbb{Q})$ optional projection of the process $\mathcal{E}(\frac{1}{^p\!\ttS}\centerdot \Alpha)\ind_{[0,\theta)}$ is equal to the process $\mathcal{E}(\frac{1}{^p\!\ttS}\centerdot \Alpha)\mathtt{S}\ind_{\{^p\!\ttS>0\}}=\ttS_0 \mnqs \ind_{\{^p\!\ttS>0\}}=\ttS_0 \cQ\ind_{\{^p\!\ttS>0\}}$. 

\item 
The process $\mathcal{E}(\frac{1}{^p\!\ttS}\centerdot \Alpha)\ind_{[0,\theta)}$ is a $(\mathbb{G},\mathbb{Q})$ local martingale on $\{^p\!\ttS>0\}$. 

\item 
If the family of the random variables $\mathcal{E}(\frac{1}{^p\!\ttS}\centerdot \Alpha)_ \sigma\ind_{\{ \sigma<\theta\}}$, for any $\mathbb{F}$ stopping time $ \sigma$ such that $[0, \sigma]\subset \{^p\!\ttS>0\}\cap[0,T]$, is $\mathbb{Q}$ uniformly integrable, then $\cQ$ is an $(\mathbb{F},\mathbb{Q})$ true martingale on $ [0,T]$.
\dbe

\el

\proof 
1) Let $\sigma$ be any $\mathbb{F}$ stopping time and $\chi$ be any bounded $\mathcal{F}_\sigma$ measurable random variable. We have
(using $[0,\theta)\subset \{^p\!\ttS>0\}$)
 %\b{by \qr{yortheta},}
\beql{proj}
&\mathbb{E}[\chi \mathcal{E}(\frac{1}{^p\!\ttS}\centerdot \Alpha)_{\sigma}\ind_{\{\sigma<\theta\}}]
=
\mathbb{E}[\chi \mathcal{E}(\frac{1}{^p\!\ttS}\centerdot \Alpha)_{\sigma}\ind_{\{\sigma<\theta\}}\ind_{\{\sigma \in \{^p\!\ttS>0\} \}}]\\
&\qqq
=
\mathbb{E}[\chi \mathcal{E}(\frac{1}{^p\!\ttS}\centerdot \Alpha)_{\sigma}\mathtt{S}_\sigma\ind_{\{\sigma \in \{^p\!\ttS>0\} \}}\ind_{\{\sigma<\infty\}}].
\eeql
Hence, the $(\mathbb{F},\mathbb{Q})$ optional projection of the process $\mathcal{E}(\frac{1}{^p\!\ttS}\centerdot \Alpha)\ind_{[0,\theta)}$ is equal to the process 
$\mathcal{E}(\frac{1}{^p\!\ttS}\centerdot \Alpha)\mathtt{S}\ind_{\{^p\!\ttS>0\}} $. Moreover, by Lemma \ref{optionalsplitting} 5), on $\{\pSigma>0\}$, we have
$$
\mathcal{E}(\frac{1}{^p\!\ttS}\centerdot \Alpha)\ \mathtt{S}
=
\mathcal{E}(\frac{1}{^p\!\ttS}\centerdot \Alpha)\ \ttS_0 \mathcal{E}(-\frac{1}{\ttS_{-}}\centerdot \Alpha)
\mathcal{E}(\frac{1}{\pSigma}\centerdot\cM)=
\ttS_{0}\mathcal{E}(\frac{1}{\pSigma}\centerdot\cM)
=
\ttS_{0}\cQ. 
$$

\noindent
2) Note that $\ttS_{0}\mathcal{E}(\frac{1}{\pSigma}\centerdot\cM)$ is an $(\mathbb{F},\mathbb{Q})$ local martingale on $\{^p\!\ttS>0\}$. Let $\sigma$ be a finite $\mathbb{F}$ stopping time 
%"\zeta_n \wedge k"
such that $[0,\sigma]\subset\{^p\!\ttS>0\}$
and $\ttS_{0}\mathcal{E}(\frac{1}{\pSigma}\centerdot\cM)^\sigma$ is an $(\mathbb{F},\mathbb{Q})$ uniformly integrable martingale. Consider any $\mathbb{G}$ stopping time $\tau$. 
Recalling that $\tau'$ is an $\mathbb{F}$ stopping time such that $\tau\wedge \theta=\tau' \wedge \theta$ (cf. Lemma \ref{optionalsplitting} 1)), we have
\beql{l:pouryan}
& \mathbb{E}[\mathcal{E}(\frac{1}{^p\!\ttS}\centerdot \Alpha)_{\tau\wedge \sigma}\ind_{\{\tau\wedge \sigma<\theta\}}]
=
\mathbb{E}[\mathcal{E}(\frac{1}{^p\!\ttS}\centerdot \Alpha)_{\tau' \wedge \sigma}\ind_{\{\tau' \wedge \sigma<\theta\}}]\\
&\qqq
=
\mathbb{E}[\mathcal{E}(\frac{1}{^p\!\ttS}\centerdot \Alpha)_{\tau' \wedge \sigma}\mathtt{S}_{\tau' \wedge \sigma}]= 
\mathbb{E}[\ttS_{0}\mathcal{E}(\frac{1}{\pSigma}\centerdot\cM)_{\tau' \wedge \sigma}]
=
\mathbb{E}[\ttS_{0}],
\eeql
where part 1) was used in the second line.
As a consequence, according to \shortciteN[Theorem 4.40]{HeWangYan92}, $(\mathcal{E}(\frac{1}{^p\!\ttS}\centerdot \Alpha)\ind_{[0,\theta)})^{\sigma}$ is a $(\mathbb{G},\mathbb{Q})$ uniformly integrable martingale, which proves the second assertion of the lemma.\vspace{6pt}

\noindent 3) By virtue of de la Vall\'ee-Poussin theorem,
there exists a nonnegative increasing convex 
%test 
function $\phi$ 
such that
$$\lim_{t\to \infty} \frac{\phi(t)}{t}=\infty
\mbox{ and }
\sup_ \sigma\mathbb{E}[\phi(\mathcal{E}(\frac{1}{^p\!\ttS}\centerdot \Alpha)_ \sigma\ind_{\{ \sigma<\theta\}})]<\infty,
$$
%for the uniform integrability of the family of the random variables $\mathcal{E}(\frac{1}{^p\!\ttS}\centerdot \Alpha)_ \sigma\ind_{\{ \sigma<\theta\}}$, 
where $ \sigma$ runs over the family of $\mathbb{F}$ stopping times such that $[0, \sigma]\subset \{^p\!\ttS>0\}\cap[0,T]$. 
Applying the part 1) of the lemma and Jensen's inequality, we obtain$$
%projection=esp cond!
\sup_ \sigma\mathbb{E}[\phi(\ttS_0\cQ_ \sigma\ind_{\{^p\!\ttS_{ \sigma } >0\}})]<\infty.
$$
Hence, another application of the de la Vall\'ee-Poussin theorem yields that $\ttS_0\cQ$ is a martingale of class $(D)$ on the set $\{^p\!\ttS>0\}\cap[0,T]$. In view of \qr{e:multdeclim}, we have
$$
 \cQ= \lim_{n\rightarrow\infty}  \monqs^{\zeta_{n}}.
$$
As $\zeta_{n}\in \{^p\!\ttS>0\}$ for every $n\in\mathbb{N}$ (cf. \qr{e:Sp}), then, on the one hand, the family of random variables $\{\ttS_{0}\monqs_{\zeta_{n}\wedge T}, n\in\mathbb{N}\}$ is uniformly integrable and, on the other hand, $\ttS_{0}\monqs^{\zeta_{n}\wedge T}$ is a uniformly integrable martingale, for every $n\in\mathbb{N}$. 
These two properties imply the two following equalities$$
\mathbb{E}[\ttS_{0}\cQ_{T}]
 = 
\lim_{n\rightarrow\infty}\mathbb{E}[\ttS_{0}\monqs_{\zeta_{n}\wedge T}]
=\mathbb{E}[\ttS_{0}],
$$ 
which proves that the nonnegative $(\mathbb{F},\mathbb{Q})$ local martingale $\ttS_{0} \cQ $
%easy: supermart->esp cond dans un sens->il n'y a plus qu'a s'assurer que l'esp incond est constante. See http://math.stackexchange.com/questions/177471/supermartingale-with-constant-expectation-is-a-martingale
 is an $(\mathbb{F},\mathbb{Q})$ true martingale on $ [0,T]$. We now write, for any $0\leq t\leq T$,$$
\ttS_{0}\mathbb{E}[\cQ_{T}|\mathcal{F}_{t}]=\mathbb{E}[\ttS_{0}\cQ_{T}|\mathcal{F}_{t}] = \ttS_{0}\cQ_{t}.
$$
As $\cQ\equiv 1$ on $\{\ttS_{0}=0\}$, we have in turn
$$
\dcb
\mathbb{E}[\cQ_{T}|\mathcal{F}_{t}]
&=&
\mathbb{E}[\cQ_{T}|\mathcal{F}_{t}]\ind_{\{\ttS_{0}>0\}}
+
\mathbb{E}[\cQ_{T}|\mathcal{F}_{t}]\ind_{\{\ttS_{0}=0\}}\\
&=&
\cQ_{t}\ind_{\{\ttS_{0}>0\}}+ \ind_{\{\ttS_{0}=0\} }
=
 \cQ_{t},\\
\dce
$$ 
which finishes the demonstration.
\ \finproof\\

We now state a sufficient condition for the true martingale property of $\cQ$.
\bethe\label{sfcnd}
If 
$
\mathbb{E}[\mathcal{E}(\ind_{\{\pSigma>0\}}\frac{1}{^p\!\ttS}\centerdot \Alpha)_{\theta\wedge T}]<\infty,
$ 
then $\cQ$
%\b{the martingale part $\cQ$ of $\ttS$ in its predictable multiplicative decomposition}
%$\monqs$
is
an $(\mathbb{F},\mathbb{Q})$ true martingale on $ [0,T].$
Assuming $\ttS_{T}$ positive, then $\theta$ is an invariance time. 
Assuming further $\theta$ positive, then 
the nonnegative $(\mathbb{G},\mathbb{Q})$ local martingale 
$\mathcal{E}(\ind_{\{\pSigma>0\}}\frac{1}{^p\!\ttS}\centerdot \Alpha)\ind_{[0,\theta)}$ (by Lemma \ref{trsf} 2)
is a $(\mathbb{G},\mathbb{Q})$ true martingale
on $[0,T]$, and an invariance measure $\P$ is provided by the restriction
to $\mathcal{F}_{T}$ of the
probability measure $\mathbb{S}$ with $\mathcal{E}(\ind_{\{\pSigma>0\}}\frac{1}{^p\!\ttS}\centerdot \Alpha)\ind_{[0,\theta)}$ 
%(or very much related quantities) 
as $\gg$ density process of $\frac{d\mathbb{S}}{d\Q}.$
\ethe

\proof 
If 
$
\mathbb{E}[\mathcal{E}(\ind_{\{\pSigma>0\}}\frac{1}{^p\!\ttS}\centerdot \Alpha)_{\theta\wedge T}]<\infty,
$ 
then the nonnegative $(\mathbb{F},\mathbb{Q})$ local martingale 
$\mathcal{E}(\frac{1}{^p\!\ttS}\centerdot \Alpha)\ind_{[0,\theta)}$ is a $(\mathbb{G},\mathbb{Q})$ martingale of class $(D)$ on $\{^p\!\ttS>0\}\cap[0,T]$. The $(\mathbb{F},\mathbb{Q})$ true martingale property of $\cQ$ on $ [0,T]$ 
is then the consequence of Lemma \ref{trsf} 3). If, in addition,
the positivity of 
 $\ttS_{T}$ is assumed, then \qr{e:sss} implies that $T<\varsigma$ and $[0,T]\subset \{\pSigma>0\},$
so that $\cQ=\mnqs>0$ holds on $[0,T]$, by Lemma \ref{optionalsplitting} 5).
Hence, Lemma \ref{lem:intmult} and Theorems \ref{A2bis}-\ref{c:mart} imply that $\theta$ is an invariance time.
%Lemma \ref{lem:intmult} and Theorem \ref{t:cerise} imply that $\cQ=\mnqs$ is positive on $[0,T].$
%Hence 
%$\theta$ is an invariance time, by Theorems \ref{A2bis}-\ref{c:mart}. 
Assuming further $\theta$ positive, i.e. $\ttS_0=1,$ 
Lemma \ref{trsf} 1) implies that the $(\mathbb{F},\mathbb{Q})$ optional projection of the $\gg$ density process of
$\frac{d\mathbb{S}}{d\mathbb{Q}}$ is $\cQ.$ But, in view of Theorems \ref{A2bis}-\ref{c:mart}, 
$\cQ$
is the $\ff$ density process of $\frac{d\mathbb{P}}{d\mathbb{Q}}$ for some invariance measure $\mathbb{P},$ which is therefore the restriction
to $\mathcal{F}_{T}$ of the
probability measure $\mathbb{S}.$ \ok\\

On top of the above sufficient condition for the true martingale property of $\cQ$ on $[0,T]$, we now look for a necessary and sufficient condition. According to \shortciteN[Theorem 8.18]{HeWangYan92} and \citeN[Lemme 1.37]{Jacod1979}, the first two conditions can always be made to hold in what follows. Hence, only the third one 
is really material for the true martingale property.

\bethe\label{ncsfcnd}
Suppose $\ttS_{T}$ positive. Then, the $(\ff,\Q)$ true martingale property of $\cQ$ on $[0,T]$ holds if and only if 
there exists a nondecreasing sequence of $\mathbb{F}$ stopping times $(\sigma_{n})_{n\in\mathbb{N}}$ such that 
\ebe
\item[-]
$\cup_{n}[0,\sigma_{n}]=[0,T]$,
\item[-]
$\mathcal{E}(\frac{1}{^p\!\ttS}\centerdot \Alpha)_{\sigma_{n}}$ is bounded,
for any $n \in\N,$
\item[-]
The family of random variables 
$
(\Alpha_{T}-\Alpha_{\sigma_{n}})\mathcal{E}(\frac{1}{^p\!\ttS}\centerdot \Alpha)_{\sigma_{n}}\sp n\in\N,
$ 
is $\mathbb{Q}$ uniformly integrable.
\dbe
In this case, the condition (A) is satisfied.
\ethe 
\proof
By the argument used at the end of the proof of Lemma \ref{trsf}, the $(\ff,\Q)$ true martingale property of $\cQ$ on $[0,T]$ is equivalent to $\mathbb{E}[\ttS_{0}\cQ_{T}]=\mathbb{E}[\ttS_{0}]$.

Let there be given a nondecreasing sequence of $\mathbb{F}$ stopping times $(\sigma_{n})_{n\in\mathbb{N}}$ satisfying the first two conditions in the above.

%\b{Since $\cQ$ is a nonnegative $(\ff,\Q)$ local martingale, the $(\ff,\Q)$ true martingale property of $\cQ$ on $[0,T]$ is equivalent to $\mathbb{E}[\cQ_{T}]=1$} or, By the argument used at the end of the proof of Lemma \ref{trsf}, the $(\ff,\Q)$ true martingale property of $\cQ$ on $[0,T]$ is equivalent to $\mathbb{E}[\ttS_{0}\cQ_{T}]=\mathbb{E}[\ttS_{0}]$.
 
On the one hand, by Lemma \ref{optionalsplitting} 5) (having assumed $\ttS_T>0$), we have
$$
\mathbb{E}[\ttS_{0}\cQ_{T}]
=\mathbb{E}[\ttS_{0}\mnqs_{T}] 
=
\mathbb{E}[\mathcal{E}(\frac{1}{^p\!\ttS}\centerdot \Alpha)_{T}\ttS_{T}]
=\lim_{n\rightarrow\infty}\mathbb{E}[\mathcal{E}(\frac{1}{^p\!\ttS}\centerdot \Alpha)_{\sigma_{n}}\ttS_{T}],
$$
by the monotone convergence theorem.

On the other hand,
we have $[0,\sigma_n]\subset [0,T]\subset\{\ttS>0\}\subset \{^p\!\ttS>0\}$ (as $\ttS_{T}>0$), so that, for any $n\in\mathbb{N}$, $\ttS_{0}\mnqs^{\sigma_{n}}$ is a well-defined $(\mathbb{F},\mathbb{Q})$ local martingale. Actually, $\ttS_{0}\mnqs^{\sigma_{n}}$ is a bounded martingale 
because, by Lemma \ref{optionalsplitting} 5), it is equal to $
\mathcal{E}(\frac{1}{^p\!\ttS}\centerdot \Alpha)^{\sigma_{n}}\ttS^{\sigma_{n}},
$ 
which is bounded by assumption.
As a consequence, \beql{e:QS}
\mathbb{E}[\ttS_{0}]=\mathbb{E}[\ttS_{0}\mnqs_{\sigma_{n}}]=\mathbb{E}[\mathcal{E}(\frac{1}{^p\!\ttS}\centerdot \Alpha)_{\sigma_{n}}\ttS_{\sigma_{n}}]
.
\eeql

Hence, $\cQ$ is an $( \mathbb{F} ,\mathbb{Q})$ true martingale on $[0,T]$ if and only if 
%\beql{e:iff}
%\lim_{n\rightarrow\infty} \mathbb{E}[\mathcal{E}(\frac{1}{^p\!\ttS}\centerdot \Alpha)_{\sigma_{n}}\ttS_{T}]
%=
%\mathbb{E}[\ttS_{0}] .
%\eeql
% is equivalent to
$$
\lim_{n\rightarrow\infty}\Big(\mathbb{E}[\mathcal{E}(\frac{1}{^p\!\ttS}\centerdot \Alpha)_{\sigma_{n}}\ttS_{\sigma_{n}}]-\mathbb{E}[\mathcal{E}(\frac{1}{^p\!\ttS}\centerdot \Alpha)_{\sigma_{n}}\ttS_{T}]\Big)=0.
$$
But, by definition of $\ttS,$
%$(\ff,\Q)$ optional projection, 
we have
\bel
&\mathbb{E}[\mathcal{E}(\frac{1}{^p\!\ttS}\centerdot \Alpha)_{\sigma_{n}}\ttS_{\sigma_{n}}]-\mathbb{E}[\mathcal{E}(\frac{1}{^p\!\ttS}\centerdot \Alpha)_{\sigma_{n}}\ttS_{T}]
=
\mathbb{E}[\mathcal{E}(\frac{1}{^p\!\ttS}\centerdot \Alpha)_{\sigma_{n}}\ind_{\{\sigma_{n}<\theta\leq T\}}]
{= 
\mathbb{E}[(\Alpha_{T}-\Alpha_{\sigma_{n}})\mathcal{E}(\frac{1}{^p\!\ttS}\centerdot \Alpha)_{\sigma_{n}}]},
\eel
by definition of $\Alpha$ as $(\ff,\Q)$ dual predictable projection of ${\ind_{\{0<\ftime\}}}\ttH$.\footnote{Cf.~\shortciteN[Theorem 5.26 2)]{HeWangYan92}.}

Hence, the true martingale property of $\cQ$ on $[0,T]$ is equivalent to the $L^1$ convergence to zero of the sequence of random variables $(\Alpha_{T}-\Alpha_{\sigma_{n}})\mathcal{E}(\frac{1}{^p\!\ttS}\centerdot \Alpha)_{\sigma_{n}}$. 
Now, as $\ttS_{T}>0$, we have $\mathcal{E}(\frac{1}{^p\!\ttS}\centerdot \Alpha)_{T}<\infty$. The random variables $(\Alpha_{T}-\Alpha_{\sigma_{n}})\mathcal{E}(\frac{1}{^p\!\ttS}\centerdot \Alpha)_{\sigma_{n}}$ converge in probability to zero. Therefore, their $L^1$ convergence to zero is equivalent to their uniform integrability,\footnote{Cf.~\shortciteN[Theorem 1.11]{HeWangYan92}.} 
 which concludes the proof of the equivalence stated in the theorem.

Moreover, having assumed $\ttS_T$ positive, Lemma \ref{optionalsplitting} 5) implies that
$\cQ=\mnqs>0$ holds on $[0,T].$ 
%whereas 
%the final part in Theorem
%\ref{t:cerise} says that $\mnqs$ is positive on $[0,T].$ Hence, 
%$\cQ$ is positive on $[0,T],$ so that
Hence, 
the condition (A) reduces to the $(\ff,\Q)$ true martingale property of $\cQ$ on $[0,T]$, by Theorems \ref{A2bis}-\ref{c:mart}.
\ \ok\\

\noindent
As shown in Lemma \ref{l:inten}, whenever $\theta$ has a $(\gg,\Q)$ intensity $\gamma$,
we have $\mathtt{D}= {\varphi}{\ttS_-} .\boldsymbol\lambda$ and $\mathcal{E}(\frac{1}{\pSigma}{\centerdot}\mathtt{D})=e^{\gamma' {\centerdot}\boldsymbol{\lambda}}$ on $[0,T]$ (assuming $\ttS_T>0$). Hence, in this case, Theorem \ref{ncsfcnd} reduces to a condition on $\gamma.$

\brem
In the case $\ttS_{T}>0$, Theorem \ref{sfcnd} can be deduced from Theorem \ref{ncsfcnd}. In fact, given
a sequence $(\sigma_{n})_{n\in\mathbb{N}}$ (which exists) satisfying the first two conditions in Theorem
 \ref{ncsfcnd}, we have the inequalities
\beql{e:lim}
(\Alpha_{T}-\Alpha_{\sigma_{n}})\mathcal{E}(\frac{1}{^p\!\ttS}\centerdot \Alpha)_{\sigma_{n}}
\leq
\int_{\sigma_{n}}^T\mathcal{E}(\frac{1}{^p\!\ttS}\centerdot \Alpha)_{s}d\Alpha_{s}
\leq
\int_{0}^T\mathcal{E}(\frac{1}{^p\!\ttS}\centerdot \Alpha)_{s}d\Alpha_{s}.
\eeql
Since $\Alpha$ is the $(\mathbb{F},\mathbb{Q})$ dual predictable projection of $\ind_{\{\theta>0\}}\ind_{[\theta,\infty)}$, we have
$$
\mathbb{E}[\mathcal{E}(\frac{1}{^p\!\ttS}\centerdot \Alpha)_{\theta\wedge T}]
=
\mathbb{E}[\ind_{\{\theta=0\}}]
+
{\mathbb{E}[\int_{0}^T\mathcal{E}(\frac{1}{^p\!\ttS}\centerdot \Alpha)_{s}d\Alpha_{s}]}
+
\mathbb{E}[\mathcal{E}(\frac{1}{^p\!\ttS}\centerdot \Alpha)_{T}\ind_{\{T\leq \theta\}}]
.
$$
Hence, the assumption of Theorem \ref{sfcnd} implies
that ${\mathbb{E}[\int_{0}^T\mathcal{E}(\frac{1}{^p\!\ttS}\centerdot \Alpha)_{s}d\Alpha_{s}]}$ is finite. As a consequence in view of \qr{e:lim}, the sequence of random variables $(\Alpha_{T}-\Alpha_{\sigma_{n}})\mathcal{E}(\frac{1}{^p\!\ttS}\centerdot \Alpha)_{\sigma_{n}}$ tends to zero in $L^1$,
by the dominated convergence theorem. One can then apply Theorem \ref{ncsfcnd} to deduce that
$\cQ$ 
is
an $(\mathbb{F},\mathbb{Q})$ true martingale on $ [0,T],$ as concluded in Theorem \ref{sfcnd}.\ \ok
\erem

\subsection{Local Martingales Under an Invariance Measure}\label{ss:carmap}

In this section we characterize
the local martingales under an invariance measure $\mathbb{P}.$ 

On top of the theoretical interest,
the following result is the key for the demonstration of the equivalence in \citeN{Crepey13a1} between the counterparty risk ``full'' $(\gg,\mathbb{Q})$ BSDE \qr{equat5} and the ``reduced'' $(\ff,\mathbb{P})$ BSDE
\qr{equat7} (cf.~\sr{s:bsdecr}).

\bt\label{c:pq} 
Assume the condition (A),
% on $[0,T],$ 
with an invariance measure 
\index{p@$\mathbb{P}$}$\mathbb{P}$.
\ebe
\item 
A process $P$ is in $\mathcal{M}_{\{\pSigma >0\}\cap[0,T]}(\mathbb{F},\mathbb{P})$
if and only if
${\pSigma }{\centerdot }P+[\cM,P]$ is in $\mathcal{M}_{\{\pSigma >0\}\cap[0,T]}(\mathbb{F},\mathbb{Q}).$

%A process $P$ is an
%$(\ff,\P)$ local martingale on $\{\pSigma >0\}\cap[0,T]$ if and only if
%${\pSigma }{\centerdot }P+[\cM,P]$ is an $(\ff,\Q)$ local martingale on $\{\pSigma >0\}\cap[0,T]$,
\item 
%In the case where $\ftime$ has a continuous compensator $\Alpha$ 
If $\mathtt{D}$ is continuous,
then 
$\pSigma =\ttS_- $ and 
the previous condition becomes 
\beql{e:conma}{\ttS_{-}}{\centerdot }P+ [\ttS,P]\in\mathcal{M}_{\{\ttS_{-} >0\}\cap[0,T]}(\mathbb{F},\mathbb{Q}).\eeql
In addition, we have
\beql{yorthetaprelsup}
 \{\ttS>0\}=\{\pSigma >0\}=\{\ttS_{-}>0\} = [0,\varsigma) ,
\eeql 
\dbe
where $\varsigma$ is the time of first zero of $\ttS$ (cf. \qr{e:Sp}).
\et

\proof 
1) On $\{\pSigma >0\}\cap[0,T]$, we have
$$
\dcb
q P
&=&
P_-{\centerdot }q +q _-{\centerdot } {P
}+[q ,P]\\

&=&P_-{\centerdot }q + q _-{\centerdot }P+q _-\frac{1}{\pSigma }{\centerdot }[\cM ,P],

\dce
$$
where the second equality comes from the formula (\ref{e:yf}) in Theorem \ref{A2bis}. 
Hence, $P$ is in
$\mathcal{M}_{\{\pSigma >0\}\cap[0,T]}(\ff,\P),$ i.e. $q P$ is in
$\mathcal{M}_{\{\pSigma >0\}\cap[0,T]}(\ff,\Q),$\footnote{Cf.~\shortciteN[Theorem 12.12]{HeWangYan92}.} 
if and only if 
$q _-{\centerdot }P+q _-\frac{1}{\pSigma }{\centerdot }[\cM , P]$ is in
$\mathcal{M}_{\{\pSigma >0\}\cap[0,T]}(\ff,\Q).$

Consider the sequence $(\zeta_n)_{n\in\mathbb{N}}$ introduced in \qr{e:Sp}. Let $(\sigma_n)_{n\in\mathbb{N}}$ be a nondecreasing sequence of $\ff$ stopping times tending to infinity and reducing $q _- $ and its inverse to bounded processes on $[0,T]$. 
By definition of a local martingale on a predictable interval (cf.~\qr{e:intint}), the above condition on $P$ can be stated as $$
 q _-{\centerdot }P^{\zeta_n \wedge \sigma_n}+q _-\frac{1}{\pSigma }{\centerdot }[\cM , P]^{\zeta_n \wedge \sigma_n}
\in\mathcal{M}_{[0,T]}(\ff,\Q)\sp \forall n\in\N .
% \mbox{ is an $(\ff,\Q)$ local martingale on $[0,T]$ for every $n.$} 
 $$ 
As $\pSigma$ and $q _- $ 
are bounded away from 0 on $[0,\zeta_n \wedge \sigma_n]$, this is equivalent to 
 $$ {\pSigma }{\centerdot }P^{\zeta_n \wedge \sigma_n}+[\cM , P]^{\zeta_n \wedge \sigma_n}\in\mathcal{M}_{[0,T]}(\ff,\Q)\sp \forall n\in\N,$$
i.e. $${\pSigma }{\centerdot }P+[\cM,P]\in\mathcal{M}_{\{\pSigma >0\}\cap[0,T]}(\ff,\Q).$$ 

\noindent
2) In the case where
$\Alpha$ is continuous, we have $[\Alpha,\cdot]=0$ and $\pSigma =\ttS_- $ (cf. \qr{e:inte}),
 so that the condition in part 1) is rewritten as \qr{e:conma}. In order to establish \qr{yorthetaprelsup}, in view of \qr{yorthetaprel}, all we need is 
showing $ \{\ttS_- >0\} \subseteq [0,\varsigma)$.
Toward this aim, we apply the local martingale characterization \qr{e:conma} to
$P = \indi{0<\varsigma_n=\varsigma<C} \ind_{[\varsigma,\infty)},$ where $(\varsigma_n)_{n\in\N}$ is the sequence 
defined in 
\qr{e:Ibis}
and $C$ is a positive constant. 
Namely, we compute
$${\ttS_{-}}{\centerdot }P+ [\ttS,P]=\ttS_{\varsigma-}\indi{0<\varsigma_n=\varsigma<C} \ind_{[\varsigma,\infty)}+
\Delta_{\varsigma} \ttS \indi{0<\varsigma_n=\varsigma<C} \ind_{[\varsigma,\infty)}=0$$ 
(since $\ttS_{\varsigma}=0$). Hence, 
$P$ satisfies \qr{e:conma}, so that it is a bounded $(\ff,\P)$ martingale on $\{\ttS_->0\}\cap[0,T]$.
Noting that $P_0=0$ and $$
P_{\varsigma_n\wedge T} = \ind_{\{0<\varsigma_n=\varsigma<C\}}\ind_{\{\varsigma_n\wedge T\ge\varsigma\}}= \ind_{\{0<\varsigma_n=\varsigma<C,\varsigma\leq T\}},
$$ 
 we conclude that
$$0 = \E^{\P}[P_{\varsigma_n\wedge T}] = \P[0<\varsigma_n=\varsigma<C,\varsigma\leq T],$$
for every positive constant $C.$
As a consequence, we have $\P[0<\varsigma_n=\varsigma\leq T]=0.$
Thus $\varsigma_n<\varsigma$ holds whenever $\varsigma\leq T$, under $\P$ as under $\Q$. Hence, in view of
\qr{e:S}:
\begin{itemize}
\item[-] On $ \{\ttS_0 >0, \varsigma\leq T\},$ we have
$$ \{\ttS_- >0\}=\cup_n[0, \varsigma_n ] \subseteq [0,\varsigma) , $$ 
\item[-] On $ \{\ttS_0=0\},$ we have
$$ \{\ttS_- >0\} =\emptyset= [0,\varsigma).$$ 
\end{itemize}
In both cases, we have $ \{\ttS_- >0\} \subseteq [0,\varsigma)$, which finishes the demonstration.~\finproof \\

\section{Invariance Times in Different Situations}\label{diffsituas}

In this section
%what follows 
we review 
a variety of
%much more examples 
%of 
situations 
involving invariance times.

\subsection{Comparison with Pseudo-Stopping Times}\l{s:inv}

In this section we study the connection between invariance times and pseudo-stopping times as of \citeN{NikeghbaliYor05}. In addition we assess the materiality of stopping before $\theta$ as opposed to at $\theta$ 
in the condition (A).
 
Consider a $(0,+\infty)$-valued random time $\theta$. 
By \citeN{NikeghbaliYor05}, it is an $(\mathbb{F},\mathbb{Q})$
pseudo-stopping time if and only if $Q^\theta$ is a local martingale for any 
%\b{could it be "local mart" instead?} 
$(\mathbb{F},\mathbb{Q})$ local martingale $Q$.
%By definition, it is an $(\mathbb{F},\mathbb{Q})$
%pseudo-stopping time if and only if 
%$\mathbb{E}[Q_\theta]=\mathbb{E}[Q_0]$
%for any bounded $(\mathbb{F},\mathbb{Q})$ martingale $Q$ or, equivalently, if
%$Q^\theta$ is a $(\mathbb{G},\mathbb{Q})$ uniformly integrable martingale for any bounded 
%%\b{could it be "local mart" instead?} 
%$(\mathbb{F},\mathbb{Q})$ martingale $Q$ (cf.~\citeN{NikeghbaliYor05}). 
Clearly, if a pseudo-stopping time $\theta$ avoids the $\ff$ stopping times, then it is an invariance time satisfying the condition (A) for any positive constant $T,$ with invariance measure $\mathbb{P}=\mathbb{Q}$. 
It is also shown in \citeN[Theorem 1 (3)]{NikeghbaliYor05} that $\theta$ is a pseudo-stopping time if and only if $\pAlph_{\infty}\equiv 1 ,$ i.e. if and only if $\ttS=1-\pAlph,$ 
where \index{a@$\pAlph$} $\pAlph$ denotes the $\mathbb{F}$ dual optional projection of {$\ttH$}.
By contrast, Proposition \ref{c:mc} 2) shows that $\mathbb{Q}$ itself is an invariance measure for any positive constant $T$ if and only if $\ttS=1-\Alpha$ (noting that $\ttS_0=1$ here, as $\theta>0$). Both conditions coincide if and only if $\pAlph=\Alpha$. 
We recall that in the case where $\ttS_0=1$ 
and $\Alpha$ is continuous (i.e., in our setup, whenever $\theta$ is a $(\mathbb{G},\Q)$ totally inaccessible stopping time), then $\pAlph=\Alpha$ if and only if $\theta$ avoids the $\mathbb{F}$ stopping times. Hence, in the case where $\ttS_0=1$ 
and $\Alpha$ is continuous, 
there are two ``orthogonal'' cases: 
%We recall that in the case where $\theta$ is a $\b{(\mathbb{G},\Q)}$ totally inaccessible stopping time (i.e. when $\ttS_0=1$ 
%and $\Alpha$ is continuous, cf. Lemma
%\ref{l:inten}), then $\pAlph=\Alpha$ if and only if $\theta$ avoids the $\mathbb{F}$ stopping times. Hence, in case $\theta$ is a $\b{(\mathbb{G},\Q)}$ totally inaccessible stopping time, 
%there are two ``orthogonal'' cases: 
\begin{itemize}
\item[-] If $\theta$ has the avoidance property, then $\theta$ is a pseudo-stopping time if and only if 
$\mathbb{Q}$ itself is an invariance measure;
\item[-] If $\theta$ is a pseudo-stopping time without the avoidance property, then $\mathbb{Q}$ itself cannot be an invariance measure. 
%\item[-] If $\theta$ does not have the avoidance property, then $\theta$ cannot be a pseudo-stopping time and $\mathbb{Q}$ itself be an invariance measure at the same time. 
This is due to the
fact that a pseudo-stopping time is defined in terms of
stopping at $\theta,$ whereas invariance
is defined in terms of stopping before $\theta$.
\end{itemize} 
%Having said this regarding the case where $\P=\Q,$ we emphasize that,

Moreover,
by comparison with pseudo-stopping times that are defined with respect to the fixed probability measure $\mathbb{Q}$, the additional flexibility of invariance times lies in the possibility to consider the martingale property under a changed measure $\P$. 
In fact, the pseudo-stopping time condition is very restrictive. By contrast Theorem \ref{sfcnd} shows that invariance times are the rule rather the exception. Actually, 
%we emphasize that 
the spirit of invariance times is not to define one more fancy
 class of random times, but rather to show that a reduction of filtration methodology can be useful very broadly, much beyond
the basic reduced-form setup of \sr{rem:bis}.

The following examples provides more insight on the relationship between pseudo-stopping times and invariance times.

\bex\label{ex1}
Fix a filtration $\mathbb{F}$ \usual. For $i=1,2,$ let $\sigma_i>0$ be a finite $\ff$ stopping time with
bounded and continuous compensator $\mathbf{v}_i.$
% \b{hence $\pSigma=\ttS_-$ (cf. \qr{e:inte})}.
{Assuming $\sigma_2>T$}, define $\theta = \ind_A \sigma_1 + \ind_{A^c} \sigma_2 $,
for some random event $A$ independent of $\F_\infty$ such that $\alpha=\Q( A)\in (0,1)$. Let $\mathbb{G}$ be the progressive enlargement of $\mathbb{F}$ with $\theta.$ 
Hence, $\theta$ intersects the $\ff$ stopping times $\sigma_i .$
By independence of $A,$ 
on $[0,T],$ we have
$$
\dcb

\ttS= \ind_{[0,\sigma_1)}\alpha + {\ind_{[0,\sigma_2)}}(1-\alpha), \\
\Alpha = \alpha\mathbf{v}_1 + (1-\alpha)\mathbf{v}_2,\\
\cM = \alpha(\ind_{[0, \sigma_1)}+\mathbf{v}_1) + (1-\alpha)(\ind_{[0, \sigma_2)}+\mathbf{v}_2).
\dce
$$
Hence, since $\sigma_2> T,$ we have $\ttS$ and in turn
$\pSigma
%\b{\ttS_- \sout{\pSigma}}
 \geq 1- \alpha>0
\mbox{ on } [0,T],
$
so that
$$\cE(\frac{1}{\pSigma}\centerdot \Alpha) = e^{\frac{1}{\pSigma}\centerdot \Alpha}\le e^{ \frac{1}{1-\alpha}\mathtt{D}}$$
is bounded
on $[0,T]$.
%where $\Gamma :=\gamma' {\centerdot}\boldsymbol{\lambda}$ (cf. Lemma \ref{l:inten}).
%
%{$\mathbf{v} =\int_0^{\cdot\wedge \theta}\frac{1}{\mathtt{S}_{s-}}d\mathtt{D}_s\leq \frac{1}{1-\alpha}\mathtt{D}$}
%
%%{$\mathbf{v} =\int_0^{\cdot\wedge \theta}\frac{1}{\mathtt{S}_{s-}}d\mathtt{D}_s\leq \frac{1}{1-\alpha}\mathtt{D}$}
%is bounded. 
Therefore the conditions of Theorem \ref{sfcnd} are fulfilled and $\theta$ is an invariance time.
In addition, for every bounded $\mathbb{F}$ optional process $K$, by independence of $A$, we have
$$\mathbb{E}[K_\theta ]=\mathbb{E}[ \ind_A K_{\sigma_1} + \ind_{A^c} K_{\sigma_2} ]=\mathbb{E}[ \alpha K_{\sigma_1} + (1-\alpha) K_{\sigma_2} ],
$$
hence $$
\pAlph
%=(\ttH)^o 
=(\ind_{[\theta,\infty)})^o=\alpha\ind_{[\sigma_1,\infty)} +(1-\alpha) \ind_{[\sigma_2,\infty)}.
$$
As the ${\sigma_i}$ are finite, it follows that $\pAlph_{\infty}\equiv 1$ and $\theta$ is a pseudo-stopping time.
\eex

\bex\def\eta{\sigma_3}
To obtain an invariance time $\theta$ intersecting $\ff$ stopping times without being a pseudo-stopping time, we set
$$\theta = \ind_{A_1} \sigma_1 + \ind_{A_2} \sigma_2 + \ind_{A_3} \eta ,$$
where $\sigma_1$ and $\sigma_2$ are as in the example \ref{ex1} and
where the finite random time $\eta>0$
is not an $\ff$ pseudo-stopping time, for a partition
$A_i, i=1,2,3,$ independent of $\F_\infty$ and of $\eta$.
Assuming
$\alpha_i=\Q(A_i)>0$,
we have
\bel&\pAlph
%=(\ttH)^o
=(\ind_{[\theta,\infty)})^o =
\alpha_1\ind_{[\sigma_1,\infty)} +\alpha_2\ind_{[\sigma_2,\infty)} +\alpha_3 (\ind_{[\eta,\infty)})^o,
%\rightarrow 1 \mbox { as } t\rightarrow \infty,
\eel
where $(\ind_{[\eta,\infty)})^o_{\infty}\neq 1,$ hence $\pAlph_{\infty}\neq 1$, with positive
$\Q$
probability. Hence,
%Thus, by the converse part in Theorem 1 (3) in \citeN{NikeghbaliYor05}, 
$\theta$ is not a pseudo-stopping time.
But the Az\'ema supermartingale of $\theta$ is given by
$$
\ttS=\ind_{[0,\sigma_1)}\alpha_1 + \ind_{[0,\sigma_2)}\alpha_2+ ^o\!(\ind_{[0,\eta)})\alpha_3\geq\alpha_2 \mbox{ on } [0,T],$$
so that $\theta$ is an invariance time, by the same arguments as in the example \ref{ex1}.
\eex

\subsection{Connection with the Survival Measure}\label{s:applic}

%\section{Invariance Transfer Formulas}

%In this section we provide an invariance time sufficiency condition and we make the connection between the invariance measure and the so-called survival measure. 
 
In the financial context of defaultable asset pricing,
for deriving an intensity-based pricing formula exempt from the no-jump condition in \shortciteN{DuffieSchroderSkiadas96}, 
an alternative to 
the reduction of filtration approach
of this paper
%Proposition \ref{t:exptransf}
is to work purely in 
the filtration $\mathbb{G},$ 
but to switch from the risk-neutral measure $\mathbb{Q}$ to
some nonequivalent pricing measure.

To establish the connection between the two frameworks, let us postulate a $\gg$ stopping time $\theta$
in an enlargement setup satisfying the condition (B) (as everywhere in this paper), with a $(\gg,\Q)$ intensity $\gamma$ 
and $\ttS_T>0$,
so that $\mathcal{E}(\frac{1}{\pSigma}\centerdot \Alpha)=\mathcal{E}(\frac{1}{\ttS_-}\centerdot \Alpha)= e^{\gamma' {\centerdot}\boldsymbol{\lambda}}$
holds on $[0,T]$, by Lemma \ref{l:inten}.
Assuming further that
$e^{\gamma {\centerdot}\boldsymbol{\lambda}_{\theta\wedge T}}$ is $\Q$ integrable (this is the basic assumption in
\shortciteN{CollinDufresneGoldsteinHugonnier2004}), then all the conditions of Theorem 
\ref{sfcnd} are satisfied.
Hence, by application of this theorem, we conclude that
 $\theta$ is an invariance time, that
the process 
$e^{\gamma {\centerdot}\boldsymbol{\lambda}}\ind_{[0,\theta)}$ 
is a $(\mathbb{G},\mathbb{Q})$ true martingale
on $[0,T]$ and that an invariance measure $\P$ is provided by the restriction
to $\mathcal{F}_{T}$ of the
``survival measure'' $\mathbb{S}$ with $e^{\gamma {\centerdot}\boldsymbol{\lambda}}
\ind_{[0,\theta)}$ 
%(or very much related quantities) 
as $\gg$ density process of $\frac{d\mathbb{S}}{d\Q}$ 
 (the ``survival measure'' idea and terminology were first introduced and used for different purposes in \citeANP{Schoenbucher99} (\citeyearNP{Schoenbucher99}, \citeyearNP{Schoenbucher04})).

Put differently, if the default time $\theta$ in \shortciteN{CollinDufresneGoldsteinHugonnier2004}
also satisfies the condition (B), then
the condition (A) is satisfied and the restriction to $\mathcal{F}_T$
of the survival measure $\mathbb{S}$ is an 
invariance measure. This establishes the connection between the notions of 
invariance and survival measures. In particular, this shows that, even though $\mathbb{S}$ and $\mathbb{Q}$ are not equivalent (not even in a basic reduced-form setup as of \sr{rem:bis}), their restrictions to $\mathcal{F}_{T}$ are equivalent.

Assuming that $e^{\gamma {\centerdot}\boldsymbol{\lambda}_{\theta\wedge T}}$ is $\Q$ integrable,  
\shortciteN{CollinDufresneGoldsteinHugonnier2004} obtain
an intensity-based $(\mathbb{G},\mathbb{S})$ pricing formula
exempt from the no-jump condition in \shortciteN{DuffieSchroderSkiadas96}. 
%However, 
%$\mathbb{S}$ is not equivalent to $\mathbb{Q},$
%not even in a basic reduced-form setup as of \sr{rem:bis}. 
%Beyond its auxiliary role in calculations,
A no-arbitrage interpretation of the survival measure $\mathbb{S}$ is given in \citeN{FisherPulidoRuf2015}, who use it to model financial assets that may potentially lose value relative to each other at time $\theta.$

}

\subsection{BSDEs of Counterparty Risk}\label{ex:cap} Invariance times
make the filtration reduction technique particularly efficient for dealing with BSDEs with random terminal time
(this was actually our initial motivation for the introduction of
invariance times, cf. \sr{s:bsdecr}). 
%for motivation purposes in \sr{s:bsdecr}. 
For instance, in the two-part paper by \citeN{BichuchCapponiSturm15}, where
$\ff$ is a Brownian filtration and $\gg$ is $\ff$ progressively enlarged by the default times of a bank and its counterparty,
the analysis 
%conducted in the two papers 
can be drastically simplified by converting, by an application of \citeN[Theorem 4.3]{Crepey13a1}, their main jump BSDE (17) (or likewise (18)) in Part I, formulated with respect to the enlarged filtration $\gg$ on the time interval $[0,\theta\wedge T],$ into a continuous BSDE with respect to the reference filtration $\ff$ on the time interval $[0, T]$. The two BSDEs are equivalent but the continuous one is of course much simpler to study. The related PDE also becomes continuous. Without jumps, the proofs of all the technical results in their work, i.e. the BSDE well-posedness and comparison theorems A.2 and A.3 in part I and the PDE unique viscosity solution theorem 3.2 in part II (or rather their continuous analogues), become a matter of referring to standard continuous BSDE and PDE results.

%a reduction of filtration approach (, specifically)
%yields the main BSDE well-posedness and comparison results there in a few lines, by application of standard Brownian BSDE theorems that apply in
%$\ff$, as opposed to lengthy developments involving PIDE arguments if one works in $\gg$. 

The reformulation of the ``full'' $(\gg,\Q)$ BSDE \qr{equat5} 
as the ``reduced'' $(\ff,\P)$ BSDE (\ref{equat7}) 
can also represent a significant improvement with numerical solutions in view.
In the above example, it implies that standard
numerical schemes without jumps can be used for solving the problem (cf. \citeN[Remark 3.4 in Part II]{BichuchCapponiSturm15}).
% realize in the Remark II.3.4 on page 12 of the second part). 

\subsection{Dynamic Copulas}\label{ss:dcm}
%Models
%of Counterparty Risk
% on credit derivatives}

%We already mentioned counterparty risk as our motivating example in \sr{s:bsdecr} and \ref{ss:BSDEsagain}.
A singular point in the field of counterparty risk is 
when the underlying contracts are credit derivatives, due to the extreme wrong-way risk effects that occur, in the form of frailty and contagion, between the
credit risk of the counterparty and the credit risk of the underlying market
(credit in this case) exposure. In models of ``instantaneous contagion'' where the counterparty and the reference credit entities may default at the same time, wrong-way risk may take the extreme form of an instantaneous gap risk, which is not only a surge in the value of credit protection, 
%(hence an increased exposure) at time $\theta$, 
but also credit protection cash flows that fail to be paid to the bank by the defaulted counterparty at time $\theta$.

To embed such credit frailty and contagion effects in their models,
market practitioners use (static) copulas for the default times of the reference entities. In the case of counterparty risk on credit derivatives, one needs to model the default time of the counterparty jointly with these, i.e. one needs a model for $(\theta_0,\theta_1,\ldots,\theta_n),$ where $\theta_0=\theta$ corresponds to the default time of the counterparty of a bank involved in credit derivatives on reference entities with the respective default times $\theta_i,$ $i=1,\ldots,n$. Moreover, as the counterparty risk and funding valuation adjustments are in fact options on these, one needs to make the model dynamic in some way.
%so that not only present but also future prices of credit derivatives can be assessed consistently and computed numerically. 
One possibility is to use
``dynamic copula models'' resulting from the introduction
of a suitable filtration on top of a static copula model. 
%\b{This idea is commonly used to model wrong-way risk in financial software.}
Related references in the literature include the seminal working paper by \citeN{SchonbucherSchubert2001}
as well as, among others,
\citeN[Section 3]{BrigoCapponiPallavicini}, \citeN{BoCapponi15}, 
\citeN{LeeCapriotti2015} 
or \citeN{CrepeySong15} (cf. also \citeN{Kusuoka1999} and \shortciteN[Section 7.3]{Bielecki2009}). 
%\b{Trioptima, Global Valuation}
% for the default times of the two parties and the reference entities.
%\begin{itemize} 
%\item[-] 
%\emph{Intensity models of counterparty credit risk} with \emph{strong (adverse) dependence} between the credit risk of a counterparty and the underlying market exposure
%\begin{itemize}
%\item[-] \emph{Wrong way and gap risk} modeling
%\begin{itemize}\item[-] \emph{Frailty and contagion} in the case of a credit derivatives exposure \end{itemize} \item[-] As opposed to factor dependence only in standard Cox (doubly stochastic) models 
%\end{itemize}
%\item[-] 
%\emph{Copula model} of $\theta_0,\theta_1,\ldots,\theta_n,$ where $\theta_0=\theta$ corresponds to the default time of the counterparty of a bank in credit derivatives on names $1,\ldots,n$
%\item[-] Counterparty risk computations:
%need \emph{make the model dynamic} by introduction of a suitable model filtration $\gg$

Now, in order to understand the nature of the credit dependence in
a dynamic copula model, one would like to address the following question: Given a stopping time $\theta$ relative to the full model filtration $\gg$, when and how can one separate the information that comes from $\theta$ from a ``market filtration'' $\ff$,
such that, for consistency and tractability, some kind of local martingale invariance property holds between $\ff$ and $\gg$? 
%%We are talking of reduction of filtration in this sense
%%(not completely unrelated with, but different from, filtration shrinkage, whereby
%%\citeN{FollmerProtter2011} study the stability of the local martingale property by projection onto a smaller filtration)
%Toward this aim, some kind of local martingale invariance property is required, but under minimal assumptions, so that the model stays as flexible as possible in view of applications. 
Invariance times are precisely designed in this spirit.
%%Specifically, we define an invariance time as a stopping time such that local martingales with respect to a smaller filtration and a possibly changed probability measure, once stopped right before that time, are local martingales with respect to the original model filtration and probability measure. 
%The possibility to change the measure and to use a full model filtration
%%$\gg$ larger than simply $\ff \vee \hh,$ 
%larger than simply the progressive enlargement of a market reference filtration by $\theta$
%provide additional degrees of freedom with respect to the basic immersion setup that is usually considered for modeling default times in applications.}
%%\begin{itemize} 
%%\item[\fl] Reduction of filtration in this sense
%%\item[-] Not unrelated with, but different from, filtration shrinkage, whereby
%%\citeN{FollmerProtter2011} project local martingales onto smaller filtrations
%%\end{itemize} 
%%\item[-] For applications, some kind of \emph{martingale invariance property is required, but under minimal assumptions}, so that the model stays as flexible
%%\item[\fl] \emph{Invariance times} 
%%\item[-] \emph{Dynamic copula models} \b{\citeN{CrepeySong15}, CreditSuisse},
%%cf. also \citeN[Section 3]{BrigoCapponiPallavicini}, \citeN{BoCapponi15}, Sch\"on\-bucher=
%We will show in \sr{s:dgc}
%and \ref{s:dmo} that {the} condition (A) holds in the DGC and DMO models , using an equivalent measure
%$\P \neq \mathbb{Q}$ in the first one, and $\P = \mathbb{Q}$ (but a well chosen subfiltration
%$\ff$) in the second one.
Specifically, it is shown in \citeN{CrepeySong15} that:
\begin{itemize} 
\item[-] In the context of the dynamic Gaussian copula model, the condition (A) is achieved by $\theta=\theta_0,$ for a suitable $\mathbb{P}\neq \mathbb{Q}$ and for $\mathbb{G}$ given as the progressive enlargement of $\mathbb{\ff}$ by $\theta.$ 
This model can be used as a ``wrong-way risk model'' of counterparty risk embedded in credit derivatives, where the default intensities of the surviving reference names and therefore the value of credit protection 
%hence the exposure process $G$ in \sr{s:bsdecr}, 
spike, as an effect of $\mathbb{P}\neq \mathbb{Q},$ at the default time of the counterparty;
\item[-] In the context of the dynamic Marshall-Olkin copula (or ``common-shock'' model, where the credit dependence stems from the possibility of simultaneous defaults), the condition (A) is achieved by $\theta=\theta_0,$ for $\mathbb{P}= \mathbb{Q}$ but $\mathbb{G}$ greater than the progressive enlargement of $\mathbb{\ff}$ by $\theta,$ reflecting various possible joint default scenarios that may prompt the default of the counterparty. 
This model can be used as a ``gap risk model'' of counterparty risk embedded in credit derivatives, where promised cash-flows fail to be paid at the counterparty default time $\theta$ in joint default scenarios.
%This can be seen as a representative example of a ``gap risk model'' where promised cash-flows fail to be paid at the counterparty default $\theta$, instantaneously contributing to make $G_\theta$ bigger
\end{itemize} 
(cf. Figure 7 in \citeN[preprint version]{CrepeySong15}).

\subsubsection{An Open Problem}
A subfiltration $\ff$ satisfying the condition (B) is considered as given everywhere in this paper. As an open problem related to invariance times, there is the question of finding a subfiltration $\ff$ of a given full model filtration $\gg$ such that the condition (B) (before considering (A)) is satisfied. 
%Building on martingale representation considerations, 
Suitable subfiltrations $\ff$ can be worked-out on a case-by-case basis in the above dynamic copula models. But, 
beyond a necessary condition which is derived in \citeN[Sect. 9]{Song16a}, 
we have no constructive methodology to offer in this regard.

\section*{Conclusion}
%\section*{Conclusion and Perspectives}

From the enlargement of filtration literature of the seventies, two facts were known. First, the study of progressive
enlargement of a filtration with a random time $\theta$ is ``easy'' before $\theta$. Second, the Jeulin-Yor formulas for
the semimartingale decomposition, in a progressively enlarged filtration, of martingales in a smaller filtration $\ff,$ are similar to the Girsanov measure change formulas, but the Jeulin-Yor and Girsanov formulas are not equivalent to each other, mainly due to integrability
reasons at time $\theta$.

In this paper we characterize
%prove a conjecture, which 
%arises naturally from 
%a counterparty risk pricing problem, regarding a characterization of 
the random times $\theta$ 
such that the local martingale property is preserved for a process stopped before $\theta$ under filtration enlargement and equivalent change of measure.
%We
%provide an explicit characterization of a class of random times $\theta$ that preserve the local martingale property of a process stopped at $\theta-$ when a filtration is enlarged
%and an equivalent change of a probability measure is made. Such 
The corresponding random times, called invariance
times, make the
filtration reduction technique particularly efficient for dealing with BSDEs with random terminal time.
 
More broadly, this paper is a contribution to two parallel progressive enlargement of filtration literatures, depending on whether a predictable or optional point of view is considered:
\begin{itemize}
\item[-] Ends of sets: predictable in \citeN{Azema72} versus optional in \citeN{JeulinYor78};
\item[-] Multiplicative decompositions of the Az\'ema supermartingale $\ttS$ of $\theta$: predictable in \citeN{Jacod1979} and
\citeN{Song16a} versus optional in \citeN{Kardaras15};
%\b{?\citeN{AcciaioFontanaKardaras}} APPLI seul; 
\item[-] Representations of $\ttS$ under the form $\frac{X}{X^*},$ for some positive local martingale $X$
and $X^* =\sup_{0\leq s\leq \cdot}X_{s}$,
following \citeN{NikeghbaliYor2006} (cf. the example \ref{ex:invBessel}): for honest times avoiding $\ff$ (optional) stopping times in \citeN{Kardaras14} versus $\ff$ predictable stopping times in \citeN{AcciaioPenner14} (and for general honest times in 
\citeN{Song16});
\item[-] Deflators: predictable in \citeN{Song16a} versus optional in  
\shortciteN{AksamitChoulliDengJeanblanc13} or \shortciteN{AcciaioFontanaKardaras};
\item[-] Local martingale invariance properties: for processes stopped before $\theta,$ based on the dual predictable projection $\Alpha$ of $\theta$, regarding the invariance times of this paper, versus processes stopped at $\theta,$ based on the optional projection $\mathtt{A}$
 of $\theta$ (and without measure change), in the case of pseudo-stopping times in \citeN{Nikeghbali2005}.
\end{itemize}
\noindent
From an application point of view, the relevance of stopping at $\theta$ (as for the study of no arbitrage for an insider observing $\theta$
in \shortciteN{AksamitChoulliDengJeanblanc13} or \shortciteN{AcciaioFontanaKardaras}) 
or before $\theta$ (as for pricing counterparty risk) depends on the problem at hand. 
At the technical level, the best approach to deal with the process $X^{\theta-}$ 
(as in the present paper) is to make use of the predictable multiplicative decomposition 
%\qr{e:multdec} 
 of the Az\'ema supermartingale
%, of the \b{??predictable} decomposition formula of \citeN{JeulinYor78} 
and more generally of the reduction of filtration methodology. By contrast, the consideration of $X^{\theta}$ 
leads naturally to optional computations.
% with an optional multiplicative decomposition of the Az\'ema supermartingale.
% as in \citeN{AcciaioFontanaKardaras} \b{?? vs \citeN{Kardaras15}} and with \b{??optional decomposition formulas as in \citeN{AksamitChoulliDengJeanblanc13}}. 

As far as the pricing of defaultable securities is concerned, the $(\mathbb{F},\mathbb{P})$ (or ``extended reduced-form'') approach based on this paper sheds light on the relation between 
three streams of mathematical finance literature, namely the 
$(\mathbb{G},\mathbb{Q})$ seminal approach by \shortciteN{DuffieSchroderSkiadas96}, the nowadays standard $(\mathbb{F},\mathbb{Q})$ approach in a basic reduced-form setup and the $(\mathbb{G},\mathbb{S})$ survival measure approach of \shortciteN{CollinDufresneGoldsteinHugonnier2004}.

\appendix

\section{About The Az\'ema Supermartingale}
\label{s:az} 

%Let \index{o@$^o$}${^{\fo }}\!\cdot$ and \index{p@$^p$}${^{p}}\!\cdot$ denote the $\mathbb{F}$ optional and predictable projections. 
%The fundamental tool to work with the condition (B) is the Az\'ema supermartingale $\ttS={^{\fo }}\! \mathtt{J} $ of $\theta$, 
%%which is defined by $\ttS $\index{s@$\ttS$},
%i.e. $\ttS_t =\mathbb{Q}(\theta>t\big|\cF_t),$ $t\in \R_+,$
%with
%canonical Doob-Meyer decomposition $\ttS=\cM-\Alpha,$
%where $\cM$ is an $(\mathbb{F},\Q)$ martingale {(with $\cM_0=\ttS_0$)}
%and $\Alpha$ is the
%$(\mathbb{F},\Q)$ dual predictable projection of ${\ind_{\{0<\ftime\}}}\ttH$.

%As everywhere else in the paper, one assumes a subfiltration $\ff$ of $\gg$ and
%a $\gg$ stopping time $\theta$ such that the condition (B) holds on $[0,T].$
This section recapitulates the most classical properties of the Az\'ema supermartingale $\ttS={^{\fo }}\! \mathtt{J} $ of a random time $\theta$, where $\ttJ=\ind_{[0,\theta)}$, 
with
canonical Doob-Meyer decomposition $\ttS= \ttS_0+ \cM-\Alpha$ (cf. \sr{cB}). For more information, see 
\citeN{Jeulin1980} and \citeN{JeulinYor78}.

We have
\beql{e:ejeu}
{^{p}}
({ \ttJ_{-} })=\ttS_- \mbox{ on } (0,\infty) \eeql
(see \citeN[page 63]{Jeulin1980}) and 
\beql{e:SS}
{\pSigma }=\cM_{-}-\Alpha=\ttS-\Delta \cM=\ttS_{-}-\Delta \Alpha\leq \ttS_{-}.
\eeql 
In particular, if $\Alpha$ is continuous, then
\beql{e:inte} {\pSigma }=\ttS_{-}.\eeql

%\noindent
The $(\gg,\mathbb{Q})$ compensator $\mathbf{v}$ of $\theta$ satisfies
\beql{e:comp}
\mathbf{v} =\int_0^{\cdot\wedge \theta}\frac{1}{\mathtt{S}_{s-}}d\Alpha_s
\eeql
 (see \citeN[Remark 4.5]{Jeulin1980}).

\bl\label{l:inten}
The $\gg$ stopping time $\theta$ is $(\gg,\Q)$ totally inaccessible if and only if $\ttS_0=1$ 
and $ \mathtt{D}$ is continuous (so that $\mathcal{E}(\pm\frac{1}{\ttS_-}{\centerdot}\mathtt{D})=e^{\pm\frac{1}{\ttS_-}{\centerdot}\mathtt{D}}$ holds
on $\{\ttS_{-}>0\}$). 
In the special case where $\theta$ has a $(\gg,\Q)$ intensity $\gamma$, 
then 
$\ind_{(0,\theta]} \frac{1}{\ttS_-}\centerdot \mathtt{D} =\gamma \centerdot \boldsymbol{\lambda} $
and
$\mathtt{D} = \gamma' \ttS_- \centerdot \boldsymbol{\lambda}$ hold on $\R_+$
(where $\gamma'$ denotes an $\ff$ predictable reduction of $\gamma$), hence $\frac{1}{\ttS_-}{\centerdot}\mathtt{D}=\gamma' {\centerdot}\boldsymbol{\lambda}$ holds
on $\{\ttS_{-}>0\}$ .
%$\frac{\nu }{S_-} = {\varphi},$ hence
%$\frac{1}{\ttS_-}{\centerdot}\mathtt{D}=\gamma' {\centerdot}\boldsymbol{\lambda}$ holds
%on $\{\ttS_{-}>0\}$ 
%$\mathcal{E}(-\frac{1}{\ttS_-}{\centerdot}\mathtt{D})=e^{-\Gamma},$ where $\Gamma= \gamma' {\centerdot}\boldsymbol{\lambda}.$
%If $\theta$ is totally inaccessible, then $\ttS_0=1,$ and
%$ \mathtt{D} = \nu  {\centerdot}\boldsymbol{\lambda}$ for some density process $\nu$ of 
%$ \mathtt{D} ,$ then $\frac{\nu }{S_-} = {\varphi}$ and 
%$\mathcal{E}(-\frac{1}{\ttS_-}{\centerdot}\mathtt{D})=e^{-\Gamma},$ where $\Gamma= \gamma' {\centerdot}\boldsymbol{\lambda}.$
\el
\proof 
By definition in this paper (cf. \sr{ss:san}), $\theta$ is $(\gg,\Q)$ totally inaccessible if and only if $\ttS_0=1$ and
the $(\gg,\Q)$ compensator $\mathbf{v}$ of $\theta$ is continuous on $[0,\theta].$ 
In view of \qr{e:comp}, this happens
if and only if $\ttS_0=1$ and $\Delta\Alpha=0$ on $[0,\theta)$ or, equivalently by the predictable version of Lemma \ref{l:Spos}, if
$\Delta\Alpha=0$ on $\{\ttS_->0\}$, i.e. if $\Alpha$ is continuous on $\R_+$ (cf. \citeN[Lemma 3.7]{Song16a}).
If $\theta$ has an intensity $\gamma$, then
\qr{e:comp} and Lemma \ref{e:ejeu} show that $\ind_{(0,\theta]} \frac{1}{\ttS_-}\centerdot \mathtt{D} =\gamma \centerdot \boldsymbol{\lambda} $
and
$\mathtt{D} = \gamma' \ttS_- \centerdot \boldsymbol{\lambda}$.~\finproof\\

Define 
\index{s@$\varsigma_n$}
\beql{e:Ibis}
\varsigma=\inf\{s>0;\ttS_s=0\},\
\varsigma_n=\inf\left\{s>0; \ttS_s\leq \frac{1}{n}\right\} ~ (n> 0). 
\eeql
Then,
\index{s@$\varsigma$}
\beql{e:I}
\varsigma=\inf\{s>0; \ttS_{s-}=0\}
\eeql
(since $\ttS$ is a nonnegative supermartingale)\footnote{Cf.~n$^\circ$17 Chapitre VI in \citeN{DM75}.}
and
\beql{e:S}
 & \varsigma=\sup_n\varsigma_n\sp \{\ttS_->0\}\cup\gr0\rg=\cup_n[0,\varsigma_n] .
\eeql 
Likewise,
$\{\pSigma >0\}$ is a predictable interval and, according to \citeN[(6.24) and (6.28)]{Jacod1979}, there exists
a nondecreasing sequence
\index{z@$\zeta_n$}
$(\zeta_n)_{n\in\mathbb{N}}$ of $\ff$ stopping times such that
\beql{e:Sp}
\{\pSigma>0\}\cup\gr0\rg=\cup_{n}[0,\zeta_n] \mbox{ and $\frac{1}{\pSigma}$ is bounded on $(0,\zeta_n]$ for every $n$}.
\eeql 
The following relations hold:
\beql{yorthetaprel}
 {[0,\varsigma)=}\{\ttS>0\}\subset \{\pSigma >0\}\subset \{\ttS_{-}>0\},
\eeql
whereas, on $(\varsigma,\infty)$, 
\beql{e:sss}
\ttS =\ttS _{-}={\pSigma }=0\sp \ \Alpha\mbox{ and } \cM \mbox{ are constant}. 
\eeql
We have
\beql{yortheta} 
\ttS_{\theta	-}>0 \mbox{ on }\{0<\theta <\infty \} 
\eeql 
(cf.~\citeN[page 63]{Yor78} and see also \citeN[Lemma 3.7]{Song16a}).

\section{Equivalent Changes of Probability Measures}\label{ss:equiv}

The following folklore result of local martingales
plays an important role in the paper.
\bl\label{orthonull}
Let $U$ be a 
%\b{??should we add: finite} ,
stopping time. Let $X$ be a local martingale on $[0,U]$ such that,
for any locally bounded local martingale $Y$ on $[0,U],$
$X$ is orthogonal to $Y$, i.e. $[X,Y]$ is a local martingale on $[0,U]$. 
Then $X=X_0$ on $[0,U]$.
\el

\proof 
Suppose without loss of generality that $X$ is uniformly integrable. 
For any stopping time $\sigma>0$, let $\epsilon$ denote the sign of $\Delta_\sigma X$. First assuming $\sigma>0$ totally inaccessible, let $\mathbf{u}$ be the compensator of $\epsilon\ind_{[\sigma,\infty)},$ so that $(\epsilon\ind_{[\sigma,\infty)}-\mathbf{u})$ is a locally bounded local martingale. By the assumed orthogonality, as $\mathbf{u}$ is continuous (cf.~\shortciteN[Corollary 5.28]{HeWangYan92}), the process $|\Delta_{\sigma} X|\ind_{[\sigma,\infty)}=[X ,\epsilon\ind_{[\sigma,\infty)}-\mathbf{u}]$ must be a local martingale on $[0,U].$ It is therefore null. Likewise, for any predictable stopping time $\sigma>0$ and bounded random variable $\chi$ such that $\mathbb{E} [\chi|\cF_{\sigma-}]=0$ (denoting the filtration by $\mathbb{F}=(\mathcal{F}_t)_{t\in\R_+}$),
 $\chi\ind_{[\sigma,\infty)}$ is a bounded martingale and the orthogonality assumption implies that  
$\Delta_{\sigma} X \chi \ind_{[\sigma,\infty)}$ is a martingale on $[0,U]$. Considering $\chi=\epsilon\ind_{\{\sigma<\infty\}}-\mathbb{E} [\epsilon\ind_{\{\sigma<\infty\}} |\cF_{\sigma-}]$, we have
\beq
&\mathbb{E} [|\Delta_{\sigma} X|\ind_{\{\sigma<\infty\}}]
=
\mathbb{E} [\Delta_{\sigma} X \chi \ind_{\{\sigma<\infty\}}]
=0,
\eeq
because $\mathbb{E} [\epsilon\ind_{\{\sigma<\infty\}} |\cF_{\sigma-}]$ is 
$\cF_{\sigma-}$ measurable and $\mathbb{E} [ \Delta_{\sigma} X |\cF_{\sigma-}]=0$, by the predictability of $\sigma.$ 
In conclusion, the local martingale $X$ is continuous at any predictable or totally inaccessible stopping time.
Hence $X$ is a continuous local martingale on $[0,U]$. As a consequence, by the orthogonality assumption applied to $X$
itself, $[X,X]$ is a continuous local martingale on $[0,U]$. Therefore $[X,X]=[X,X]_0=0$ on $[0,U]$ and the lemma is proved.~\finproof \\

In the rest of this section we derive measure change results of independent interest
(unrelated to enlargement of filtration), used for our second proof of Theorem \ref{part1} in \sr{proofinGJY}. 
Given a probability measure $\mathbb{P}$ equivalent to the probability measure $\mathbb{Q}$
on $\mathcal{F}_T,$
we use the notation $\monp,\thenu,\monq,\therho$
%for densities
introduced in
\qr{e:pq}. 
%Superscripts \index{o@$^o$}${\cdot}^{o \cdot \mathbb{P}}$ and \index{p@$^p$}${\cdot}^{p\cdot \mathbb{P}}$ are used for the $(\mathbb{\ff},\mathbb{P})$ optional or predictable dual projections, whereas \index{o@$^o$}${\cdot}^{o}$ and \index{p@$^p$}${\cdot}^{p}$ denote as before the $(\mathbb{\ff},\mathbb{Q})$ ones.
%Superscripts \index{o@$^o$}${\cdot}^{\mathbb{Q}\cdot o}$ and \index{p@$^p$}${\cdot}^{\mathbb{Q}\cdot p}$ are used for the $(\mathbb{\ff},\mathbb{Q})$ optional or predictable dual projections, and similar definitions are used with respect to the probability measure $\mathbb{P}$. 
%\b{By definition,
%if one argument is continuous (resp. discontinuous), then brackets (whether predictable or square) coincide with the brackets (in the same sense) of the continuous (resp. discontinuous) parts. Moreover a square bracket coincides with the predictable bracket and is continuous whenever one argument is continuous. Finite variation predictable processes do not contribute to predictable brackets against local martingales (this can be seen as a consequence of Yoeurp's formula).}
The $(\mathbb{F},\mathbb{P})$ predictable bracket is denoted by \index{�@$\cro{\cdot,\cdot}$}$\cro{\cdot,\cdot}$, whereas the $(\mathbb{F},\mathbb{Q})$ predictable bracket
is denoted as before by $\langle\cdot,\cdot\rangle$ (of course, when one argument is continuous, the two brackets coincide and we typically write $\langle\cdot,\cdot\rangle$). 
%The $(\mathbb{F},\mathbb{Q})$ and $(\mathbb{F},\mathbb{P})$ predictable brackets are denoted respectively by \index{�@$\croc{\cdot,\cdot}$}$\croc{\cdot,\cdot}$ 
%and \index{�@$\cro{\cdot,\cdot}$}$\cro{\cdot,\cdot}$ (or sometimes simply $\langle\cdot,\cdot\rangle$ when it makes no difference, e.g.~when one argument is continuous). 
{The continuous and purely discontinuous parts (starting from 0) of a local martingale 
are denoted by \index{c@${\cdot}^c$}${\cdot}^c$ and \index{d@${\cdot}^d$}${\cdot}^d$.}

There are two forms of the Girsanov theorem.\footnote{Cf.~\qr{e:go} and \qr{e:gp} and see \shortciteN[Theorem 12.18]{HeWangYan92}.} 
The optional bracket Girsanov formula states that,
for any $\thisX$ in $\mathcal{M}(\mathbb{F},\mathbb{P}),$ 
\begin{equation}\label{Gf1}
\thisX-{\monq}\centerdot [\monp,\thisX]
\end{equation}
is in $\mathcal{M}_{[0,T]}(\mathbb{F},\mathbb{Q}).$ The predictable bracket Girsanov formula states that, if
$\thisX$ in $\mathcal{M}(\mathbb{F},\mathbb{P})$ is such that 
 $[\monp,\thisX]$ is $(\mathbb{F},\mathbb{P})$ locally integrable,\footnote{{Cf.~\shortciteN[Theorem 12.13]{HeWangYan92}}.} 
then
\index{é@$\widetilde{\cdot}$}
\beql{e:girs}
\girs{\thisX}
:=\thisX-{\monq _-}\centerdot \cro{\monp ,\thisX}
=\thisX-\cro{\thenu,\thisX}
\eeql
is in $\mathcal{M}_{[0,T]}(\mathbb{F},\mathbb{Q}).$ 
Lemmas \ref{Gdense} and \ref{cdisc} study the class of $(\mathbb{F},\mathbb{Q})$ local martingales issued from predictable bracket Girsanov transforms.

\bl\label{Gdense}
Let $\thisY$ be an $(\mathbb{F},\mathbb{Q})$ uniformly integrable martingale such that $Q_0=0$ and, for any bounded $(\mathbb{F},\mathbb{P})$ martingale $\thisX$ starting from 0, $\girs{\thisX}\thisY$ is an $(\mathbb{F},\mathbb{Q})$ local martingale on $[0,T]$. Then $\thisY= 0$ on $[0,T]$.
\el

\proof For any bounded $(\mathbb{F},\mathbb{P})$ martingale $\thisX$ starting from 0 and any $\mathbb{F}$ stopping time $\theV \leq T$ reducing the involved processes to integrability, we have
\beql{e:lGdense}
&0=\E [\girs{\thisX}_\theV \thisY_\theV ]	
=\EP [(\thisX-{\monq _-}\centerdot \cro{\monp ,\thisX})_\theV \thisY_\theV \monp _\theV ]\\
&\qqq=\EP [[\thisY\monp ,\thisX]_\theV -\thisY_-\centerdot \cro{\monp ,\thisX}_\theV ]
=\EP [[\thisY\monp -\thisY_-\centerdot \monp ,\thisX]_\theV ],
\eeql
where the $(\mathbb{F},\mathbb{P})$ martingale properties of $P^\theV$ and $(\thisY\monp)^\theV $ were used to pass to the second line, in combination with the following predictable projection formula:\footnote{Cf.~\shortciteN[Remark 5.3 and Theorem 5.26]{HeWangYan92}.}
%\s{only use of $^{\mathbb{P}\cdot p}$}
$$
\EP [ {\monq _-}\centerdot \cro{\monp ,\thisX}_\theV \thisY_\theV \monp _\theV ]
=
\EP [ ^{ p\cdot\mathbb{P}}(\thisY_\theV \monp _\theV){\monq _-}\centerdot \cro{\monp ,\thisX}_\theV ]
=
\EP [ \thisY_-\centerdot \cro{\monp ,\thisX}_\theV ],
$$
where \index{p@$^p$}${\cdot}^{p\cdot \mathbb{P}}$ denotes the $(\mathbb{\ff},\mathbb{P})$ predictable dual projection.

Hence, according to \shortciteN[Theorem 4.40]{HeWangYan92}, $[\thisY\monp -\thisY_-\centerdot \monp ,\thisX]$ is in $\mathcal{M}_{[0,T]}(\ff,\P)$.
%a local martingale on $[0,T]$ under $\mathbb{P}$. 
Since this must hold for every bounded $P$ in $\mathcal{M}_{[0,T]}(\ff,\P)$,
Lemma \ref{orthonull} implies 
\beql{e:zero}
 0=\thisY_0\monp_0 = \thisY\monp -\thisY_-\centerdot \monp
=\thisY\monp - (\thisY\monp )_- \centerdot ({\monq _-}\centerdot \monp )
 \mbox{ on } [0,T].
\eeql 
Consequently, $\thisY\monp=\thisY_0\monp_0\mathcal{E}({\monq _-}\centerdot \monp)=0$ 
%on $[0,T]$ 
(having assumed $\thisY_0=0$).
\finproof

\bl\label{cdisc}
For any $(\mathbb{F},\mathbb{P})$ local martingale $\thisX$ starting from 0 such that $[\monp,\thisX]$ is $(\mathbb{F},\mathbb{P})$ locally integrable, $\girs{\thisX^c}$ and $\girs{\thisX^d}$ are a continuous $(\mathbb{F},\mathbb{Q})$ martingale and a purely discontinuous $(\mathbb{F},\mathbb{Q})$ martingale on $[0,T]$, respectively. Hence,
$\girs{\thisX^c}=(\girs{\thisX})^c$ and $\girs{\thisX^d}=(\girs{\thisX})^d$ hold on $[0,T]$.
\el

\proof 
The predictable bracket Girsanov formula \qr{e:girs} shows that $\girs{\thisX^c}$ is a continuous local martingale on $[0,T]$. Regarding $\girs{\thisX^d}$, it is enough to prove that, for any continuous $(\mathbb{F},\mathbb{Q})$ martingale {$\thisY$} starting from 0,
%starting from 0, 
${\girs{\thisX^d }} \thisY $ is in $\mathcal{M}_{[0,T]}(\mathbb{F},\mathbb{Q}).$ \footnote{Cf.~\shortciteN[Theorem 7.34]{HeWangYan92}.}

In fact, for any $\mathbb{F}$ stopping time $\theV \leq T$ reducing the concerned processes to integrability, as in the proof of Lemma \ref{Gdense}, we can write:
$$
\EQ [ {\girs{\thisX^d}_\theV} \thisY_\theV ]
=\EP [[\thisY\monp -\thisY_-\centerdot \monp ,\thisX^d]_\theV ],
$$
where, by the integration by parts formula {on $[0,T]$},
$$
[\thisY\monp -\thisY_-\centerdot \monp ,\thisX^d]
=[\monp _-\centerdot \thisY+[\thisY,\monp ],\thisX^d]=0,
$$
because $\monp _-\centerdot \thisY+[\thisY,\monp ]$ is continuous, like $Q$. By \shortciteN[Theorem 4.40]{HeWangYan92}, this proves that 
${\girs{\thisX^d }} \thisY $ is in $\mathcal{M}_{[0,T]}(\ff,\Q).$ 
%for any continuous $(\mathbb{F},\mathbb{Q})$ martingale {$\thisY$} starting from 0. 
\finproofs

\bl\label{Apre}
We have the following relationships between the measure change density processes $\monq, \monp$
and their stochastic logarithms $\therho, {\thenu}:$ 
%We have on $[0,T]$ 
\beql{e:idq}
\therho^{\thec}=-\girs{\thenu^c} \sp
\Delta\therho^{\thed}=-{\monq }\Delta \monp \mbox{ on $[0,T]$ }.
%\r{\Delta\therho^{\thed}=-{\monq }\Delta \monp}.
\eeql
\el
\proof 
Note that, for any bounded $\mathbb{F}$ predictable process $K$ and
$\mathbb{F}$ predictable stopping time $\theV \leq T$,
$$
\EQ[K_\theV \monq _\theV \Delta_\theV \monp ]
%]
=\EP[K_\theV \Delta_\theV \monp ]=0.
$$
As a consequence, by Theorem 7.42 in \shortciteN{HeWangYan92}, there exists an $(\mathbb{F},\mathbb{Q})$ purely discontinuous local martingale $\mathbf{q}$\index{q@$\mathbf{q}$} on $[0,T]$ such that \beql{e:bfq}\Delta_s\mathbf{q}={\monq _s}\Delta_s\monp .\eeql

By Lemma 3.4 in \citeN{KaratzasKardaras07}, we have the following
relationship between
$\therho$ and $\thenu $:
\beql{e:kk}
&\therho=
-\thenu +\cpq{\thenu ^c,\thenu ^c}+\sum_{s\leq\cdot}\frac{(\Delta_s\thenu )^2}{1+\Delta_s\thenu }.
\eeql
Moreover, on $[0,T]$, we have
\begin{eqnarray}\label{e:qfracp}
\Delta_t\mathbf{q}=\frac{\Delta_t\monp }{\monp _t}
=\frac{\monp _{t-}\Delta_t \thenu }{\monp _{t-}+\Delta_t\monp }
=\frac{\Delta_t \thenu }{1+\Delta_t\thenu }\mbox{ , hence }
[\mathbf{q},\thenu ^d]=
\sum_{s\leq\cdot}\frac{(\Delta_s\thenu )^2}{1+\Delta_s\thenu }.
\end{eqnarray}
Using \qr{e:kk} and \qr{e:qfracp}, we obtain
$$
\dcb
\therho
&=&
-\thenu +\cpq{\thenu ^c,\thenu ^c} +[\mathbf{q},\thenu ^d]=
-\thenu ^c+\cpq{\thenu ^c,\thenu ^c}-\thenu ^d+[\mathbf{q},\thenu ^d]=-\girs{\thenu ^c} -\thenu ^d+[\mathbf{q},\thenu ^d]
\dce
$$
on $[0,T]$ (cf.~\qr{e:girs}).
As a consequence together with Lemma \ref{cdisc}, 
%he first identity in follows.
we have $\therho^{\thec}=-\girs{\thenu ^c}.$
In addition, by \qr{e:kk},
$$
\Delta_t\therho^{\thed}=\Delta_t\therho=-\Delta_t\thenu +\frac{(\Delta_t\thenu )^2}{1+\Delta_t\thenu }=-\frac{\Delta_t\thenu }{1+\Delta_t\thenu }=-\Delta_t\mathbf{q},
$$
by the first part in \qr{e:qfracp}.
Since $\therho^{\thed}$ and $(-\mathbf{q})$ are both $(\mathbb{F},\mathbb{Q})$ purely discontinuous local martingales, they coincide on $[0,T],$ by Corollary 7.23 in \shortciteN{HeWangYan92}. In view of \qr{e:bfq},
this proves 
%his finishes the proof of \qr{e:idq}
that $\Delta\therho^{\thed}=-{\monq }\Delta \monp$.~\finproofs

\section{Characterization of Invariance Measures 
via the Girsanov and Jeulin-Yor Formulas}\label{proofinGJY}
 
%Our original proof of Theorem \ref{part1}, using the reduction 
%methodology based on Lemma \ref{optionalsplitting} 4),
%does not directly explain how a Girsanov drift can compensate a Jeulin-Yor drift. Given the importance of that matter for the purpose of this work, we provide in this final section of the paper
%an alternative proof of Theorem \ref{part1} directly based on this compensation.

In this section, we combine the measure change lemmas \ref{Gdense} 
through \ref{Apre} (still using the notation introduced in \sr{ss:equiv}) with enlargement of filtration computations for obtaining our second 
proof of Theorem \ref{part1}. 
The goal is to relate the invariance measure property to (\ref{e:cor}). 
%Let $\thisX$ be any $(\mathbb{F},\mathbb{P})$ local martingale. 
Passing from $(\mathbb{F},\mathbb{P})$ to $(\mathbb{F},\mathbb{Q})$ creates a Girsanov's drift while passing from $(\mathbb{F},\mathbb{Q})$ to $(\mathbb{G},\mathbb{Q}),$ in combination with stopping before $\theta$, creates a Jeulin-Yor's drift. Therefore, to have the invariance probability property, the two drifts must cancel out each other. The projection methodology, a classical approach systematically used in \citeN{Jeulin1980}, can then be used then to transform this cancellation condition in the filtration $\mathbb{G}$ into a cancellation condition in the filtration $\mathbb{F}$, which, with the help of Lemmas \ref{Gdense} through \ref{Apre}, will be seen to be equivalent to (\ref{e:cor}).

Recall $\ttJ=\ind_{[0,\theta)}$. 
For any $\thisX\in\mathcal{M}(\mathbb{F},\mathbb{P}),$ 
the Girsanov formula (\ref{Gf1}) says that $\thisX-\monq\centerdot[\monp ,\thisX]\in\mathcal{M}_{[0,T]}(\mathbb{F},\mathbb{Q})$. 
The Jeulin-Yor formula
\qr{e:jyt} applied to $Q=\thisX-\monq\centerdot[\monp ,\thisX]$ then yields that $$
\dcb
&(\thisX-\monq\centerdot[\monp ,\thisX])^{\ftime -}
-\ttJ_{-}\frac{1}{\tSigma _-}\centerdot \croc{\cM, {\thisX}-\monq \centerdot[\monp ,\thisX]}\\
&\qqq=
\thisX^{\ftime -}-\big(\ttJ \monq \centerdot[\monp ,\thisX]
+\ttJ_{-}\frac{1}{\tSigma _-}\centerdot \croc{\cM, {\thisX}-\monq \centerdot[\monp ,\thisX]}\big) \in \mathcal{M}_{[0,T]}(\mathbb{G},\mathbb{Q}).
\dce
$$ 
%\b{Hence, the composition of the Girsanov and of the Jeulin-Yor formulas reveals a $(\mathbb{G},\mathbb{Q})$ local martingale on $[0,T]$
%stopped before $\theta$ generically associated with any $(\mathbb{F},\mathbb{P})$ local martingale $\thisX$, namely the right-hand side above. 
Therefore the invariance measure property that 
the first term $\thisX^{\ftime -}$ in the right-hand-side is in $\mathcal{M}_{[0,T]}(\mathbb{G},\mathbb{Q})$ is equivalent to the condition that the second term (in parentheses) is in $\mathcal{M}_{[0,T]}(\mathbb{G},\mathbb{Q}),$ for any $P\in\mathcal{M} (\mathbb{F},\mathbb{P})$. 
%Projecting this condition on $\ff$ will, by a density argument, yield (\ref{e:cor}).
In other words, we conclude from the above that
%Hence,
$\mathbb{P}$ is an invariance measure 
%(i.e. the condition $(A)$ is satisfied) 
if and only if, for any $\thisX\in\mathcal{M}(\mathbb{F},\mathbb{P}),$ 
\begin{equation}\label{themartingale}
\ttJ \monq \centerdot[\monp ,\thisX]
+\ttJ_{-}\frac{1}{\tSigma _-}\centerdot \croc{\cM, {\thisX}-\monq \centerdot[\monp ,\thisX]}\in \mathcal{M}_{[0,T]}(\mathbb{G},\mathbb{Q}).
\end{equation} 

In the next steps, we derive, by projection, an $\mathbb{F}$ counterpart of the condition (\ref{themartingale}). For doing so, we need an expression of the condition (\ref{themartingale}) in the form of expectations. 
In fact, (\ref{themartingale})
is equivalent to 
%the following integral exists and 
\beql{themartingaleaux}
&\mbox{$\exists$ a nondecreasing sequence of $\mathbb{G}$ stopping times $(\tau_n)_{n\in\mathbb{N}}$ tending to}\\& \mbox{infinity such that, $\forall$ bounded $\mathbb{G}$ predictable process $L$ and $n\in\mathbb{N}$,}\\
&\EQ[L\ttJ \monq \centerdot[\monp ,\thisX]_{\tau_n\wedge T}
+L\ttJ_{-}\frac{1}{\tSigma _-}\centerdot \croc{\cM, {\thisX}-\monq \centerdot[\monp ,\thisX]}_{\tau_n\wedge T}]
=0.
\eeql

By Lemma \ref{optionalsplitting} 1), the nondecreasing sequence of $\mathbb{G}$ stopping times $(\tau_n)_{n\in\mathbb{N}}$
is associated by reduction with a nondecreasing sequence of $\mathbb{F}$ stopping times $(\sigma_n)_{n\in\mathbb{N}}$ such that, for every $n\in\mathbb{N}$, $\theta\wedge \sigma_n=\theta\wedge \tau_n$. We have necessarily $\cup_{n\in\mathbb{N}}[0,\sigma_n]\supseteq [0,\theta]$. By \citeN[Lemma (4,3)]{Jeulin1980}, $
\{{^{p}}(\ind_{(\theta,\infty)})=1\}
$
is the largest $\mathbb{F}$ predictable set contained in $(\theta,\infty)$. As a consequence, $
\{{^{ p}}(\ind_{(\theta,\infty)})=1\}
 \supseteq ( \R_+\times\Omega)\setminus (\cup_{n\in\mathbb{N}}[0,\sigma_n])
$
or, equivalently 
%\s{??why $<\infty$}
\begin{equation}\label{S-U}
\{0\}\cup \{\ttS_->0\}
=
\{0\}\cup \{1-\ttS_-<1\}
=
\{{^{p}}(\ind_{(\theta,\infty)})<1\}\subset\cup_{n\in\mathbb{N}}[0,\sigma_n].
\end{equation}
Conversely, if (\ref{S-U}) holds for a nondecreasing sequence of $\mathbb{F}$ stopping times $(\sigma_n)_{n\in\mathbb{N}}$, then, 
in view of (\ref{yortheta}),
%\beql{yortheta} 
%\ttS_{\theta	-}>0, \mbox{ i.e. } \theta<\varsigma, \mbox{ on }\{0<\theta\} 
%\eeql 
the nondecreasing sequence of $\mathbb{G}$ stopping times $\tau_{n}=(\sigma_n)_{\{\sigma_n<\theta\}}$ tends to infinity.
In addition, by the condition (B), the bounded $\mathbb{G}$ predictable processes $L$ are associated with the bounded $\mathbb{F}$ predictable processes $K$ through the reduction identity $K\ind_{(0,\theta]}=L\ind_{(0,\theta]}$. 
%there exists a bounded $\mathbb{F}$ predictable process $K$ such that $K\ind_{(0,\theta]}=L\ind_{(0,\theta]}$. 

Through these correspondences and the projection identities ${^{o}}(\ttJ )=\ttS$ and ${^{p}}(\ttJ_- )=\ttS_-$ on $(0,\infty)$, 
%the expectation in 
\qr{themartingaleaux} can be rewritten in terms of 
$\ff$ stopping times and adapted processes, as
\beql{KUN=0}
&\mbox{$\exists$ a nondecreasing sequence of $\mathbb{F}$ stopping times $(\sigma_n)_{n\in\mathbb{N}}$ satisfying}\\& \mbox{\qr{S-U} such that, $\forall$ bounded $\mathbb{F}$ predictable process $K$ and $n\in\mathbb{N}$,}\\&
\EQ[K\ttS \monq \centerdot[\monp ,\thisX]_{\sigma_n\wedge T}
+K\centerdot\croc{\cM, {\thisX}-\monq \centerdot[\monp ,\thisX]}_{\sigma_n\wedge T}]
=0.
\eeql

We are now transferred into the filtration $\mathbb{F}$. To establish the desired connection with (\ref{e:cor}), we interpret \qr{KUN=0} as a local martingale condition in $\ff$. In fact, (\ref{KUN=0}) is equivalent to the following $\ff$ counterpart of (\ref{themartingale}):
%that
%Hence, \qr{themartingale} is equivalent to the existence of a nondecreasing sequence of $\mathbb{F}$ stopping times $(\sigma_n)_{n\in\mathbb{N}}$ satisfying the condition (\ref{S-U}) such that, for any bounded $\mathbb{F}$ predictable process $K$ and $n\in\mathbb{N}$, \qr{KUN=0} is satisfied, i.e.
\beql{KUN=1}&\ttS \monq \centerdot[\monp ,\thisX]
+\croc{\cM, {\thisX}-\monq \centerdot[\monp ,\thisX]}
%\mbox{ and } \ttS \monq \centerdot[\monp ,\thisX]+[\cM, {\thisX}-\monq \centerdot[\monp ,\thisX]]
 \in\mathcal{M}_{\{\ttS_->0\}\cap[0,T]}(\mathbb{F},\mathbb{Q}),\eeql
which is equivalent to
%This analysis shows that the probability measure $\mathbb{P}$ is an invariance measure if and only if \qr{KUN=1}, \b{i.e.} 
\beql{KUN=2}&\ttS \monq \centerdot[\monp ,\thisX]
+[\cM, {\thisX}-\monq \centerdot[\monp ,\thisX]]\in\mathcal{M}_{\{\ttS_->0\}\cap[0,T]}(\mathbb{F},\mathbb{Q}). \eeql 
%holds for any $\thisX\in\mathcal{M}(\mathbb{F},\mathbb{P}).$ 
Operating separately on the continuous parts $P^c$ and the purely discontinuous parts $P^d$ of the $(\mathbb{F},\mathbb{P})$ local martingales $\thisX,$ 
we conclude that $\mathbb{P}$ is an invariance measure if and only if, for any $\thisX\in\mathcal{M}(\mathbb{F},\mathbb{P}),$
\begin{equation}\label{SqpP}
\left\{
\dcb
\ttS \monq \centerdot\langle\monp^c ,\thisX^c\rangle
+\langle \cM^c, {\thisX}^c-\monq \centerdot\langle\monp ,\thisX^c\rangle\rangle\in\mathcal{M}_{\{\ttS_->0\}\cap[0,T]}(\mathbb{F},\mathbb{Q}),\\
\\
\ttS \monq \centerdot[\monp^d ,\thisX^d]
+[\cM^d, {\thisX}^d-\monq \centerdot[\monp ,\thisX^d]]\in\mathcal{M}_{\{\ttS_->0\}\cap[0,T]}(\mathbb{F},\mathbb{Q}).
\dce
\right.
\end{equation}

We are now in a position to deduce the equivalent condition (\ref{e:cor}).
Using
the identities
${\pSigma }=\ttS-\Delta \cM $
and $\therho^{\thec}=-\girs{\thenu^c}$
(cf. \qr{e:SS} and Lemma \ref{Apre}), by continuity, we have
$$
\dcb
&&\ttS \monq \centerdot\langle\monp^c ,\thisX^c\rangle
+\langle \cM^c, {\thisX}^c-\monq \centerdot\langle\monp ,\thisX^c\rangle\rangle
=
\pSigma \centerdot\langle\thenu^c,\thisX^c\rangle
+\langle \cM^c, {\thisX}^c-\monq \centerdot\langle\monp ,\thisX^c\rangle\rangle\\
&&\qqq=
\pSigma \centerdot\langle\girs{\thenu^c} ,\girs{\thisX^c}\rangle
+\langle \cM^c, \girs{\thisX^c}\rangle
=
\langle -\pSigma \centerdot\therho^c +\cM^c, \girs{\thisX^c}\rangle.
\dce
$$ 
Hence,
the first line in (\ref{SqpP}) means that
$$
\dcb
&\langle -\pSigma \centerdot\therho^c +\cM^c, \girs{\thisX^c}\rangle\in\mathcal{M}_{\{\ttS_->0\}\cap[0,T]}(\mathbb{F},\mathbb{Q}).
\dce
$$ 
This holding for all the $(\mathbb{F},\mathbb{P})$ local martingales $\thisX,$ including the bounded ones for which 
$$
\langle -\pSigma \centerdot\therho^c +\cM^c, \girs{\thisX^c}\rangle
=\langle -\pSigma \centerdot\therho^c +\cM^c, \girs{\thisX}\rangle
$$
(cf.~Lemma \ref{cdisc}),
is equivalent by Lemma \ref{Gdense} to
\begin{equation}\label{continEEE}
-\pSigma \centerdot\therho^c +\cM^c =0 \mbox{ on } \{\ttS_->0\}\cap[0,T].
\end{equation} 
Likewise,
we have
$$
\dcb
&&\ttS \monq \centerdot[\monp^d ,\thisX^d]
+[\cM^d, {\thisX}^d-\monq \centerdot[\monp ,\thisX^d]]
=
\sum(\ttS \monq \Delta\monp \Delta\thisX
+\Delta\cM (\Delta{\thisX} - \monq \Delta\monp \Delta\thisX))\\
&&\quad = 
\sum\big((\ttS-\Delta \cM\big) \monq \Delta\monp \Delta\thisX
+\Delta\cM \Delta{\thisX})
=
-\pSigma \centerdot[\therho^d,\thisX]
+[\cM^d,{\thisX}]
=
[-\pSigma \centerdot\therho^d+\cM^d,{\thisX}], 
%\r{\sum\big((\ttS-\Delta \cM\big) \monq \Delta\monp \Delta\thisX
%+\Delta\cM \Delta{\thisX})
%=
%-\pSigma \centerdot[\therho^d,\thisX]
%+[\cM^d,{\thisX}]
%=
%[-\pSigma \centerdot\therho^d+\cM^d,{\thisX}],} 
\dce
$$
where the identities
${\pSigma }=\ttS-\Delta\cM $
and $\Delta\therho^{\thed}=-{\monq }\Delta \monp$
(cf. \qr{e:SS} and Lemma \ref{Apre}) are used in the next-to-last equality.
Hence, 
the second line in (\ref{SqpP}) means that
$$[-\pSigma \centerdot\therho^d+\cM^d,{\thisX}]\in\mathcal{M}_{\{\ttS_->0\}\cap[0,T]}(\mathbb{F},\mathbb{Q}).$$ For $\thisX$ bounded
(so that the predictable bracket Girsanov formula \qr{e:girs} is applicable),
in view of Yoeurp's lemma, this means that
$$
[-\pSigma \centerdot\therho^d+\cM^d,\girs{\thisX}]\in\mathcal{M}_{\{\ttS_->0\}\cap[0,T]}(\mathbb{F},\mathbb{Q}).
$$
Using Lemma \ref{Gdense},
we conclude that the second line in (\ref{SqpP}) holds 
for all the $(\mathbb{F},\mathbb{P})$ local martingales $\thisX$ (including the bounded ones)
if and only if
\begin{equation}\label{discontinEEE}
-\pSigma \centerdot\therho^d+\cM^d =0\mbox{ on } \{\ttS_->0\}\cap[0,T].
\end{equation} 
%Conversely, if (\ref{discontinEEE}) holds, then the second equation in (\ref{SqpP}) holds, which are equivalent. 
 
%Putting \qr{continEEE}
%and \qr{discontinEEE} together, we conclude that $\mathbb{P}$ is an invariance measure if and only if
%$$-\pSigma \centerdot\therho^c +\cM^c =-\pSigma \centerdot\therho^d+\cM^d =0\mbox{ on }\{\ttS_->0\}\cap[0,T].$$
By Lemma \ref{QS=0}, $\cM$ is constant on $\{\ttS_-=0\}\subset \{\pSigma=0\}$.
Hence, (\ref{continEEE}) and (\ref{discontinEEE}) are respectively the continuous and 
purely discontinuous parts of (\ref{e:cor}), so that
Theorem \ref{part1} is proved.

%Putting \qr{continEEE}
%and \qr{discontinEEE} together, we conclude that $\mathbb{P}$ is an invariance measure if and only if
%$$-\pSigma \centerdot\therho^c +\cM^c =-\pSigma \centerdot\therho^d+\cM^d =0\mbox{ on }\{\ttS_->0\}\cap[0,T].$$
%By Lemma \ref{QS=0}, $\cM$ is constant on $\{\ttS_-=0\}\subset \{\pSigma=0\}$.
%Hence, (\ref{continEEE}) and (\ref{discontinEEE}) are respectively the continuous and 
%purely discontinuous parts of (\ref{e:cor}), so that
%Theorem \ref{part1} is proved.

%Finally, the combination of (\ref{continEEE}) and (\ref{discontinEEE}) is equivalent to (\ref{e:cor}) (as $\cM$ is constant on $\{\ttS_-=0\}$), which proves Theorem \ref{part1}. 
% 

\subsection{A By-Product}

Inspecting the above proof of Theorem \ref{part1}, we see that the implication from the 
invariance measure property 
to (\ref{e:cor}), which is based on Lemma \ref{Gdense}, only makes use of the bounded $(\mathbb{F},\mathbb{P})$ martingales. 
We can therefore introduce a seemingly weaker condition.\\

\noindent\textbf{Condition \index{a@(A')}(A')}. 
There exists 
a probability measure 
\index{p@$\mathbb{P}$}$\mathbb{P}$ equivalent to $\mathbb{Q}$ on {${\cal F}_{T}$} 
such that, for any bounded $({\ff},\mathbb{P})$ martingale $\thisN$,
{$\thisN^{\theta-}$}
is a $(\gg,\mathbb{Q})$ local martingale {on $[0,T]$.}

\begin{cor}
The condition $(A)$ is equivalent to the condition $(A')$.
\ecor

\end{document}